\documentclass[fleqn,usenatbib]{mnras}

\usepackage{newtxtext,newtxmath}
\usepackage{fancyhdr}

\usepackage[T1]{fontenc}
\usepackage{ae,aecompl}

\usepackage{graphicx}	            
\usepackage{amsmath}	            
\usepackage{caption}                
\usepackage{pdfpages}               
\usepackage{float}                  
\usepackage{longtable}
\usepackage{lscape}

\title[Dense core alignment with magnetic fields]{Alignment of dense molecular core morphology and velocity gradients with ambient magnetic fields}

\author[A. Pandhi et al.]{
A. Pandhi$^{1, 2}$,
R. K. Friesen$^{1}$,
L. Fissel$^{3}$,
J. E. Pineda$^{4}$,
P. Caselli$^{4}$,
M. C-Y. Chen$^{3}$,
\newauthor
J. Di Francesco$^{5, 6}$,
A. Ginsburg$^{7}$,
H. Kirk$^{5,6}$,
P. C. Myers$^{8}$,
S. S. R. Offner$^{9}$,
\newauthor
A. Punanova$^{10}$,
F. Quan$^{11}$,
E. Redaelli$^{4}$,
E. Rosolowsky$^{12}$,
S. Scibelli$^{13}$,
Y. M. Seo$^{14}$,
\newauthor
Y. Shirley$^{13}$
\\
$^{1}$David A. Dunlap Department of Astronomy and Astrophysics, University of Toronto, 50 St. George Street, Toronto, ON M5S 3H4, Canada\\
$^{2}$Dunlap Institute for Astronomy and Astrophysics, University of Toronto, 50 St. George Street, Toronto, ON M5S 3H4, Canada\\
$^{3}$Department of Physics, Engineering Physics \& Astronomy, Queen's University, 64 Bader Lane, Kingston, ON K7L 3N6, Canada\\
$^{4}$Max-Planck-Institut f\"ur extraterrestrische Physik, Giessenbachstrasse 1, 85748 Garching, Germany\\
$^{5}$Herzberg Astronomy and Astrophysics Research Centre, National Research Council of Canada, 5071 West Saanich Road, Victoria, BC, V9E 2E7, Canada\\
$^{6}$Department of Physics and Astronomy, University of Victoria, 3800 Finnerty Road, Victoria, BC, V8P 5C2, Canada\\
$^{7}$Department of Astronomy, University of Florida, P.O. Box 112055, Gainesville, FL, USA\\
$^{8}$Harvard-Smithsonian Center for Astrophysics, 60 Garden St., Cambridge, MA 02138, USA\\
$^{9}$Department of Astronomy, The University of Texas at Austin, TX 78712, USA\\
$^{10}$Onsala Space Observatory, Chalmers University of Technology, Observatorievägen 90, Råö, 439 92 Onsala, Sweden\\
$^{11}$Department of Physical and Environmental Sciences, University of Toronto at Scarborough, Toronto, ON M1C 1A4, Canada\\
$^{12}$Department of Physics, University of Alberta, Edmonton, Alberta, T6G2E1, Canada\\
$^{13}$Steward Observatory, University of Arizona, Tucson, AZ 85721, USA\\
$^{14}$Jet Propulsion Laboratory, California Institute of Technology, 4800 Oak Grove Dr. Pasadena, CA, 91109, USA}

\begin{document}
\maketitle
\fancyhf{} 
\renewcommand{\headrulewidth}{0pt}
\thispagestyle{fancy}
\rhead{Compiled using MNRAS \LaTeX\ style file v3.0}

\begin{abstract}
Studies of dense core morphologies and their orientations with respect to gas flows and the local magnetic field have been limited to only a small sample of cores with spectroscopic data. Leveraging the Green Bank Ammonia Survey alongside existing sub-millimeter continuum observations and \textit{Planck} dust polarization, we produce a cross-matched catalogue of 399 dense cores with estimates of core morphology, size, mass, specific angular momentum, and magnetic field orientation. Of the 399 cores, 329 exhibit 2D $\mathrm{v}_\mathrm{LSR}$ maps that are well fit with a linear gradient, consistent with rotation projected on the sky. We find a best-fit specific angular momentum and core size relationship of $J/M \propto R^{1.82 \pm 0.10}$, suggesting that core velocity gradients originate from a combination of solid body rotation and turbulent motions. Most cores have no preferred orientation between the axis of core elongation, velocity gradient direction, and the ambient magnetic field orientation, favouring a triaxial and weakly magnetized origin. We find, however, strong evidence for a preferred anti-alignment between the core elongation axis and magnetic field for protostellar cores, revealing a change in orientation from starless and prestellar populations that may result from gravitational contraction in a magnetically-regulated (but not dominant) environment. We also find marginal evidence for anti-alignment between the core velocity gradient and magnetic field orientation in the L1228 and L1251 regions of Cepheus, suggesting a preferred orientation with respect to magnetic fields may be more prevalent in regions with locally ordered fields. 
\end{abstract}

\endkeywords

\section{Introduction} \label{intro}
Investigating the processes that underlie star formation is imperative for understanding stellar and planetary evolution. Stars are formed in local over-densities within molecular clouds (MCs) called dense cores, with typical radii $\lesssim 0.1~\mathrm{pc}$ and masses of $\sim 0.1 - 10~M_{\odot}$ \citep[for reviews, see][and references therein]{2000prpl.conf...59A, 2007prpl.conf...17D, 2022arXiv220503935P}. The combination of gravity, magnetic fields, and turbulence are the foremost drivers of the dynamics of dense cores \citep[e.g.,][]{1987ARA&A..25...23S, 2007prpl.conf...63B, 2007ARA&A..45..565M}. If magnetic fields are important in the dynamic evolution of MCs, then we expect to see some level of preferential alignment between cloud structure, and potentially also core kinematics, with the magnetic field \citep{2022arXiv220311179P}. On the MC scale, the cloud-scale ambient magnetic field appears to be preferentially oriented perpendicular to high-density filamentary structure and parallel to low-density striations, indicating its importance in regulating gas flows towards high-density MC regions \citep{2012ARA&A..50...29C, 2013A&A...550A..38P, 2016A&A...586A.138P}. In the transition to smaller scales, there is some observational evidence in the Serpens South region to show that magnetic fields warp and return to a parallel alignment in the highest density regions of the filament \citep{pillai_2020}. On core scales, the magnetic field is expected to remove angular momentum from collapsing core systems via magnetic braking and could restrain, or even prevent, proto-planetary disk formation \citep{2008ApJ...681.1356M} and suppress binary formation \citep[e.g.,][]{2004MNRAS.347.1001H, 2005A&A...435..385Z, 2006A&A...457..371F, 2007MNRAS.377...77P}. For a review of the role of magnetic fields in the formation of protostars and disks, see \cite{2022arXiv220913765T}. In the presence of ambipolar diffusion, especially important if very small grains are removed via adsorption onto larger grains \citep[e.g.,][]{2020A&A...641A..39S}, the magnetic flux decreases within the core allowing the formation of rotationally supported proto-planetary disks \citep[e.g.,][]{2016MNRAS.460.2050Z, 2018MNRAS.473.4868Z}. Observations of relative alignment of core orientation and rotation with the local magnetic field, however, have only been studied for a small sample of cores \citep[e.g.,][]{2020A&A...640A.111A, 2020MNRAS.494.1971C, 2021ApJ...907...33Y, 2022MNRAS.517.1138S}, underscoring the open question of the role played by magnetic fields on core-scales. Uncovering whether magnetic fields remain dynamically important in the transition from cloud scales to core scales, or if their role is diminished with respect to gravity and turbulence, will lead to a more complete picture of MC fragmentation into cores and the early stages of star formation.

\cite{2014prpl.conf...27A} emphasize the correlation between star-forming MCs and filamentary structure in \textit{Herschel} observations. The physical scenario in which large-scale supersonic flows compress the gas into filaments and then gravity dominates in the densest regions and creates dense cores is consistent with many simulations of turbulent clouds and their internal star formation \citep[e.g.,][]{2014ApJ...791..124G, 2015ApJ...806...31G}. Velocity gradients have been observed at both cloud and core scales and can be attributed to the rotation of clouds or cores, or other ordered flows, such as accretion or collapse. Multiple studies have previously measured the specific angular momentum $J/M$ as a function of core size $R$ and observed an apparent loss of specific angular momentum from cloud scales to core scales \citep[e.g.,][]{1993ApJ...406..528G, 2002ApJ...572..238C, 2003A&A...405..639P, 2007ApJ...669.1058C, 2011ApJ...740...45T, 2015ApJ...799..193Y, 2019MNRAS.490..527C}. Further, \cite{2019ApJ...882..103P} study the specific angular momentum radial profile within individual cores on 800 to 10,000~au scales. For $\gtrsim 100$ gravitationally bound cores in 3D, turbulent MHD simulations, \cite{2018ApJ...865...34C} show that the power-law relation between $J/M$ and $R$ extends down to $\sim 10^{-3}~\mathrm{pc}$. Though limited in sample size, this set of observations and simulations highlight the unresolved problem of how angular momentum is redistributed during dense core gravitational collapse and fragmentation. 

Estimating the magnetic field strength on core scales is challenging as it requires either observations of the Zeeman effect \cite[e.g.,][]{1993ApJ...407..175C} which are resource intensive and restricted to specific spectral lines, or the presence of a coincident, background polarized transient source inducing Faraday rotation \citep[e.g., using pulsars or fast radio bursts][]{2020MNRAS.496.2836N, 2022MNRAS.516.4739P} which is unlikely. We can, however, readily measure the magnetic field direction in the plane of the sky, averaged along the line of sight, from existing observations. Elongated dust grains can have their longer axes preferentially oriented perpendicular to the local magnetic field \citep{1951ApJ...114..206D, 2007JQSRT.106..225L} and therefore the sub-mm polarized thermal emission from dust is an excellent tracer of the local plane-of-sky magnetic field $B_{\perp}$ morphology \citep[e.g.,][]{1993prpl.conf..279H}. On large scales, the \textit{Planck} Collaboration produced all-sky maps of the linearly polarized dust emission at $353$~GHz with a resolution of $1^\circ$ \citep{2015A&A...576A.104P}, and provide the $B_{\perp}$ geometry for cloud-scale magnetic fields.

There is a wealth of dense core MHD simulations examining the relative alignment between core major axis orientation and the ambient magnetic field. \cite{2017ApJ...834..201L} look at 3D MHD simulations of low-mass cores and show that the outflow direction (tracing the angular momentum direction) of weakly magnetized cores tends to align randomly with the local magnetic field, while more strongly magnetized cores do exhibit alignment between the outflow and local magnetic field vectors. \cite{2018ApJ...865...34C} see that cores tend to be triaxial with the core-scale magnetic field most perpendicular to the core major axis and no preferred alignment between the magnetic field and core angular momentum orientation. In MHD simulations, \cite{2020MNRAS.494.1971C} examine dense structures in cloud-cloud collisions (specifically a magnetized shocked layer produced by plane-parallel converging flows; see their Section 2.1 for details) and find a marginal preference for cores to align with the core-scale magnetic field$-$indicating material flows along field lines and is accreted onto cores$-$but no preferred orientation relative to the cloud-scale magnetic field. This result would imply a disconnect between cloud- and core-scale magnetic field structure. Using sink-patch implementation in the ideal MHD simulations, \cite{2020ApJ...893...73K} see that magnetic fields assist in depositing material onto dense cores by collimating low-density gas flows, but the mass accretion is highly sporadic with short-term variability and no long-term growth of the specific angular momenta. Further, they find that the relative angle between the spin axis of the core and the local core-scale magnetic field is consistent with being randomly distributed at a $99.5\%$ confidence level.
 
Observations of the relative alignment of cores and magnetic fields tend to suffer from small sample sizes (on the order of tens of cores), making it difficult to draw strong, statistically robust conclusions. There have been, however, some recent observational efforts on this front. In general, cores tend to be largely randomly aligned with the cloud-scale magnetic field \citep[][]{2020MNRAS.494.1971C, 2022MNRAS.517.1138S}, although the relative alignment may vary between regions, with clouds containing ordered magnetic fields showing preferential anti-alignment \citep[e.g., as seen in Taurus by][]{2020MNRAS.494.1971C}. \cite{2022ApJ...941...81X} find that across a sample of 200 protostellar outflows, the plane-of-sky outflow direction tends to align with the \textit{Planck} cloud-scale magnetic field. The difference between these results may hint that preferred alignment with the cloud-scale magnetic field varies with the evolutionary stage of dense cores. Some studies have measured the core-scale magnetic field orientation using high-resolution polarimetry with the Atacama Large Millimeter Array (ALMA) and the James Clerk Maxwell Telescope (JCMT) B-fields In STar-forming Region Observations (BISTRO) survey. With these observations, we see a trend for the core-scale magnetic field to be aligned with the observed core major-axis \citep{2020A&A...640A.111A} and misaligned or randomly aligned with the outflow direction \citep{2020A&A...640A.111A, 2021ApJ...907...33Y}.

In this paper, we undertake a systematic analysis of the gas kinematics, specific angular momenta, and relative alignments of core elongation, velocity gradient direction, and local cloud-scale magnetic field orientation for 399 dense cores, across seven clouds, identified in the Green Bank Ammonia Survey (GAS) and cross-matched with continuum observations from \textit{Herschel} and JCMT. These continuum data provide core orientation, size, and mass, while the GAS NH$_3$ spectral line observations provide line-of-sight velocity information of gas resolved on core scales. The Planck observations, however, provide magnetic field orientation information on cloud scales. This combined data set allows us to build up a catalogue of velocity gradients across cores in the Gould Belt and characterize their rotation and specific angular momenta. Such analysis has been, as of yet, only conducted on a limited amount of data or in simulations \citep[e.g.,][]{1993ApJ...406..528G, 2002ApJ...572..238C, 2003A&A...405..639P, 2007ApJ...669.1058C, 2011ApJ...740...45T, 2015ApJ...799..193Y, 2018A&A...617A..27P, 2019MNRAS.490..527C, 2020ApJ...893...73K, 2022ApJ...925...78A}. Moreover, we analyze whether the relative alignments of core elongation, velocity gradient direction, and ambient magnetic field orientation have any similarities or differences globally, in individual regions, or as a function of the core type classification. These findings are compared to prior work on MHD simulations \citep[e.g.,][]{2017ApJ...834..201L, 2018ApJ...865...34C, 2020MNRAS.494.1971C, 2020ApJ...893...73K} and observations \citep[e.g.,][]{2020MNRAS.494.1971C, 2021ApJ...907...33Y, 2022MNRAS.517.1138S}.

The remainder of this paper is structured as follows. In Section \ref{data}, we discuss the spectral line, continuum and dust polarization data, the derived dense core properties from these data, and the cross-matching process between the spectral line and continuum dense core catalogues. The results of the cross-matching, core rotation fitting routine, angular momenta estimation, and relative orientations of core elongation, velocity gradient, and ambient magnetic field direction are presented in Section \ref{results}. In Section \ref{discussion}, we interpret the results and compare them to previous observations and simulations. We summarize our findings and identify avenues for future work in Section \ref{conclusions}. A summarized version of the cross-matched catalogue is presented in Appendix \ref{app:catalog} and instructions for accessing the full catalogue are provided in Section \ref{data_avail}.

\section{Data and derived properties} \label{data}
\subsection{Spectral line core catalogue} \label{data_specline}
The first data set we employ is the NH$_3$ (1, 1) and (2,2) inversion transition emission observations from GAS, which have a beam size of $\sim 32''$ full width at half maximum (FWHM), therefore probing $\sim 0.02 - 0.07~\mathrm{pc}$ at the distances of $\sim 130 - 450~\mathrm{pc}$ to the clouds in our sample. NH$_3$ is excited at densities of $n \geq 2 \times 10^3~\mathrm{cm}^{-3}$ \citep[for gas at $10$~K;][]{2015PASP..127..299S}, which makes it particularly effective at tracing the kinematics of nearby regions with high volume densities similar to those we expect to see in and around dense cores. The NH$_3$ (1, 1) through (3, 3) inversion line observations were taken using the K-Band Focal Plane Array at the Robert C. Byrd Green Bank Telescope (GBT), with the Versatile GBT Astronomical Spectrometer, between $23.7$ and $23.9$~GHz. The spectral resolution is $5.7$~kHz ($\sim 0.07~\mathrm{km}~\mathrm{s}^{-1}$at 23.7~GHz). Most GAS maps were observed in $10' \times 10'$ footprints, with a scan rate of $6.2''~\mathrm{s}^{-1}$, and combined to provide a total sky coverage of $\sim 4~\mathrm{deg}^2$. For a complete description of the GAS observations, data reduction, data products, and analysis, we refer the reader to \cite{2017ApJ...843...63F}. Here, we focus on the details regarding the dense core identification and line-of-sight velocity measurements from the GAS data products.

We classify the various layers of hierarchical structure present in the GAS integrated NH$_3$ (1,1) intensity maps using the {\tt astrodendro} package\footnote{\href{http://www.dendrograms.org/}{http://www.dendrograms.org/}.} to create dendrograms \cite[e.g.,][]{2008ApJ...679.1338R}. This algorithm identifies structure at successive isocontour levels in the intensity map and tracks the flux values at which they merge into neighbouring emission structures. Specifically, isocontour regions with the highest flux densities that contain no further sub-structure are classified as \textit{leaves}. In the context of the GAS integrated NH$_3$ intensity maps, structures classified as \textit{leaves} represent dense cores, while large-scale filamentary structure is captured by lower isocontour levels. Within each region of our sample (see Table \ref{tb:cross_match} for a list of all $17$ regions), the cores receive a unique identifier in the form of a dendrogram index, starting at $0$. While {\tt astrodendro} can identify structure in 3D (position-position-velocity; PPV) data sets, we instead use the integrated NH$_3$ intensity maps as they have higher signal-to-noise than the individual channel maps, and because the hyperfine structure of the NH$_3$ inversion lines leads to spurious structures identified in PPV space. While it is possible that this approach may combine structures that would be separated along the velocity axis, previous studies suggest that most spectra are well characterised with a single velocity component in NH$_3$ in these regions on these scales \citep{pineda_2010,sokolov_2020,2020ApJ...891...84C}. 

\begin{figure*}
    \centering
    \includegraphics[width=0.48\textwidth]{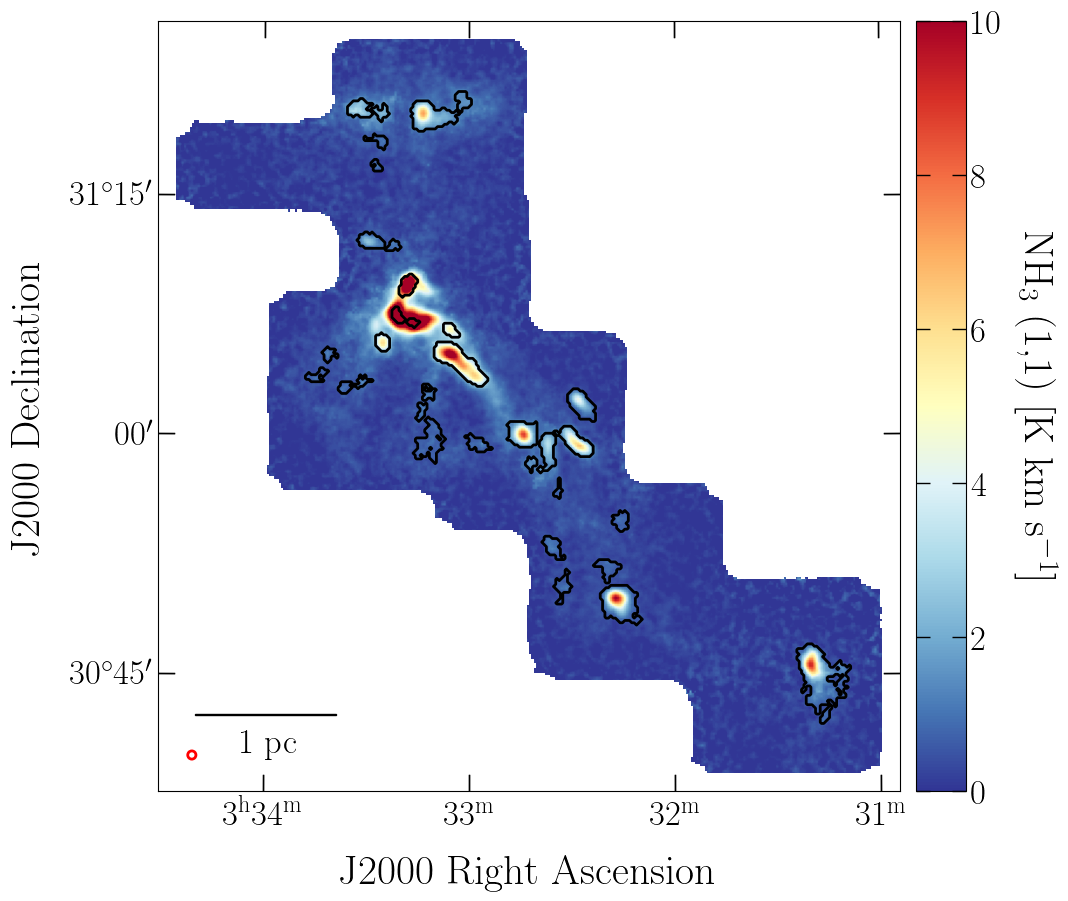}
    \includegraphics[width=0.48\textwidth]{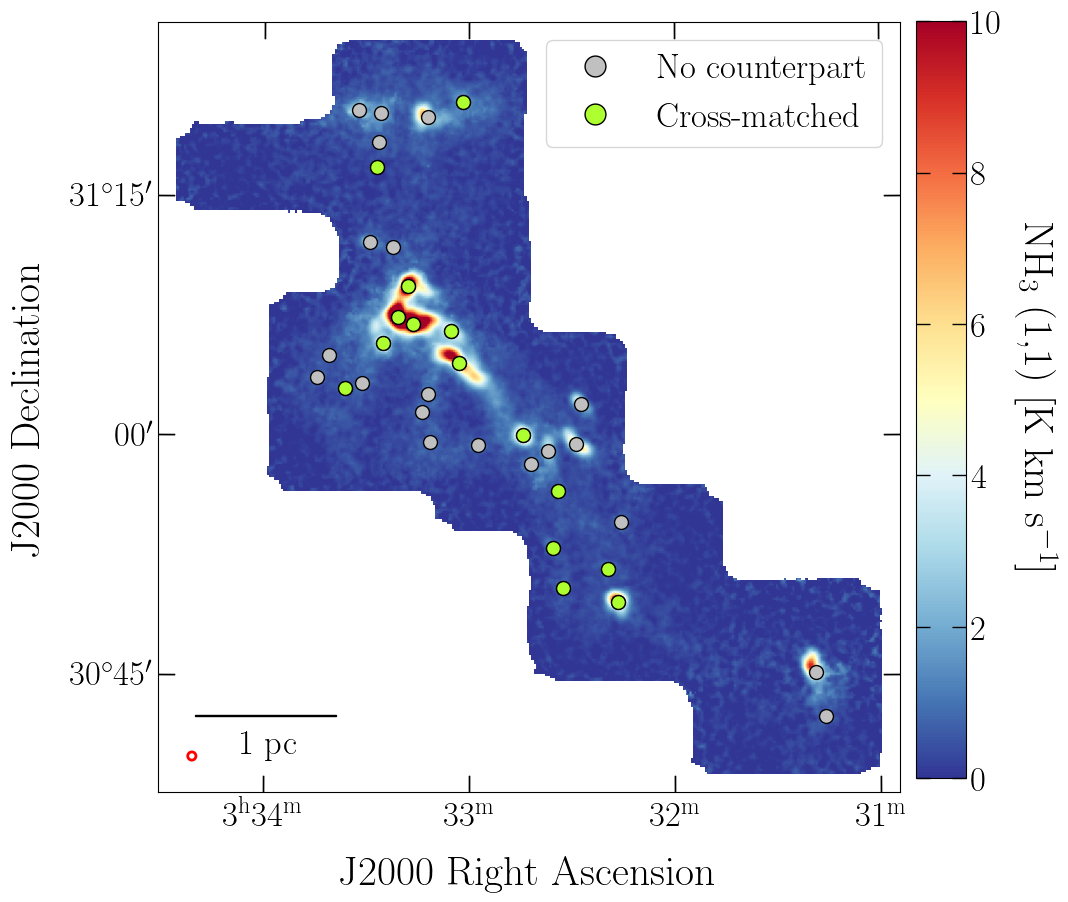}
    \caption{(Left) NH$_3$ (1,1) integrated intensity map of the B1 Perseus region with identified dense cores (classified as \textit{leaves} in {\tt astrodendro}) outlined in black. There are 35 identified cores in this region of the GAS data. A scale bar depicting the angular size of $1$~pc at the distance of the B1 Perseus region is presented in the bottom left corner. Also in the bottom left corner is a red ellipse representing the 32$''$ beam size of the GAS data. (Right) The same NH$_3$ (1,1) integrated intensity map as on the left is presented after applying the cross-matching algorithm. NH$_3$ cores that were successfully cross-matched are overlaid as green circles and cores with no identified continuum counterpart are plotted as grey circles. In this region, 15 out of 35 dense cores identified in GAS were uniquely cross-matched with a continuum counterpart in HGBS data.}
    \label{fig:b1_cores}
\end{figure*}

For most regions, we apply an emission threshold of $9~\sigma$ and isocontour levels at $3~\sigma$ intervals, where $\sigma$ is the width of the skewed Gaussian profile fit to the rms noise distribution, similar to \cite{2020MNRAS.494.1971C}. Due to larger variation in noise properties, the L1688 and L1689 regions of Ophiuchus use $5~\sigma$ contour intervals \citep[for a more detailed description of the noise properties, see][ and Pineda et al. in preparation]{2017ApJ...843...63F}. The first panel of Figure \ref{fig:b1_cores} illustrates the GAS core identification in the B1 region of Perseus. The identified cores are outlined as black contours overlaid upon the integrated NH$_3$ intensity. 

Based on the NH$_3$ line fits, we obtain a 2D map of $\mathrm{v}_\mathrm{LSR}$ across each core with an angular resolution of $32^{\prime\prime}$ and a pixel scale of $10^{\prime\prime}$ \citep[for a complete description of the line fitting routine, see Section 3.1 by][]{2017ApJ...843...63F}. To estimate the specific angular momentum $J/M$ of these cores, we require modeling of their observed velocity distribution. Previous studies measuring $J/M$ of cores (e.g., see Figure \ref{fig:full_j} and the references therein) have accomplished this by fitting the observed velocity distribution in the plane of the sky as a linear velocity gradient, which would be expected of a core rotating as a solid body about an axis at some angle relative to our line of sight. To remain consistent with previous work, we fit a 2D velocity gradient across each core, following the method described by \cite{1993ApJ...406..528G}. The 2D velocity distribution is of the form:
\begin{equation}
v(x,y) = c_0 + c_1 x + c_2 y\,, \label{eq:v_fit}
\end{equation}
where $(x, y)$ represent the position of each pixel in the map and $(c_0, c_1, c_2)$ are constant coefficients which are fit using a least squares analysis. 

In reality, not all cores will have a velocity gradient dominated by their rotation \citep[e.g., there may be contributions from multiple local gradients on core scales;][]{2002ApJ...572..238C, 2007A&A...470..221C}. There may also be contributions to the velocity gradient from mass flows along filaments towards dense cores \citep[e.g., as seen in Perseus on $\gtrsim 0.2$~pc scales;][Chen et al. submitted]{2020ApJ...891...84C}, from infall motions \citep[e.g., as suggested by core-scale gradients around a core in B5;][]{2022ApJ...935...57C}, and from protostellar outflows \citep[e.g., as observed in a massive star-forming region G31.41+0.31;][]{2013A&A...549A.122M, 2021A&A...648A.100B}. The cores could be dominated by turbulent motions or may be rotating but are viewed pole-on. We discuss the potential impacts of different velocity contributions on our analysis in Section \ref{sec4.1}, but note that \citet{2000ApJ...543..822B} show that the distribution of core specific angular momenta derived via a 2D linear gradient is accurate even if turbulence is a significant contributor to the core motions.

As long as rotating cores are not completely pole-on, a velocity gradient would still be measured in the line of sight velocity. The linear velocity gradient vector $\mathcal{G}$ is:
\begin{equation}
\mathcal{G} = (c_1, c_2)\,, \label{eq:v_grad}
\end{equation}
which has a magnitude, $|\mathcal{G}| = \left[c_1^2 + c_2^2\right]^{1/2}$, and an orientation, $\theta_{\mathcal{G}} = \rm{arctan}(c_2/c_1)$ measured west of south in celestial coordinates. The errors in this velocity gradient fit are determined by propagating the uncertainties from the NH$_3$ line fitting. Note that the velocity in neighbouring pixels may be correlated which could cause the uncertainty in the velocity gradient fit to be underestimated.

\begin{figure*}
    \centering
    \includegraphics[width=0.48\textwidth]{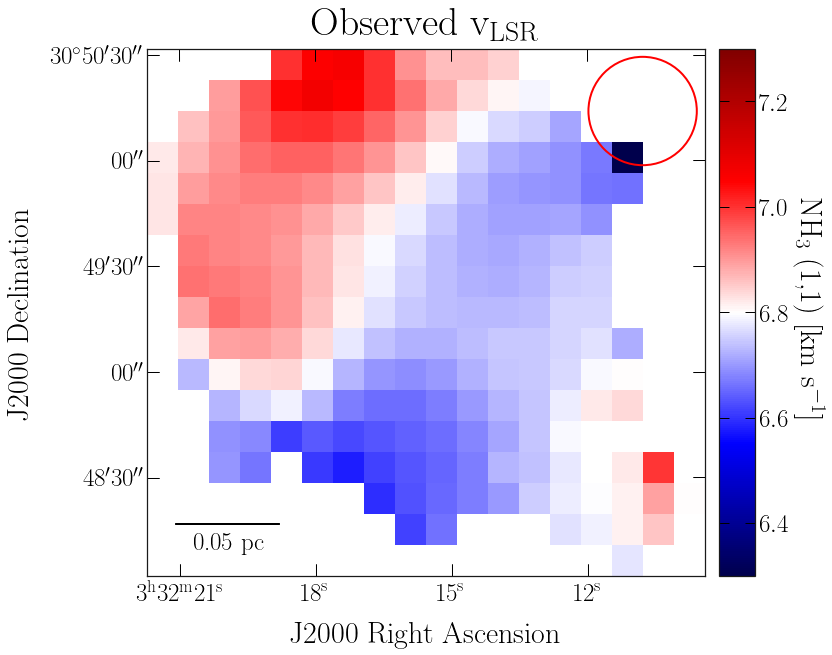}
    \includegraphics[width=0.48\textwidth]{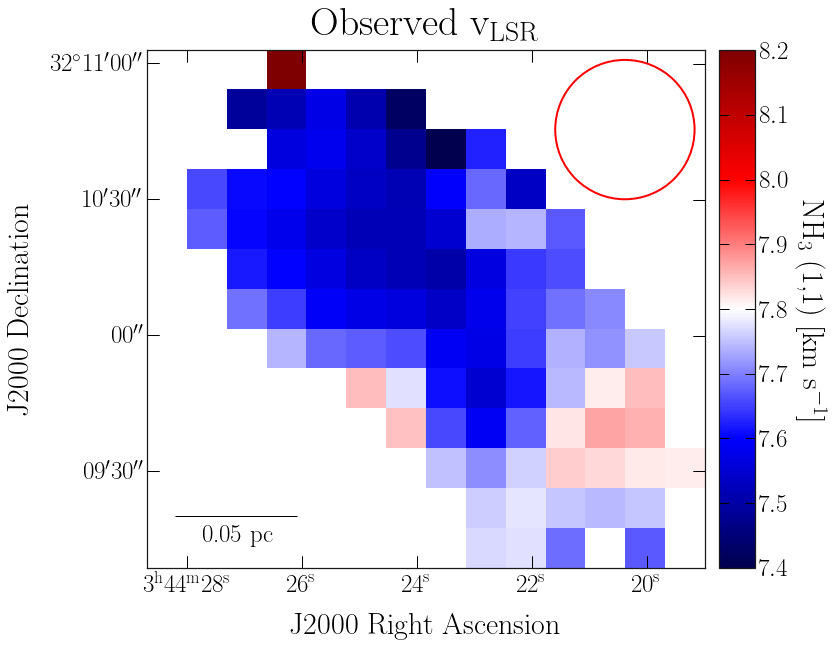}
    \includegraphics[width=0.48\textwidth]{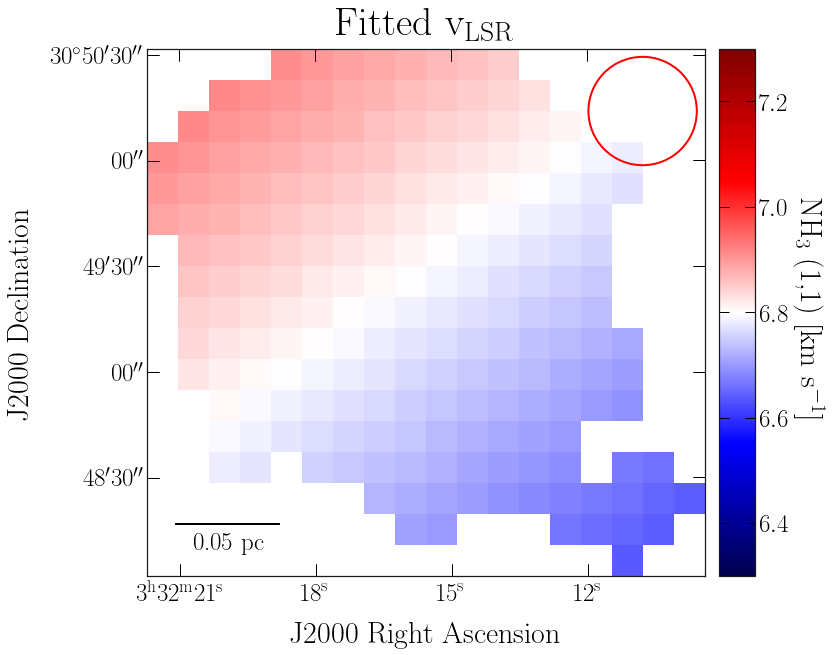}
    \includegraphics[width=0.48\textwidth]{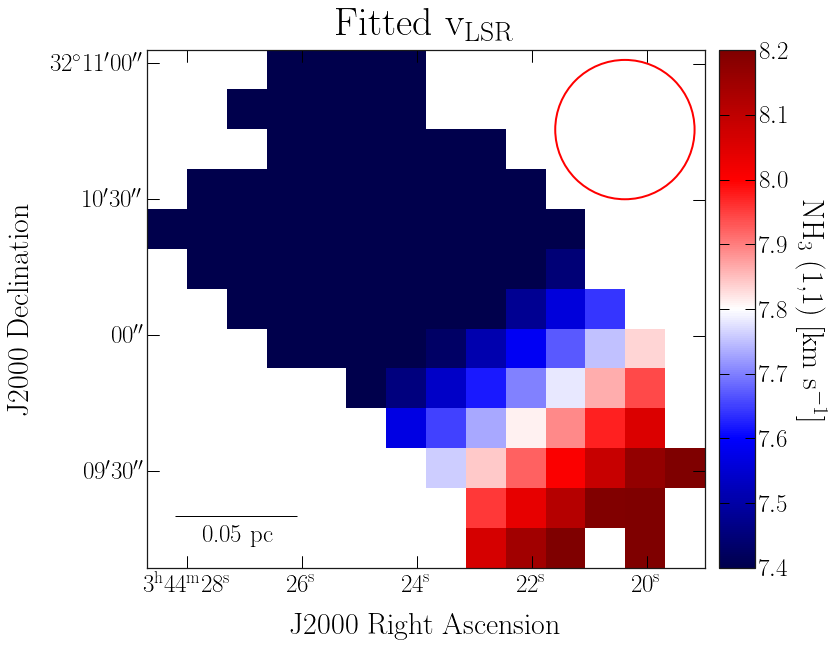}
    \includegraphics[width=0.48\textwidth]{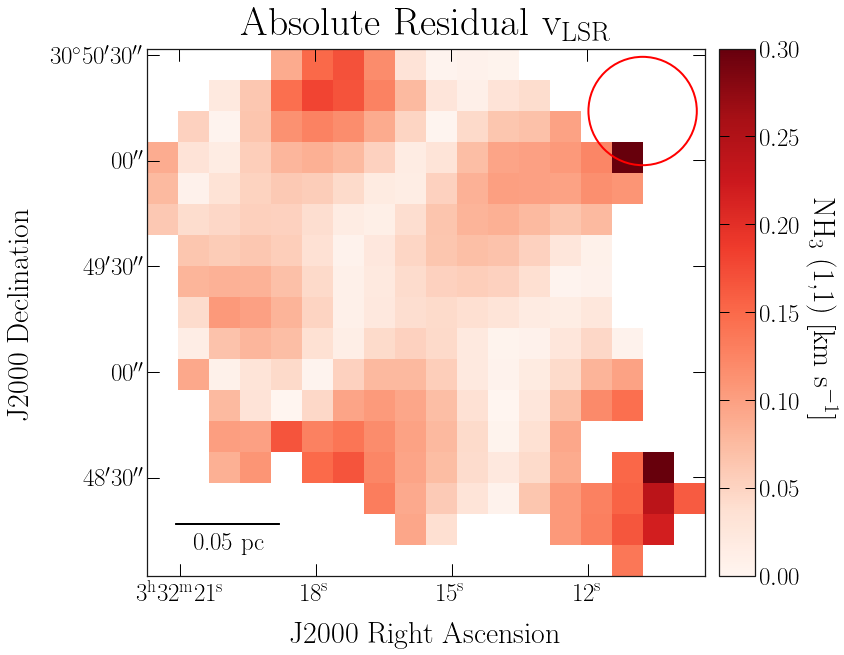}
    \includegraphics[width=0.48\textwidth]{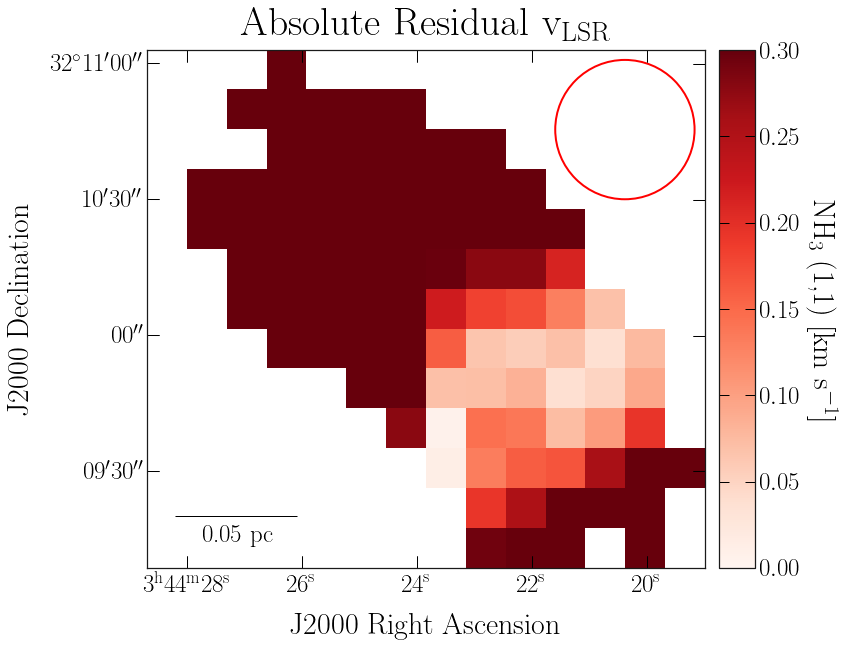}
    \caption{(Top left) Observed NH$_3$ (1,1) emission $\mathrm{v}_\mathrm{LSR}$ for a core in the Perseus B1 region (with the unique GAS dendrogram index identifier of 3). A scale bar depicting the size of $0.05$~pc is displayed at the bottom left of the plot. The red ellipse at the top right of the plot shows the 32$''$ GAS beam size. (Middle left) The best fit 2D velocity distribution following Equation \ref{eq:v_fit} for the same core. This core passes the velocity gradient criteria (Equation \ref{eq:error_cond}). (Bottom left) The absolute residual between the observed and best fit $\mathrm{v}_\mathrm{LSR}$. (Right) An example of a different core in the Perseus IC348 region (GAS dendrogram index of 32) which does not pass the velocity gradient criteria (Equation \ref{eq:error_cond}). For the core on the left $|\mathcal{G}| = 1.0~\mathrm{km}~\mathrm{s}^{-1}~\mathrm{pc}^{-1}$ and $\sigma_{\mathcal{G}} = 0.12~\mathrm{km}~\mathrm{s}^{-1}~\mathrm{pc}^{-1}$ and for the core on the right $|\mathcal{G}| = 9.1~\mathrm{km}~\mathrm{s}^{-1}~\mathrm{pc}^{-1}$ and $\sigma_{\mathcal{G}} = 4.6~\mathrm{km}~\mathrm{s}^{-1}~\mathrm{pc}^{-1}$.}
    \label{fig:ex_vlsr_combine}
\end{figure*}

We use the {\tt statsmodels}\footnote{\href{https://www.statsmodels.org/stable/index.html}{https://www.statsmodels.org/stable/index.html}.} Python package to obtain the best fit $(c_0, c_1, c_2)$ parameters and corresponding uncertainty in the fit. Following \cite{1993ApJ...406..528G}, we impose a significance threshold to identify those cores with significant velocity gradients:
\begin{equation}
|\mathcal{G}| \geq 3\sigma_{\mathcal{G}}\,, \label{eq:error_cond}
\end{equation}
where $\sigma_{\mathcal{G}}$ is the uncertainty in the 2D linear velocity gradient fit. This criterion ensures that the velocity gradient is not dominated by noise. An example of the observed $\mathrm{v}_\mathrm{LSR}$ and velocity gradient fitting routine is illustrated in the left column of Figure \ref{fig:ex_vlsr_combine} for a dense core in the Perseus B1 region (with the unique GAS dendrogram index identifier of 3) that passes the $|\mathcal{G}|$ criteria from Equation \ref{eq:error_cond}. On the other hand, the right column of Figure \ref{fig:ex_vlsr_combine} provides an example of a different core in the Perseus IC348 region (GAS dendrogram index of 32) that does not satisfy Equation \ref{eq:error_cond}, and is therefore not consistent with rotation projected onto the sky. For the two cores presented in Figure \ref{fig:ex_vlsr_combine}, the respective $|\mathcal{G}|$ and $\sigma_{\mathcal{G}}$ are $1.0~\mathrm{km}~\mathrm{s}^{-1}~\mathrm{pc}^{-1}$ and $0.12~\mathrm{km}~\mathrm{s}^{-1}~\mathrm{pc}^{-1}$ for the left column, and $|\mathcal{G}| = 9.1~\mathrm{km}~\mathrm{s}^{-1}~\mathrm{pc}^{-1}$ and $\sigma_{\mathcal{G}} = 4.6~\mathrm{km}~\mathrm{s}^{-1}~\mathrm{pc}^{-1}$ for the right column.

In total, there are 585 dense cores identified in NH$_3$ across regions in the Perseus, Ophiuchus, Cepheus, and Orion that overlap with existing continuum observations (see Section \ref{data_continuum}). For each of these dense cores, we retain the following properties from the GAS data and velocity gradient fitting: the J2000 Right Ascension ($\alpha$) in degrees, the J2000 declination ($\delta$) in degrees, the linear velocity gradient magnitude ($\mathcal{G}$) and the corresponding uncertainty of the fit ($\sigma_{\mathcal{G}}$) in units of km~s$^{-1}$~pc$^{-1}$, and the velocity gradient direction $\theta_{\mathcal{G}}$ measured west of south in celestial coordinates in degrees.

\subsection{Continuum core catalogue} \label{data_continuum}
For each of the cores identified in GAS data, we can obtain corresponding information on the radius, mass, and major axis orientation by cross-matching with existing continuum surveys (see Section \ref{crossmatch} for details regarding the cross-matching process). Crucially, the radius and mass estimates from continuum data are used in conjunction with $\mathrm{v}_{\mathrm{LSR}}$ measurements to compute angular momenta for the cores in Section \ref{results}. We use continuum derived core catalogues from the \textit{Herschel} Gould Belt Survey \citep[HGBS at $70-500~\mu\mathrm{m}$ with a $36''$ beam size;][]{2010A&A...518L.102A} in the Perseus, Ophiuchus, Serpens, Cepheus, and Orion B regions \citep{2021A&A...645A..55P, 2020A&A...638A..74L, 2015A&A...584A..91K, 2020ApJ...904..172D, 2020A&A...635A..34K} and a continuum derived core catalogue of the Orion A region (Pattle et al. in preparation) from the JCMT Gould Belt Survey \citep[JCMT GBS at $850~\mu\mathrm{m}$ with a $14.1''$ beam size;][]{2007PASP..119..855W, 2013MNRAS.430.2534D}. For the HGBS cores, we use the mass estimates provided in the respective continuum catalogues. We determine the masses of the JCMT GBS cores in Orion A using the total flux at 850~$\mu$m, assuming $T_d = 15$~K and typical dust properties, following Section 3.1 by \cite{2016MNRAS.461.4022M}.

In total, there are 1328 dense cores from HGBS and 607 dense cores from JCMT GBS that overlap with the GAS sky coverage. For each of these dense cores, we save the following properties from the continuum data: $\alpha$ in degrees, $\delta$ in degrees, the core type \citep[i.e., ``starless'', ``prestellar'', or ``protostellar'', following][]{2010A&A...518L.102A}, $\sigma_\mathrm{major}$ in arcseconds, $\sigma_\mathrm{minor}$ in arcseconds, the core orientation ($\theta_\mathrm{C}$) measured east of north in celestial coordinates in degrees, the core radius ($R$) in parsecs, and the core mass ($M$) measured in units of solar mass $M_\odot$. Note that the final catalogue only contains this information for continuum cores that are cross-matched with cores in the GAS data set. For details about the cross-matching process, see Section \ref{crossmatch}. 

The HGBS catalogues identify cores as ``starless,'' ``prestellar,'' and ``protostellar.'' Prestellar cores appear gravitationally unstable based on a comparison of the measured core mass with the local Jeans or Bonnor-Ebert mass, in contrast to starless cores, while protostellar cores show compact continuum detections at 70~\micron\ \citep{2010A&A...518L.102A}. For the JCMT GBS catalogue, we classify cores as ``protostellar'' if the core center is within 32\arcsec\ (one beam) of a young stellar object (YSO) identified in \citet{megeath_2012}, who identify $3479$ YSOs by using their mid-infrared colors as indicators of reprocessed light from dusty disks or infalling envelopes. To assess whether a core is starless or prestellar, we compute the critical Bonnor-Ebert mass assuming a temperature of $T = 15$~K and a mean molecular weight per free particle $\mu_p = 2.37$ \citep[see Appendix A by][]{2008A&A...487..993K}:
\begin{equation}
M_\mathrm{BE,crit} = 2.4 \left( \frac{c_s^2 R_\mathrm{core}}{G} \right)\,, \label{eq:M_BE}
\end{equation} 
where $c_s$ is the sound speed and $G$ is the gravitational constant. Following \citet{2010A&A...518L.102A}, we classify cores with $\alpha_\mathrm{BE} = M_\mathrm{BE,crit} / M_\mathrm{core} \leq 2$ as prestellar and those with $\alpha_\mathrm{BE} > 2$ as starless. 

When comparing the relative alignment of $\theta_\mathrm{C}$ to other vectors, we want to ensure that each core is elongated enough so that $\theta_\mathrm{C}$ is robust. To this end, we enforce a maximum threshold of 
\begin{equation}
\sigma_\mathrm{minor} / \sigma_\mathrm{major} \leq 0.9\,. \label{eq:elong_cond}
\end{equation} 
In this study, we opt to use the core radius and major axis orientation from the continuum data rather than the NH$_3$-derived properties for several reasons. The continuum emission from dust traces all the core material along the line-of-sight. While NH$_3$ (1,1) is an excellent tracer of cold, dense gas, it can also be affected by variations in temperature and chemical abundances. For example, NH$_3$ can be offset from the continuum peak in protostellar cores \citep{2011ApJ...740...45T}. In addition, the continuum catalogues used here apply robust techniques to dissociate the core emission from the background \citep[e.g., {\tt getsources};][]{2012A&A...542A..81M}. Hence, all subsequent mentions of core orientation $\theta_\mathrm{C}$ will refer to the orientation as derived from continuum HGBS or JCMT GBS observations rather than those derived using {\tt astrodendro} with the GAS data.

\subsection{Dust polarization} \label{data_dustpol}
To infer the magnetic field orientation, in the plane of the sky, we utilize $353$~GHz dust polarization observations from the Planck collaboration \citep{2015A&A...576A.104P}. We use Planck maps smoothed to a resolution of $6~\mathrm{arcminutes}$ FWHM, following the procedure described by \cite{2015A&A...576A.104P} that is also applied by \cite{2016A&A...586A.138P} and \cite{2019A&A...629A..96S}. This smoothing allows us to achieve sufficiently high sensitivity for deriving the polarization properties of our dense core sample while also preserving the small-scale variations in magnetic field orientation coincident with the MCs. These observations trace $\sim 0.25 - 0.8$~pc scale magnetic fields for the MCs analyzed in this work. From the Planck dust polarization observations, we obtain maps of Stokes $Q$ and $U$ and compute the total linearly polarized intensity as:
\begin{equation}
P = \left[Q^2 + U^2\right]^{1/2}\,. \label{eq:P}
\end{equation}
We obtain the uncertainty in the linearly polarized intensity $\sigma_P$ from the Planck covariance maps and impose a minimum threshold for detection 
\begin{equation}
P \geq 3\sigma_P\,, \label{eq:PI_cond}
\end{equation}
\citep[the same as the threshold used by][]{2021MNRAS.503.5006S} towards each dense core. This threshold ensures that we have a significant detection of $P$ towards each dense core. The magnetic field orientation projected on the plane of the sky ($\theta_{\mathrm{B}_\perp}$) can be inferred from the Stokes $Q$ and $U$ as well, i.e.,
\begin{equation}
\theta_{\mathrm{B}_\perp} = \frac{1}{2} \rm{arctan}(-U,Q)\,. \label{eq:b_dir}
\end{equation}
Using the Planck data, the magnetic field orientation $\theta_{\mathrm{B}_\perp}$ is measured counter-clockwise from Galactic north in Galactic coordinates in units of degrees. Note that we are tracing magnetic fields that are on scales larger than the typical size of cores in our data (the median core radius in our sample is $\sim 0.03$~pc), thus for each core we measure $\theta_{\mathrm{B}_\perp}$ at the core center.

\subsection{Cross-matching} \label{crossmatch}
For 17 distinct regions of the Gould Belt, we compile a cross-matched core catalogue between GAS observations (beam size of $32''$) and at least one of the two sets of continuum observations. The JCMT GBS data (beam size of $14.1''$) are used for only the Orion A and Orion A South regions, while the HGBS derived core catalogues are used for the remaining 15 regions. Based on the angular resolution of the 500~$\mu$m HGBS data, we define a cross-match cut-off radius of $36''$. Even though the JCMT GBS has a finer angular resolution, we also use a $36''$ cross-match cut-off radius for the Orion A and Orion A South regions to maintain a consistent framework across all the regions. Hence, for a core to be classified as a successful cross-match, its on-sky positions between the spectral and continuum catalogues must agree to within $36''$. In addition, we account for scenarios in which multiple continuum cores are matched with the same NH$_3$ core by considering only the nearest one and discarding the remaining matches. It is possible, however, that through this process we discard matches that have very similar separations and may have been physically correlated. As such, we add a supplementary step to our cross-matching algorithm to generate a flag when the difference in separation between two matches is less than half of the separation of the nearest match. An analogous process is applied where a continuum core has multiple NH$_3$ core matches, but no such cases are found in our data. 

The right panel of Figure \ref{fig:b1_cores} depicts the same region as the left after applying our cross-matching algorithm and labels cores that were successfully cross-matched with a continuum counterpart or not. Subsequently, we compute 2D linear velocity gradient fits for each of the cross-matched cores and apply the significance criteria described in Equation \ref{eq:error_cond}. The velocity gradient is computed within the NH$_3$-defined {\tt astrodendro} contour for each core, where the NH$_3$ traces specifically the kinematics of the high density gas.

\begin{figure}
    \centering
    \includegraphics[width=0.48\textwidth]{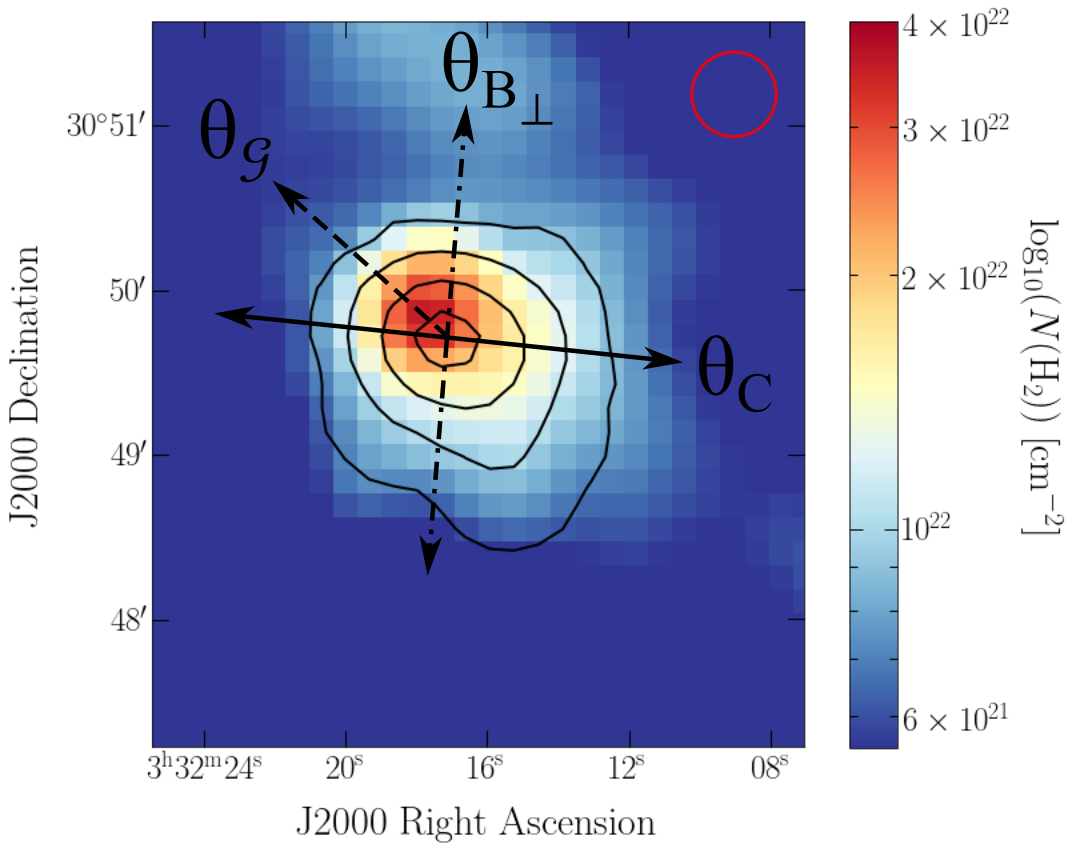}
    \includegraphics[width=0.48\textwidth]{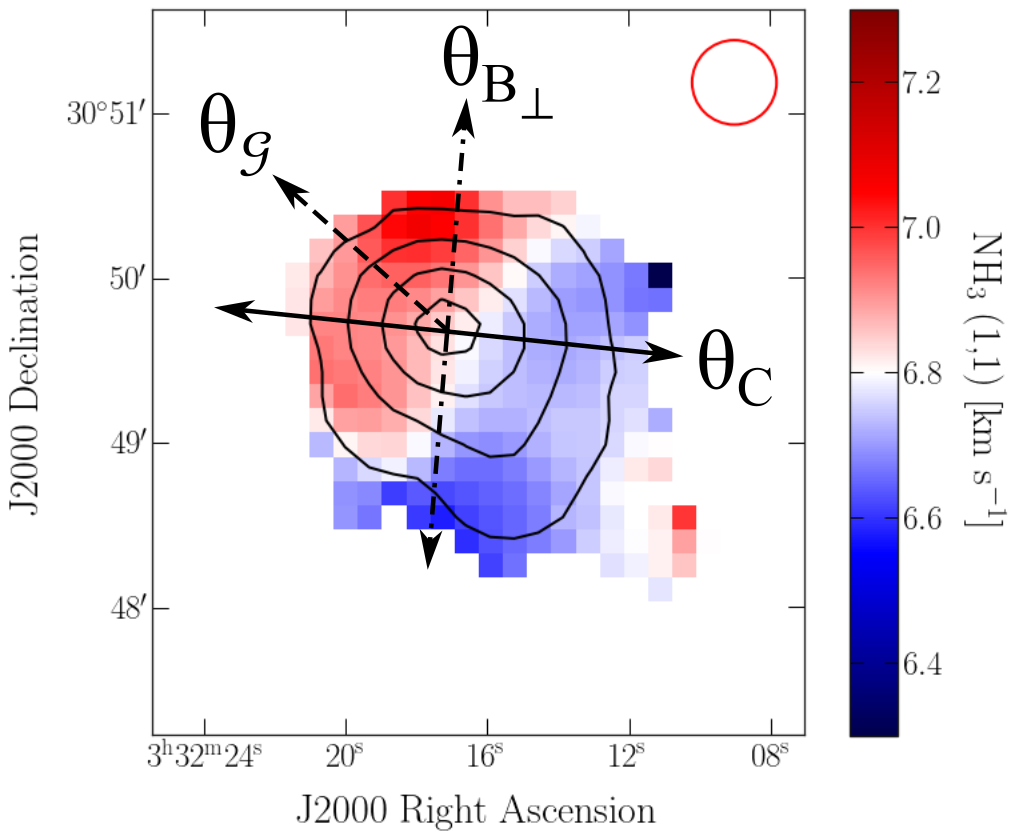}
    \caption{(Top) The same core presented in the top panel of Figure \ref{fig:ex_vlsr_combine} from the B1 region of Perseus (GAS dendrogram index of 3). The colour map depicts the logarithmic $N$(H$_2$) column density in units of cm$^{-2}$ from the HGBS continuum data. Black contours of integrated NH$_3$ (1,1) intensity from the GAS data are overplotted at levels of $1$, $3$, $6$, and $9~\mathrm{K}~\mathrm{km}~\mathrm{s}^{-1}$, respectively. The core elongation axis ($\theta_\mathrm{C}$; see Section \ref{data_continuum}), velocity gradient direction ($\theta_\mathcal{G}$; see Section \ref{data_specline}), and ambient magnetic field orientation ($\theta_{\mathrm{B}_\perp}$; see Section \ref{data_dustpol}) are overlaid as a solid, dashed, and dashed-dotted black vector, respectively. The 36$''$ beam for the HGBS data is overplotted as a red ellipse in the top right corner. Note that the continuum core properties, including core radius, mass, and $\theta_\mathrm{C}$, are derived after background subtraction which has not been applied to the $N$(H$_2$) column density presented here. (Bottom) The same core but with a colour map representing the observed NH$_3$ (1,1) emission $\mathrm{v}_\mathrm{LSR}$ to highlight the velocity gradient across the core. A mask is applied to only show the $\mathrm{v}_\mathrm{LSR}$ for pixels within the boundary of the core, as defined by {\tt astrodendro}. The contours are the same as the panel above. The $\theta_\mathrm{C}$, $\theta_\mathcal{G}$, and $\theta_{\mathrm{B}_\perp}$ vectors are the same as the top panel. The 32$''$ beam for the GAS data is overplotted as a red ellipse in the top right corner. Note that, in both panels, the $\theta_\mathrm{C}$ and $\theta_{\mathrm{B}_\perp}$ vectors have a $180^\circ$ ambiguity.}
    \label{fig:ex_vectors}
\end{figure}

We compare the relative core orientation ($\theta_\mathrm{C}$), velocity gradient across the core ($\theta_\mathcal{G}$), and ambient magnetic field direction ($\theta_{\mathrm{B}_\perp}$) to determine whether any preferential alignments exist between these three parameters. These vectors, however, are not initially measured in the same coordinate system. To correct for this difference, we first shift $\theta_{\mathcal{G}}$ for all cores counter-clockwise by $180^\circ$, which puts it in the same celestial reference frame as $\theta_\mathrm{C}$. Then, we transform the $\theta_{\mathcal{G}}$ and $\theta_\mathrm{C}$ vectors into Galactic coordinates with $0^\circ$ being towards the Galactic north pole and increasing counter clockwise, which is the reference frame used for $\theta_{\mathrm{B}_\perp}$. Once in the same coordinate system, we take the absolute difference between the vectors to measure their relative alignment. An example of a core in the B1 region of Perseus over-plotted with its measured $\theta_\mathrm{C}$, $\theta_{\mathcal{G}}$, and $\theta_{\mathrm{B}_\perp}$ is shown in Figure \ref{fig:ex_vectors}. Note that in the online catalogue (see Section \ref{data_avail}), $\theta_\mathrm{C}$, $\theta_\mathcal{G}$, and $\theta_{\mathrm{B}_\perp}$ are provided in both the original coordinate system in which they were measured and also in the common frame that is measured counter-clockwise from Galactic north, as mentioned above. Further, we would like to highlight that $\theta_\mathrm{C}$ and $\theta_{\mathrm{B}_\perp}$ have a $180^\circ$ ambiguity (e.g., values of $5^\circ$ and $185^\circ$ would be equivalent). So, while we refer to $\theta_\mathrm{C}$, $\theta_\mathcal{G}$, and $\theta_{\mathrm{B}_\perp}$ all as ``vectors'', only $\theta_\mathcal{G}$ spans a full $360^\circ$ range.

In total, we find 399 unique spectral$-$continuum cross-matches across the 17 star forming regions. Of the 399 cross-matched cores, 355 pass the core elongation cut (Equation \ref{eq:elong_cond}), 329 pass the velocity gradient significance criteria (Equation \ref{eq:error_cond}), and 385 pass the polarized intensity threshold (Equation \ref{eq:PI_cond}). A summary of the cross-match statistics for each region is presented in Table \ref{tb:cross_match}. The region name and their corresponding distance and $\alpha$ and $\delta$ coverage are presented in columns 1, 2, 3, and 4, respectively. Column 5, 6, and 7 list the number of GAS, HGBS, and JCMT sources identified, respectively, within the given $\alpha$ and $\delta$ range. Column 8 provides the number of cores that are successfully cross-matched between spectral and continuum data. The number of cores from column 8 that pass the core elongation, velocity gradient, and polarized intensity cuts are given in columns 9, 10, and 11, respectively. The summed totals for columns 5$-$11 are calculated in the final row of the table. We have made the full cross-matched core catalogue publicly accessible online (see Section \ref{data_avail} for details).

\section{Results} \label{results}
In this section, we provide results for the velocity gradient fitting, angular momentum estimation, and the relative alignments of core orientation, velocity gradient, and ambient magnetic field for our sample of 399 cross-matched cores. The results provided in this section are grouped with respect to the host cloud of each dense core.

\begin{table*}
\begin{center}
\caption{Summary of cross-match statistics between GAS, HGBS, and JCMT core catalogues in 17 regions of the Gould Belt. Assumed distances to each region are derived from the following works: $^\mathrm{a}$\protect\cite{2018ApJ...869...83Z}; $^\mathrm{b}$\protect\cite{2018ApJ...865...73O}; $^\mathrm{c}$\protect\cite{2008PASJ...60...37H}; $^\mathrm{d}$\protect\cite{2018ApJ...869L..33O};  $^\mathrm{e}$\protect\cite{2019A&A...624A...6Y}; $^\mathrm{f}$\protect\cite{2017ApJ...834..142K}; $^\mathrm{g}$\protect\cite{2008hsf1.book..483M}.} \label{tb:cross_match}
\begin{tabular}{ccccccccccc} 
\hline
\hline
Region & Distance & $\alpha$ range & $\delta$ range & GAS & HGBS & JCMT & Cross-matched & $\frac{\sigma_\mathrm{minor}}{\sigma_\mathrm{major}}$ & $|\mathcal{G}| \geq$ & $P \geq$\\
 & (pc) & (deg) & (deg) & cores & cores & cores & cores & $\leq 0.9$ & $3\sigma_{\mathcal{G}}$ & $3\sigma_P$\\
\hline
Perseus B1 & 301$^\mathrm{a}$ & 52.82 $-$ 53.44 & 30.71 $-$ 31.35 & 35 & 62 & $-$ & 15 & 15 & 12 & 15\\
Perseus B1E & 293$^\mathrm{a}$ & 54.00 $-$ 54.04 & 31.18 $-$ 31.22 & 1 & 2 & $-$ & 1 & 1 & 1 & 1\\
Perseus NGC1333 & 293$^\mathrm{b}$ & 52.13 $-$ 52.47 & 31.08 $-$ 31.65 & 38 & 102 & $-$ & 26 & 21 & 24 & 26\\
Perseus IC348 & 320$^\mathrm{b}$ & 55.72 $-$ 56.32 & 31.97 $-$ 32.17 & 22 & 49 & $-$ & 17 & 15 & 12 & 17\\
Perseus L1448 & 288$^\mathrm{a}$ & 51.30 $-$ 51.46 & 30.71 $-$ 30.75 & 4 & 3 & $-$ & 1 & 1 & 1 & 1\\
Perseus L1451 & 279$^\mathrm{a}$ & 51.11 $-$ 51.38 & 30.31 $-$ 30.41 & 6 & 15 & $-$ & 4 & 4 & 4 & 0\\
Perseus L1455 & 293$^\mathrm{c}$ & 51.76 $-$ 52.03 & 29.98 $-$ 30.26 & 16 & 34 & $-$ & 11 & 9 & 11 & 10\\
Ophiuchus L1688 & 138$^\mathrm{d}$ & 246.50 $-$ 247.32 & $-$24.74 $-$ $-$24.28 & 53 & 243 & $-$ & 38 & 35 & 37 & 34\\
Ophiuchus L1689 & 144$^\mathrm{d}$ & 247.89 $-$ 248.70 & $-$25.06 $-$ $-$24.42 & 17 & 89 & $-$ & 12 & 11 & 11 & 11\\
Serpens W40 & 436$^\mathrm{d}$ & 277.18 $-$ 278.11 & $-$2.59 $-$ $-$1.45 & 133 & 420 & $-$ & 103 & 91 & 84 & 99\\
Serpens MWC297 & 436$^\mathrm{d}$ & 277.03 $-$ 277.04 & $-$3.80 $-$ $-$3.78 & 2 & 9 & $-$ & 2 & 2 & 2 & 2\\
Cepheus L1228 & 346$^\mathrm{e}$ & 314.18 $-$ 314.51 & 77.56 $-$ 77.72 & 4 & 14 & $-$ & 1 & 1 & 0 & 1\\
Cepheus L1251 & 346$^\mathrm{e}$ & 335.35 $-$ 339.91 & 75.07 $-$ 75.31 & 19 & 118 & $-$ & 9 & 9 & 9 & 9\\
Orion B NGC2023 & 420$^\mathrm{f}$ & 85.28 $-$ 85.46 & $-$2.48 $-$ $-$1.71 & 23 & 166 & $-$ & 19 & 19 & 15 & 19\\
Orion B NGC2068 & 388$^\mathrm{f}$ & 86.52 $-$ 86.71 & $-$0.25 $-$ 0.12 & 11 & 56 & $-$ & 10 & 9 & 8 & 10\\
Orion A & 450$^\mathrm{g}$ & 83.65 $-$ 84.22 & $-$6.84 $-$ $-$4.90 & 150 & $-$ & 533 & 97 & 82 & 74 & 97\\
Orion A South & 450$^\mathrm{g}$ & 84.68 $-$ 85.36 & $-$8.00 $-$ $-$6.98 & 51 & $-$ & 74 & 33 & 30 & 24 & 33\\
\hline
\textbf{Total} & \textbf{$-$} & \textbf{$-$} & \textbf{$-$} & \textbf{585} & \textbf{1328} & \textbf{607} & \textbf{399} & \textbf{355} & \textbf{329} & \textbf{385}\\
\hline
\hline
\end{tabular}
\end{center}
\end{table*}

\subsection{Velocity gradient and angular momentum} \label{kinematics_results}
Of our sample of 399 cross-matched cores, 329 pass the velocity gradient significance criteria (Equation \ref{eq:error_cond}). The large majority ($\sim 82$\%) of cores have a velocity gradient consistent with rotation projected on the plane of the sky and are not dominated by small-scale turbulent gas motions. For the 70 cores that fail the velocity gradient significance criteria, we find that their average $|\mathcal{G}|$ is comparable to the average $|\mathcal{G}|$ of the 329 cores that pass. The $\sigma_\mathcal{G}$ of the cores that fail the cut, however, are on average $\sim 7$ times larger than the cores that satisfy the cut. For these 70 cores, their velocity gradients may be dominated by non-rotation motions such as turbulence or infalling gas. We note that resolution may affect the overall fraction of cores that show significant velocity gradients: nearby clouds have a higher fraction of cores that satisfy Equation \ref{eq:error_cond} than cores in more distant clouds (e.g., $96\%$ of cores in Ophiuchus satisfy Equation \ref{eq:error_cond}, while the number in Serpens is $82\%$). Further, there is not a significant difference in the core type classification (i.e., starless, prestellar, or protostellar) for those cores that pass or fail the velocity gradient significance criteria.

\begin{figure*}
    \centering
    \includegraphics[width=0.33\textwidth]{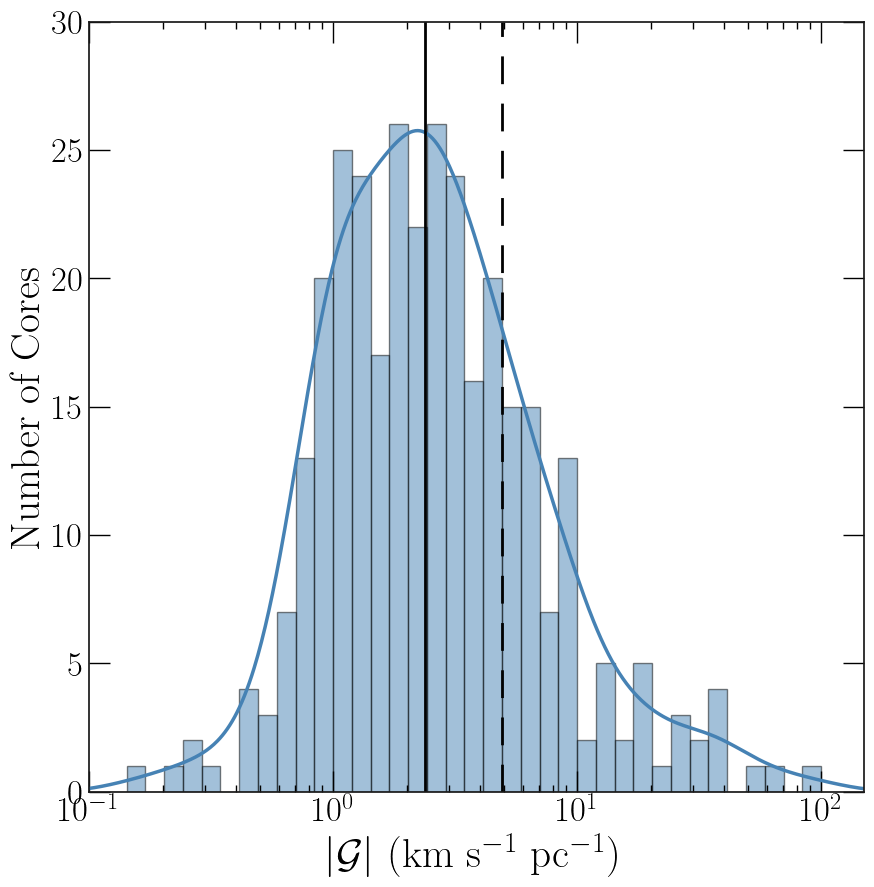}
    \includegraphics[width=0.33\textwidth]{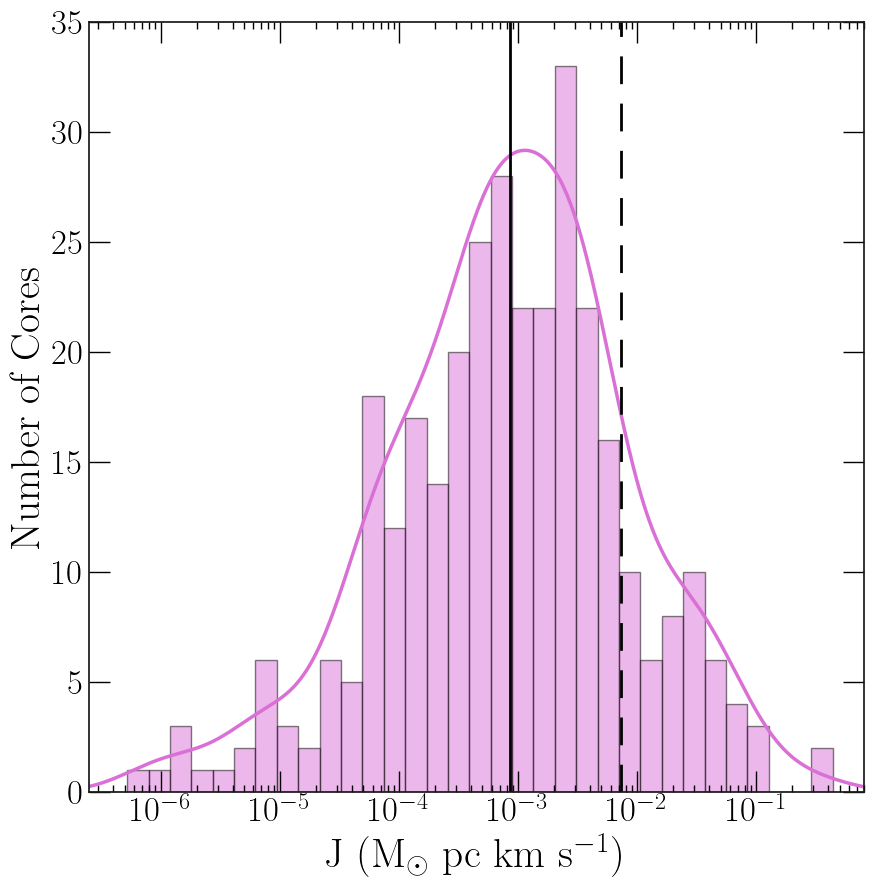}
    \includegraphics[width=0.33\textwidth]{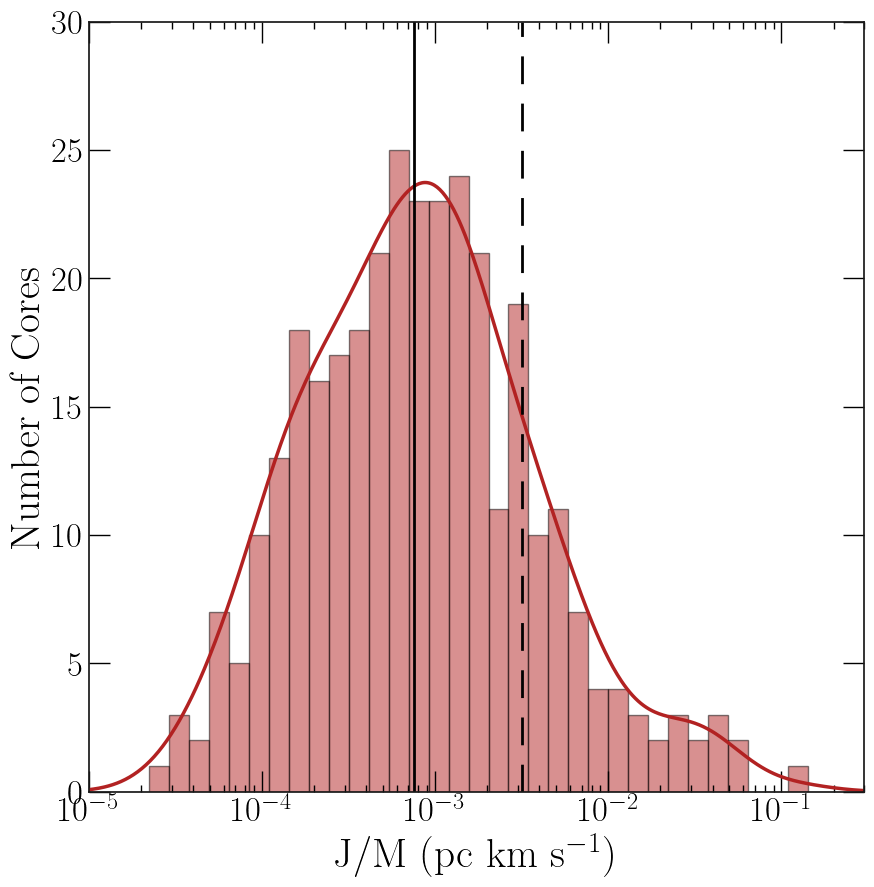}
    \caption{The distribution of the velocity gradient magnitude $\mathcal{G}$ (left), total angular momentum $J$ (middle), and specific angular momentum $J/M$ (right) for all 329 cross-matched dense cores that satisfy Equation \ref{eq:error_cond}. A Gaussian kernel density estimate of the distribution is overplotted for each distribution. The dashed black line corresponds to the mean value of the respective distribution and the solid black line corresponds to the median. The typical uncertainties are $1.0~\mathrm{km}~\mathrm{s}^{-1}~\mathrm{pc}^{-1}$, $1.7 \times 10^{-3}~\mathrm{M}_\odot~\mathrm{pc}~\mathrm{km}~\mathrm{s}^{-1}$, and $6.9 \times 10^{-4}~\mathrm{pc}~\mathrm{km}~\mathrm{s}^{-1}$ for $\mathcal{G}$, $J$, and $J/M$, respectively.}
    \label{fig:dists}
\end{figure*}

\begin{figure}
    \centering
    \includegraphics[width=0.475\textwidth]{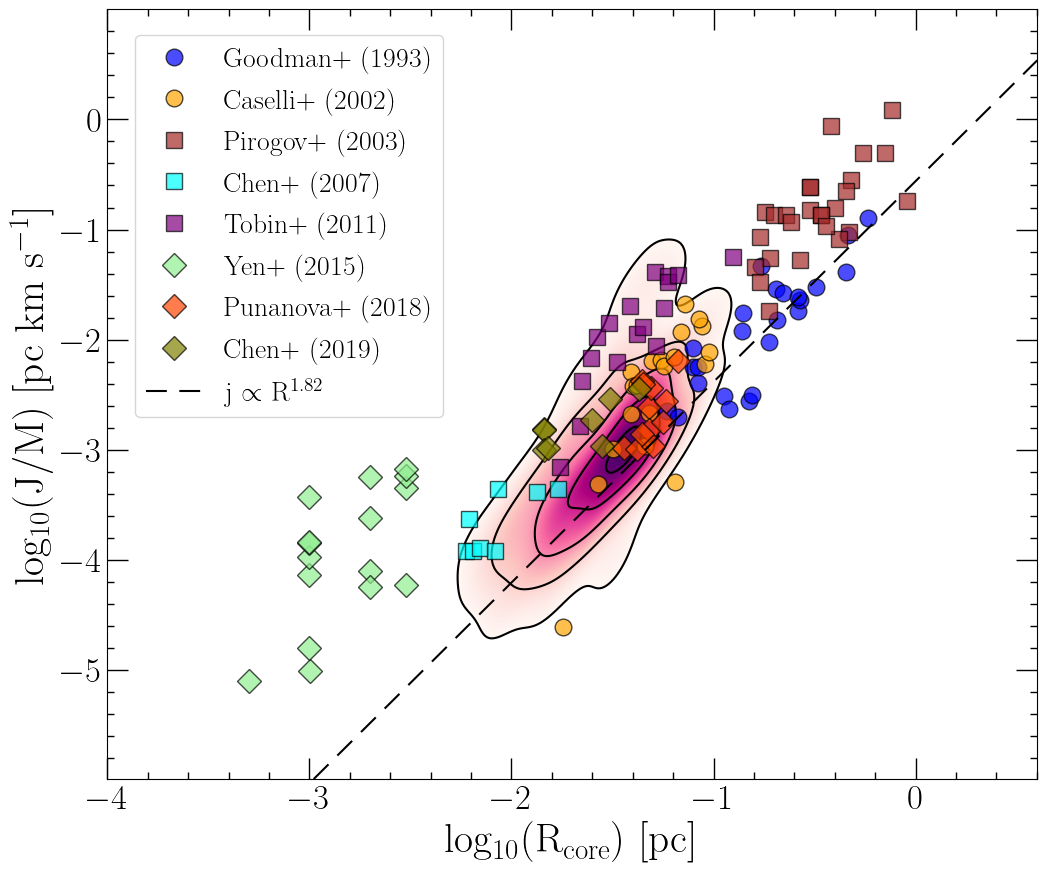}
    \caption{For our sample of $329$~cores that satisfy Equation \ref{eq:error_cond}, the specific angular momentum $J/M$ versus core radius distribution is plotted as a 2D kernel density estimate that is normalized to 1. The black contour lines represent density levels of 0.05, 0.25, 0.5, 0.75, and 0.95. The best fit model (Equation \ref{eq:j_fit}) for our sample of $329$~cores is presented as a dashed black line. Another $131$~cores with previously estimated $J/M$ are over-plotted with our sample as: blue circles \protect\citep{1993ApJ...406..528G}, orange circles \protect\citep{2002ApJ...572..238C}, brown squares \protect\citep{2003A&A...405..639P}, cyan squares \protect\citep{2007ApJ...669.1058C}, purple squares \protect\citep{2011ApJ...740...45T}, green diamonds \protect\citep{2015ApJ...799..193Y}, orange-red diamonds \protect\citep{2018A&A...617A..27P}, and olive diamonds \protect\citep{2019MNRAS.490..527C}. The data are presented in log$_{10}$ space on both axes. This Figure is adapted from Figure 13 by \protect\cite{2022arXiv220503935P}.}
    \label{fig:full_j}
\end{figure}

For the 329 cores that satisfy Equation \ref{eq:error_cond}, we can then compute their specific angular momenta. Recall that we derive the velocity gradient magnitude as $|\mathcal{G}| = \left[c_1^2 + c_2^2\right]^{1/2}$. Ideally, when computing the angular velocity $\omega$, we would factor in the inclination angle $i$ to our line of sight:
\begin{equation}
\omega = \frac{|\mathcal{G}|}{\mathrm{sin}(i)} \hat{\omega}\,, \label{eq:ang_vel}
\end{equation}
where $\hat{\omega}$ is the direction in which $\omega$ points. There is, however, no way to derive observationally $i$ from spectral line data for any given core. Thus, we approximate the angular velocity as $|\omega| \simeq |\mathcal{G}|$, pointing in the direction of $\theta_{\mathcal{G}}$. Here, we could opt to use an expected average inclination angle of $i=60^\circ$ across our sample, however, we use the $|\omega| \simeq |\mathcal{G}|$ to match previous studies (see Figure \ref{fig:full_j} and references therein). Applying an average $i=60^\circ$ across our sample would only increase $J/M$ by a factor of $\sim 1.15$ but not affect any of our results regarding $R$ scaling (see Equations \ref{eq:velmag_fit} and \ref{eq:j_fit}). To compute the angular momenta, we will make the assumption that these dense cores are modelled by spheres of constant density, thus having a moment of inertia defined as 
\begin{equation}
I = \frac{2}{5} M R^2\,. \label{eq:mom_inertia}
\end{equation}
The mass $M$ and radius $R$ of a given core in our cross-matched sample is derived via their continuum properties (see Section \ref{data_continuum} and references therein for detail). Finally, the angular momentum of a core is given by $J = I \omega$ and its specific angular momentum is $J/M$. The distributions of $\mathcal{G}$, $J$, and $J/M$ for our sample are presented in Figure \ref{fig:dists} with the median and mean value of each distribution marked as a solid black line and a dashed black line, respectively. The typical uncertainties for $\mathcal{G}$, $J$, and $J/M$, respectively, are $0.8~\mathrm{km}~\mathrm{s}^{-1}~\mathrm{pc}^{-1}$, $1.5 \times 10^{-3}~\mathrm{M}_\odot~\mathrm{pc}~\mathrm{km}~\mathrm{s}^{-1}$, and $5.8 \times 10^{-4}~\mathrm{pc}~\mathrm{km}~\mathrm{s}^{-1}$.

In Figure \ref{fig:full_j}, we plot the distribution of specific angular momentum $J/M$ versus core size $R$ for all 329 cross-matched cores which satisfy Equation \ref{eq:error_cond}. For visual clarity, we avoid plotting each individual data point and instead display the distribution as a 2D kernel density estimate that is normalized to 1 with contour levels (indicated as black lines) of 0.05, 0.25, 0.5, 0.75, and 0.95. In the same Figure, we overlay $J/M$ estimates for $131$~cores from the literature \citep{1993ApJ...406..528G, 2002ApJ...572..238C, 2003A&A...405..639P, 2007ApJ...669.1058C, 2011ApJ...740...45T, 2015ApJ...799..193Y, 2018A&A...617A..27P, 2019MNRAS.490..527C}. These previous works obtained velocity gradient measurements using NH$_3$ (1,1) and (2,2) \citep{1993ApJ...406..528G, 2011ApJ...740...45T}, N$_2$H$^\mathrm{+}$ (1-0) \citep{2002ApJ...572..238C, 2003A&A...405..639P, 2007ApJ...669.1058C, 2011ApJ...740...45T, 2018A&A...617A..27P, 2019MNRAS.490..527C}, and C$^{18}$O (2-1) \citep{2015ApJ...799..193Y} observations, in a similar manner to our work. Our sample provides a factor of $\sim 3$ increase in total number of observed cores with $J/M$ estimates. Fitting power-law relations between $|\mathcal{G}|$ and $J/M$ and $R$\footnote{The fitting was doing using the {\tt scipy} package V1.10.1, specifically using the {\tt scipy.optimize.curve\_fit} module with least squares optimization \citep{2020SciPy-NMeth}.}$-$weighted by the observational uncertainties$-$we find the best fits:
\begin{equation}
|\mathcal{G}| = 10^{-0.16 \pm 0.46} R^{-0.18 \pm 0.10}~\mathrm{km}~\mathrm{s}^{-1}~\mathrm{pc}^{-1}\,, \label{eq:velmag_fit}
\end{equation}
\begin{equation}
J/M = 10^{-0.56 \pm 0.46} R^{1.82 \pm 0.10}~\mathrm{pc}~\mathrm{km}~\mathrm{s}^{-1}\,. \label{eq:j_fit}
\end{equation}
The best fit for the $J/M$ and $R$ relation is drawn in Figure \ref{fig:full_j} as a black dashed line. It should be noted that there is a large scatter in $\mathcal{G}$ values and the $\mathcal{G}$ scaling with core radius is marginal based on the best fit. Our $J/M$ and $R$ scaling relation is consistent, within the best-fit uncertainties, with that of \cite{1993ApJ...406..528G} who found $J/M \propto R^{1.6 \pm 0.2}$, and our results are steeper and have a larger scatter than the $J/M \propto R^{1.5}$ relation seen in MHD simulations \citep{2000ApJ...543..822B, 2018ApJ...865...34C}.

\subsection{Relative alignment between core elongation, velocity gradient and magnetic field orientation} \label{alignment_results}
\begin{figure*}
    \centering
    \includegraphics[width=0.885\textwidth]{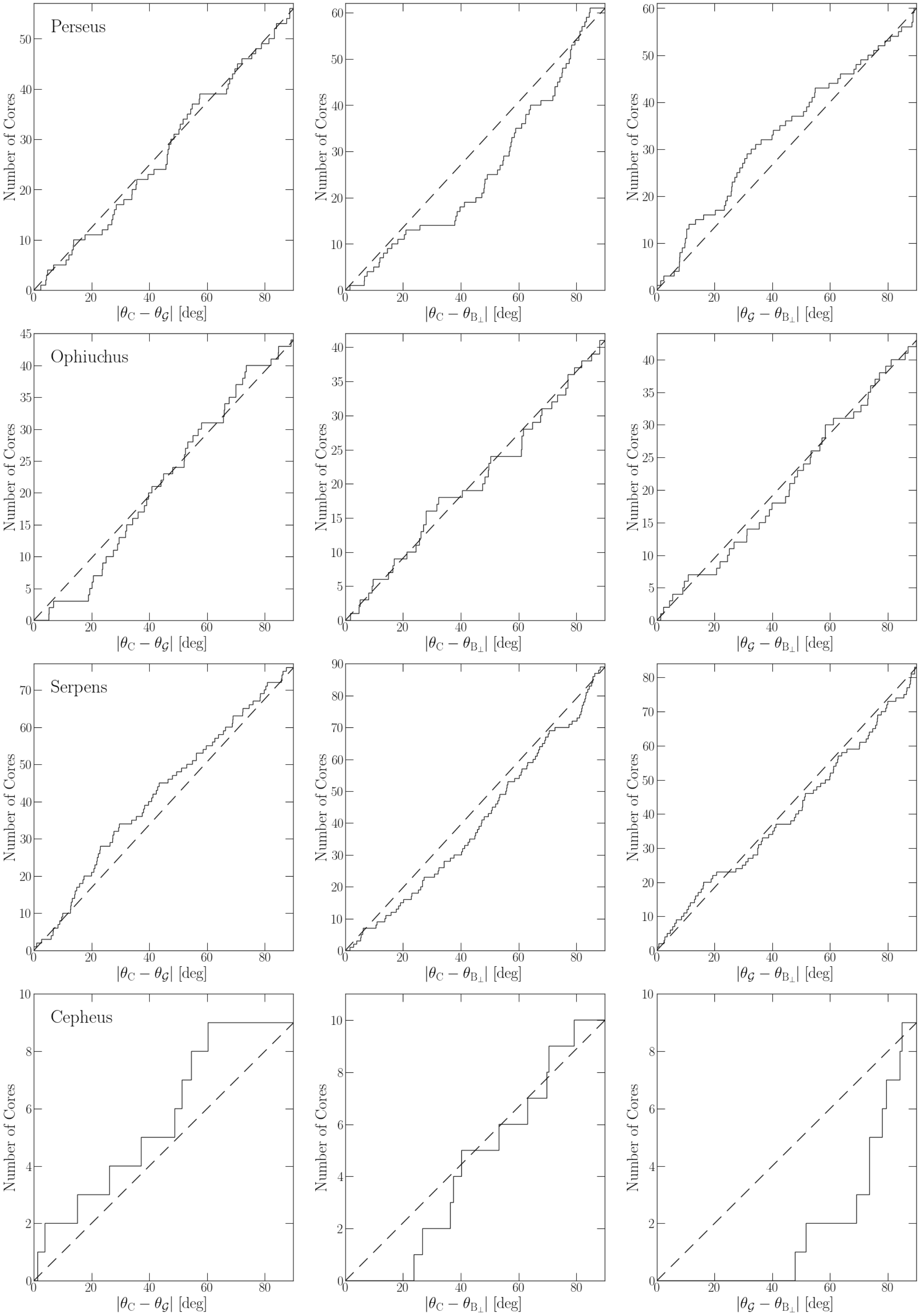}
    \caption{Cumulative distribution function of the absolute difference between core orientation $\theta_\mathrm{C}$ and velocity gradient direction $\theta_\mathcal{G}$ (left), $\theta_\mathrm{C}$ and the plane of the sky magnetic field orientation $\theta_{\mathrm{B}_\perp}$ (middle), and $\theta_\mathcal{G}$ and $\theta_{\mathrm{B}_\perp}$ (right), for the Perseus, Ophiuchus, Serpens, and Cepheus clouds, respectively. The dashed line in each plot depicts a completely random alignment between the respective vectors.}
    \label{fig:vectors_1}
\end{figure*}

\begin{figure*}
    \centering
    \includegraphics[width=0.885\textwidth]{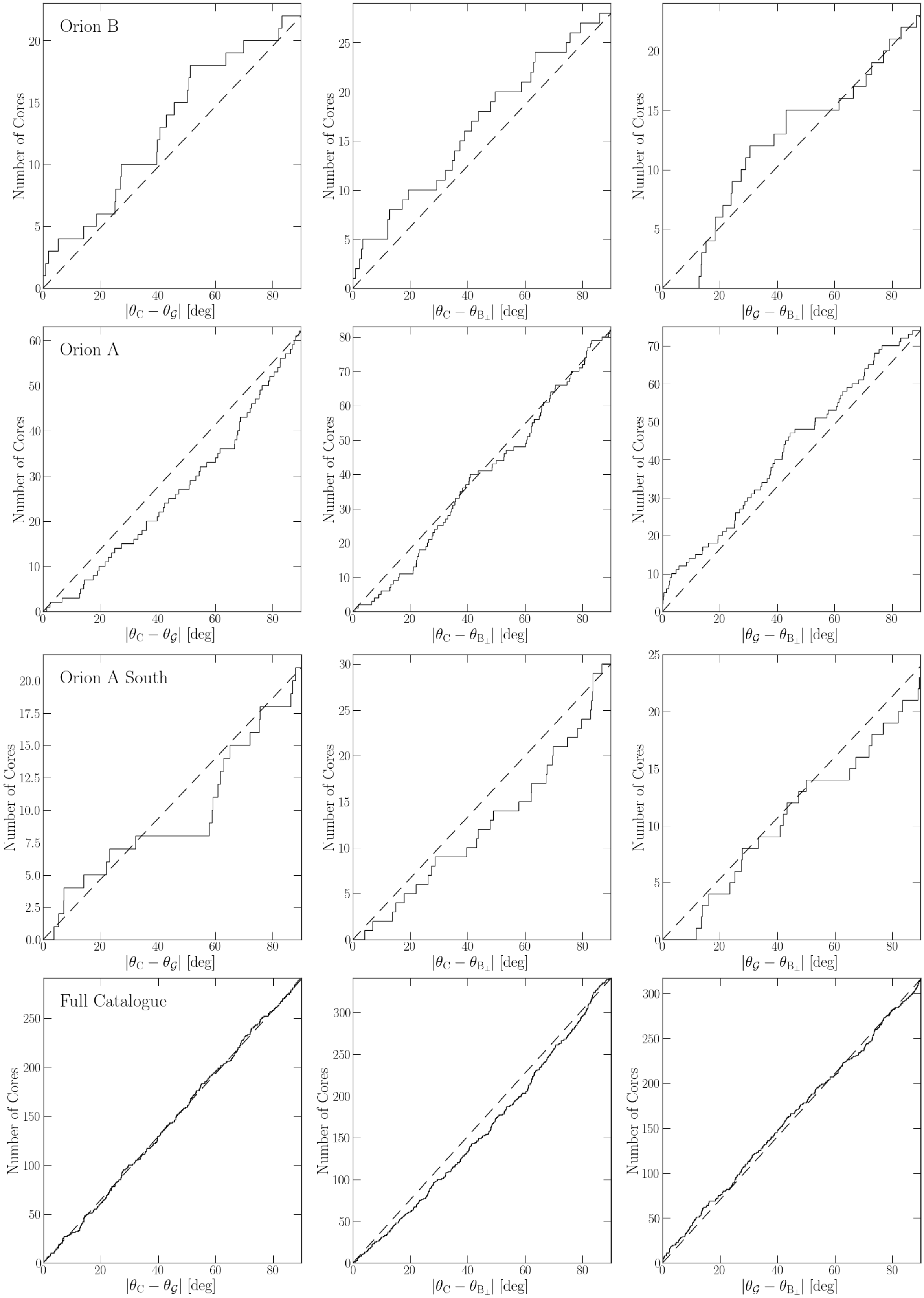}
    \caption{A continuation of Figure \ref{fig:vectors_1}, with the same format, for the Orion B, Orion A, and Orion A South regions. The final row collates the results from all seven clouds and presents the relative alignment between each pair of vectors for the full sample after applying the relevant cuts (i.e., Equations \ref{eq:error_cond}, \ref{eq:elong_cond}, and \ref{eq:PI_cond}).}
    \label{fig:vectors_2}
\end{figure*}

When analyzing the $\theta_\mathrm{C}$, $\theta_\mathcal{G}$, and $\theta_{\mathrm{B}_\perp}$ vectors, we consider not only global trends across all cores in our cross-matched catalogue but also group the data into seven physically associated molecular cloud structures (Perseus, Ophiuchus, Serpens, Cepheus, Orion B, Orion A, and Orion A South). Organizing the results in this format enables us to identify common trends in the orientations of $\theta_\mathrm{C}$, $\theta_\mathcal{G}$, and $\theta_{\mathrm{B}_\perp}$ across the full sample, as well as in specific clouds. After applying the relevant significance criteria (Equations \ref{eq:error_cond}, \ref{eq:elong_cond}, and \ref{eq:PI_cond}) and adjusting $\theta_\mathrm{C}$ and $\theta_\mathcal{G}$ to be in the same Galactic coordinate system as $\theta_{\mathrm{B}_\perp}$ (see Section \ref{crossmatch} for details), we directly compare the degree of their similarity by computing the distributions of $|\theta_\mathrm{C} - \theta_\mathcal{G}|$, $|\theta_\mathrm{C} - \theta_{\mathrm{B}_\perp}|$, and $|\theta_\mathcal{G} - \theta_{\mathrm{B}_\perp}|$, respectively. Figures \ref{fig:vectors_1} and \ref{fig:vectors_2} show the cumulative distribution functions (CDFs) for these absolute differences between the pairs of vectors as solid black curves, organized with respect to the host cloud, with the last row in Figure \ref{fig:vectors_2} showing the CDFs of the full sample of cross-matched cores. In each panel of these plots, a dashed black line represents the expected result if the relative alignment between the respective pair of vectors is completely random. If the CDF falls above or below the black dashed line, it signifies a preferred alignment or anti-alignment between the vectors, respectively. We also look at the $\theta_\mathcal{G}$ distributions of cores to ascertain whether there is any preferred orientation of core-scale gas motions across clouds.

To test rigorously whether the relative alignment or anti-alignment between pairs of vectors is significantly different than one drawn from a random distribution, we employ the Anderson-Darling \citep{ADtest} and Kolmogorov-Smirnov tests \citep{KStest}. We utilize the Python implementation of these tests ({\tt anderson\_ksamp} and {\tt ks\_2samp}, respectively) in the {\tt scipy.stats} (V1.9.3) package \citep{2020SciPy-NMeth}. For each cloud and each pair of vectors, we perform both tests on the observed distributions of relative angles and a random distribution with the same sample size. The Anderson-Darling test statistic and significance level (floored at $0.001$ and capped at $0.25$) and Kolmogorov-Smirnov test statistic and p-value (assuming a two-sided null hypothesis) are reported for each pair of vectors and each cloud in Table \ref{tb:vector_stats}. In Table \ref{tb:vector_stats}, we also provide the test statistics and p-value from comparing the $\theta_\mathcal{G}$ distributions of cores against a uniform distribution for a given cloud. We set a threshold for significance at $0.05$ for each test, meaning that we only conclude preferential alignment or anti-alignment if the distribution rejects the null hypothesis of being drawn from a random distribution to $\geq 95\%$ (i.e., an Anderson-Darling test significance level and Kolmogorov-Smirnov test p-value are $\leq 0.05$, respectively) for at least one test. 

In most cases, the Anderson-Darling test significance level tends to be smaller than the Kolmogorov-Smirnov test p-value. The Kolmogorov-Smirnov statistic measures the supremum between the observed and a uniform CDF and, as a consequence, it is more sensitive to sharp deviations from uniformity. Conversely, the Anderson-Darling statistic is more sensitive to smaller but more persistent deviations over a large range of relative angles. In our results, we encounter the latter scenario more frequently and therefore believe that the Anderson-Darling test is the more appropriate of the two tests. We still provide the results of both tests as there could be scenarios in our data in which the Kolmogorov-Smirnov is more appropriate and, further, it is useful for comparing to previous studies that chose to use only the Kolmogorov-Smirnov test to look at the relative alignment of magnetic fields with core properties.

When considering the entire sample, we find that the distribution of $|\theta_\mathrm{C} - \theta_\mathcal{G}|$, $|\theta_\mathrm{C} - \theta_{\mathrm{B}_\perp}|$, and $|\theta_\mathcal{G} - \theta_{\mathrm{B}_\perp}|$ are all consistent with being drawn from a random distribution. Specifically, $|\theta_\mathrm{C} - \theta_\mathcal{G}|$ and $|\theta_\mathcal{G} - \theta_{\mathrm{B}_\perp}|$ are both very close to being randomly distributed, while $|\theta_\mathrm{C} - \theta_{\mathrm{B}_\perp}|$ has a small deviation towards being anti-aligned but not at a statistically significant level. For example, the Anderson-Darling significance level is $0.16$ and the Kolmogorov-Smirnov p-value is $0.32$. Therefore, we conclude that there is no preferential alignment or anti-alignment between $\theta_\mathrm{C}$, $\theta_\mathcal{G}$, and $\theta_{\mathrm{B}_\perp}$. 

Separating cores by host cloud, we find that almost all clouds have distributions of $|\theta_\mathrm{C} - \theta_\mathcal{G}|$, $|\theta_\mathrm{C} - \theta_{\mathrm{B}_\perp}|$, and $|\theta_\mathcal{G} - \theta_{\mathrm{B}_\perp}|$ that are consistent with being drawn from a random distribution, except for two cases. The first exception is in the Perseus cloud, where $\theta_\mathrm{C}$ and $\theta_{\mathrm{B}_\perp}$ show a marginal preference for anti-alignment, which is supported by the Anderson-Darling (with a significance level of $0.15$) and Kolmogorov-Smirnov (with a p-value of $0.19$) tests. While the deviation from a random distribution is apparent by eye in Figure \ref{fig:vectors_1}, it does not pass our conservative statistical criteria for either test and is therefore deemed to be not statistically significant. The second exception is in the Cepheus cloud, in which $|\theta_\mathcal{G} - \theta_{\mathrm{B}_\perp}|$ shows a marginal preference for anti-alignment (with an Anderson-Darling test significance level of $0.04$ and Kolmogorov-Smirnov test p-value of $0.13$). In this case, the significance criteria is met for the Anderson-Darling test but not for the Kolmogorov-Smirnov test. We note that the GAS data only covers two regions of Cepheus (L1228 and L1251) and, thus, only nine dense cores are included in the $|\theta_\mathcal{G} - \theta_{\mathrm{B}_\perp}|$ distribution of Cepheus, making it difficult to draw a strong conclusion.

Further, we find no evidence for preferential alignment of velocity gradient vectors $\theta_\mathcal{G}$ of cores in any cloud or in the sample as a whole. In both cases, the distribution of $\theta_\mathcal{G}$ is consistent with being drawn from a random distribution between $-180^{\circ} \leq \theta_\mathcal{G} \leq 180^{\circ}$ based on Anderson-Darling and Kolmogorov-Smirnov tests.

\begin{figure}
    \centering
    \includegraphics[width=0.475\textwidth]{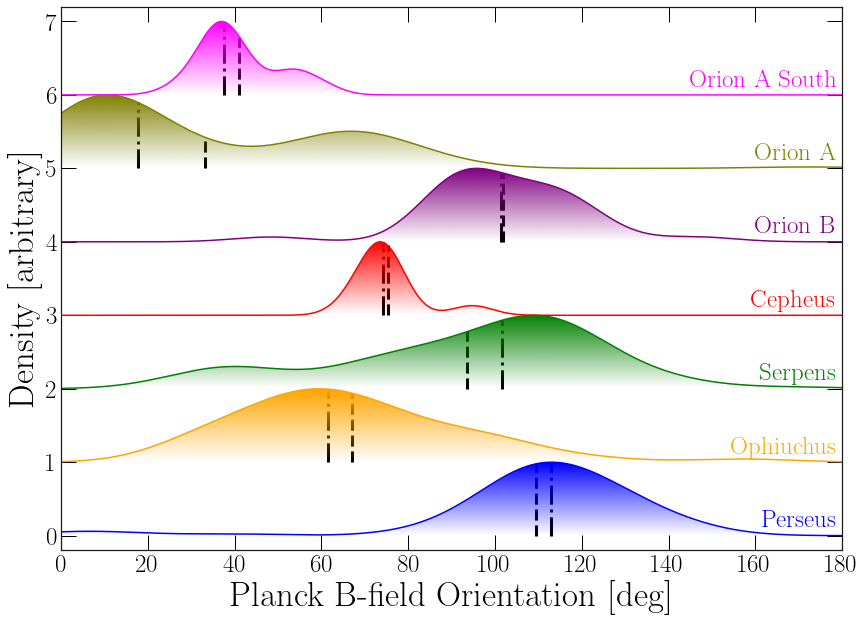}
    \caption{$\theta_{\mathrm{B}_\perp}$ distributions for cores colored with respect to the host cloud and presented as kernel density estimates. Here, $\theta_{\mathrm{B}_\perp}$ is measured in degrees counter-clockwise from Galactic north. Each distribution is normalized and the y-axis units are arbitrary. The mean and median of each distribution are shown as a dashed line and a dashed-dotted line, respectively.}
    \label{fig:b_dist}
\end{figure}

Kernel density estimates of the \textit{Planck} $\theta_{\mathrm{B}_\perp}$ distributions for cores in each cloud (distinguished by color) are presented in Figure \ref{fig:b_dist}. The mean (dashed black line) and median (dashed-dotted black line) are overlaid with each respective distribution. The mean ($\mu_{\mathrm{B}_\perp}$), median ($\eta_{\mathrm{B}_\perp}$), and standard deviation ($\sigma_{\mathrm{B}_\perp}$) of the \textit{Planck} $\theta_{\mathrm{B}_\perp}$ distributions for cores in each region are given in Table \ref{tb:b_dist}. The Figure shows that the magnetic field orientation at the positions of cores within each region is not random, although varying amounts of spread around the mean orientation is seen across regions. In Section \ref{mag_by_region}, we discuss the interpretation of these distributions and the potential impact the observed dispersions in $\theta_{\mathrm{B}_\perp}$ may have on the relative alignment between $\theta_\mathrm{C}$, $\theta_\mathcal{G}$, and $\theta_{\mathrm{B}_\perp}$.

\begin{table*}
\begin{center}
\caption{Summary of Anderson-Darling and Kolmogorov–Smirnov tests on whether $\theta_\mathcal{G}$ is consistent with being drawn from a uniform distribution between $-180^\circ \leq \theta \leq 180^\circ$ and whether the relative alignment between the $\theta_\mathrm{C}$, $\theta_\mathcal{G}$, and $\theta_{\mathrm{B}_\perp}$ vectors are consistent with being drawn from a uniform distribution between $0^\circ \leq \theta \leq 90^\circ$. The results are organized with respect to the host cloud of each core. For the Anderson-Darling test, the statistic is presented with the significance level given in parentheses. The Kolmogorov-Smirnov test results show the statistic with the associated p-value given in parentheses. Results that pass the $\geq 95\%$ significance threshold are presented in boldface.} \label{tb:vector_stats}
\begin{tabular}{ccccccccc} 
\hline
\hline
 & \multicolumn{4}{c}{\textbf{Anderson-Darling Test}} & \multicolumn{4}{c}{\textbf{Kolmogorov–Smirnov Test}}\\
\hline
Cloud & $\theta_\mathcal{G}$ & $|\theta_\mathrm{C} - \theta_\mathcal{G}|$ & $|\theta_\mathrm{C} - \theta_{\mathrm{B}_\perp}|$ & $|\theta_\mathcal{G} - \theta_{\mathrm{B}_\perp}|$ & $\theta_\mathcal{G}$ & $|\theta_\mathrm{C} - \theta_\mathcal{G}|$ & $|\theta_\mathrm{C} - \theta_{\mathrm{B}_\perp}|$ & $|\theta_\mathcal{G} - \theta_{\mathrm{B}_\perp}|$\\
\hline
Perseus & $-1.12$ ($\geq 0.25$) & $-1.09$ ($\geq 0.25$) & $0.82$ ($0.15$) & $-0.41$ ($\geq 0.25$) & $0.06$ ($1.0$) & $0.09$ ($0.98$) & $0.20$ ($0.19$) & $0.13$ ($0.66$)\\
Ophiuchus & $0.55$ ($0.20$) & $-0.57$ ($\geq 0.25$) & $-1.17$ ($\geq 0.25$) & $-1.14$ ($\geq 0.25$) & $0.21$ ($0.25$) & $0.16$ ($0.64$) & $0.10$ ($0.99$) & $0.07$ ($1.00$)\\
Serpens & $-0.92$ ($\geq 0.25$) & $-0.40$ ($\geq 0.25$) & $-0.21$ ($\geq 0.25$) & $-0.86$ ($\geq 0.25$) & $0.08$ ($0.94$) & $0.12$ ($0.66$) & $0.11$ ($0.63$) & $0.07$ ($0.98$)\\
Cepheus & $-0.34$ ($\geq 0.25$) & $-0.30$ ($\geq 0.25$) & $-0.39$ ($\geq 0.25$) & $\boldsymbol{2.14}$ \textbf{($\boldsymbol{0.04}$)} & $0.33$ ($0.73$) & $0.33$ ($0.73$) & $0.30$ ($0.79$) & $0.56$ ($0.13$)\\
Orion B & $-0.84$ ($\geq 0.25$) & $-0.30$ ($\geq 0.25$) & $-0.25$ ($\geq 0.25$) & $-0.56$ ($\geq 0.25$) & $0.13$ ($0.99$) & $0.27$ ($0.39$) & $0.18$ ($0.77$) & $0.17$ ($0.89$)\\
Orion A & $-0.32$ ($\geq 0.25$) & $0.22$ ($\geq 0.25$) & $-0.66$ ($\geq 0.25$) & $0.40$ ($0.23$) & $0.11$ ($0.78$) & $0.16$ ($0.40$) & $0.10$ ($0.83$) & $0.14$ ($0.51$)\\
Orion A South & $-1.11$ ($\geq 0.25$) & $-0.77$ ($\geq 0.25$) & $-0.22$ ($\geq 0.25$) & $-0.59$ ($\geq 0.25$) & $0.08$ ($1.0$) & $0.24$ ($0.60$) & $0.20$ ($0.59$) & $0.17$ ($0.90$)\\
\hline
All cores & $0.16$ ($\geq 0.25$) & $-1.18$ ($\geq 0.25$) & $0.73$ ($0.16$) & $-0.84$ ($\geq 0.25$) & $0.06$ ($0.51$) & $0.03$ ($1.0$) & $0.07$ ($0.32$) & $0.03$ ($0.99$)\\
\hline
\hline
\end{tabular}
\end{center}
\end{table*}

\begin{table}
\begin{center}
\caption{Mean, median, and standard deviation for each of the $\theta_{\mathrm{B}_\perp}$ distributions shown in Figure \ref{fig:b_dist}.} \label{tb:b_dist}
\begin{tabular}{cccc} 
\hline
\hline
Cloud & $\mu_{\mathrm{B}_\perp}$ (deg) & $\eta_{\mathrm{B}_\perp}$ (deg) & $\sigma_{\mathrm{B}_\perp}$ (deg)\\
\hline
Perseus & 109.51 & 112.88 & 26.28\\
Ophiuchus & 66.96 & 61.45 & 25.56\\
Serpens & 93.67 & 101.56 & 29.78\\
Cepheus & 75.29 & 74.26 & 7.07\\
Orion B & 101.88 & 101.49 & 17.02\\
Orion A & 33.08 & 17.72 & 30.82\\
Orion A South & 41.07 & 37.46 & 8.87\\
\hline
\hline
\end{tabular}
\end{center}
\end{table}

\subsection{Physically distinct regions in Orion A and Serpens} \label{orion_serpens_results}
Within our sample, the two clouds with the most dense cores (Serpens and Orion A), may be composed of multiple physically distinct regions of star formation that are projected near one another on the sky. For example, the Serpens region mapped by GAS and HGBS includes both the H{\small II} region W40 and the young, filamentary cluster-forming region Serpens South. \citet{tahani_2022} show that magnetic fields in bubbles like W40 may be pushed toward a tangential morphology relative to the bubble. Serpens South, however, shows clear anti-alignment of the dense filaments with the larger-scale magnetic field \citep{sugitani_2011, kusune_2019}, although \citet{pillai_2020} show that on smaller scales the field bends toward parallel alignment in the southern filament. For the Orion A region, \cite{2015A&A...577L...6S} find a progression in the steepness of the column density probability distribution function from north to south, indicating that the Orion A region may be comprised of multiple distinct star-forming regions at different evolutionary stages.

To explore the possibility that sub-structures in Orion A or Serpens may have distinct physical properties and varying dense core formation and evolution, we decompose the clouds into two sub-regions and analyze the relative alignment of $\theta_\mathrm{C}$, $\theta_\mathcal{G}$, and $\theta_{\mathrm{B}_\perp}$ in each, respectively. In Orion A, we divide the cloud along a declination of $\delta = -5.5^{\circ}$ and, in Serpens, we apply a divider along a Right Ascension of $\alpha = 277.67^{\circ}$, based on the expected sub-structure in each cloud \citep{sugitani_2011, 2015A&A...577L...6S, kusune_2019, pillai_2020, tahani_2022}.

\begin{figure*}
    \centering
    \includegraphics[width=0.98\textwidth]{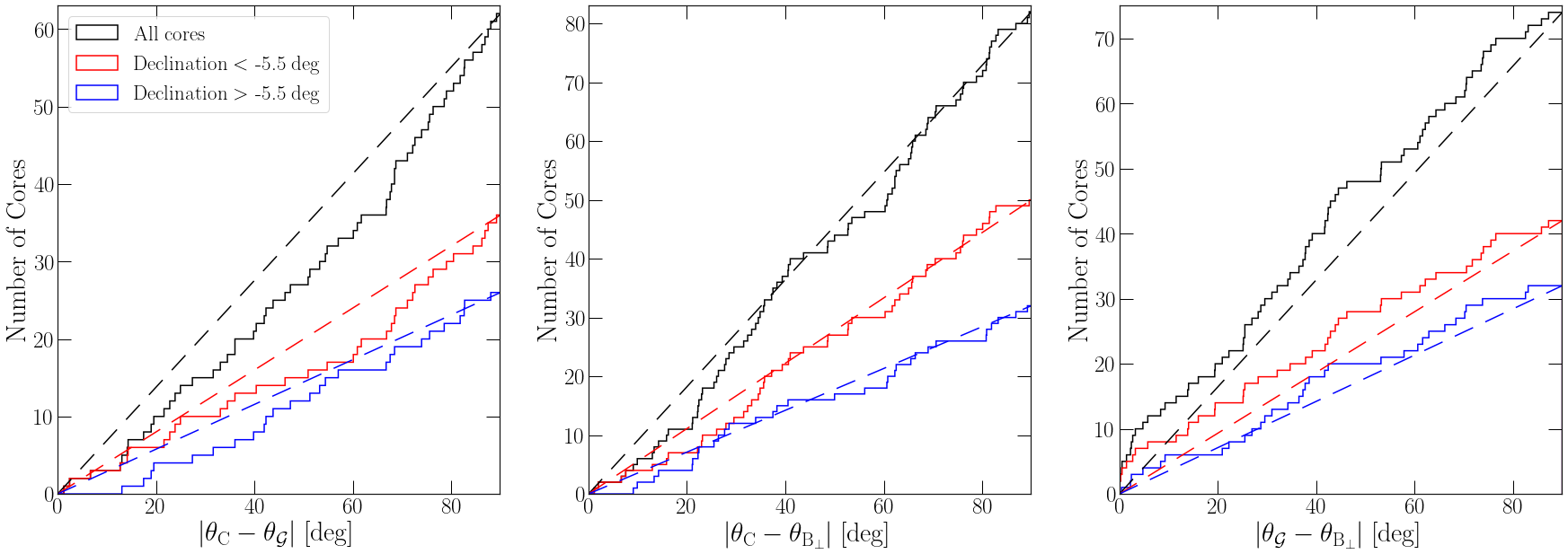}
    \caption{Relative alignment of $\theta_\mathrm{C}$, $\theta_\mathcal{G}$, and $\theta_{\mathrm{B}_\perp}$ presented in the same format as Figures \ref{fig:vectors_1} and \ref{fig:vectors_2}. The black solid curve represents all the cores that pass our respective cuts (Equations \ref{eq:error_cond}, \ref{eq:elong_cond}, and \ref{eq:PI_cond}) in the Orion A cloud. The red (blue) solid curve represent the sub-set of cores that lie below (above) $\delta = -5.5^{\circ}$. The red (blue) dashed line depicts a completely random alignment between the respective vectors for the cores below (above) $\delta = -5.5^{\circ}$.}
    \label{fig:vectors_ora}
\end{figure*}

\begin{figure*}
    \centering
    \includegraphics[width=0.98\textwidth]{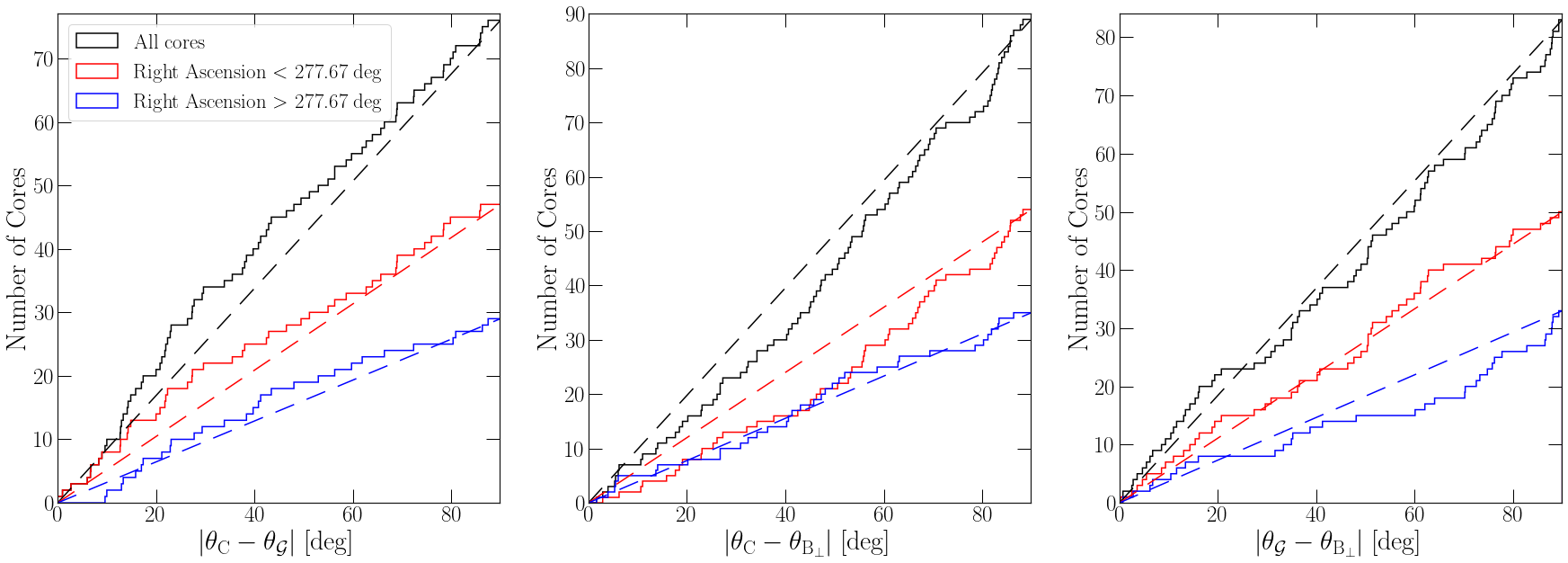}
    \caption{Relative alignment of $\theta_\mathrm{C}$, $\theta_\mathcal{G}$, and $\theta_{\mathrm{B}_\perp}$ following the same formatting as Figure \ref{fig:vectors_ora}. In this case, the red or blue solid curve and dashed line represent cores below or above $\alpha = 277.67^{\circ}$ in Serpens, respectively.}
    \label{fig:vectors_serpens}
\end{figure*}

Figure \ref{fig:vectors_ora} depicts the CDF of the absolute difference in each pair of vectors for: all cores in the Orion A region (as a black curve), the cores at $\delta < -5.5^{\circ}$ (as a red curve), and the cores at $\delta > -5.5^{\circ}$ (as a blue curve). Similarly, Figure \ref{fig:vectors_serpens} shows the results after dividing the Serpens cores into those at $\alpha < 277.67^{\circ}$ (red curve) and those at $\alpha > 277.67^{\circ}$ (blue curve). The summary statistics from applying the Anderson-Darling and Kolmogorov-Smirnov tests to these sub-regions is presented in Table \ref{tb:vector_stats_add}. We find no evidence for significantly preferential vector alignments between the sub-regions in Orion A for any of the three vector pairs. There appears to be a marginal preference (Anderson-Darling test significance level of $0.19$ and Kolmogorov-Smirnov test p-value of $0.31$) for $\theta_\mathrm{C}$ and $\theta_{\mathrm{B}_\perp}$ to be preferentially anti-aligned for cores at $\alpha < 277.67^{\circ}$ in Serpens while the cores at $\alpha > 277.67^{\circ}$ are better matched with a random distribution. This slight difference could point towards distinct properties between the two regions but we note that the deviation from a random distribution for $\alpha < 277.67^{\circ}$ cores is minor and not statistically significant. We find no evidence for different vector alignments for $|\theta_\mathrm{C} - \theta_\mathcal{G}|$ or $|\theta_\mathcal{G} - \theta_{\mathrm{B}_\perp}|$ in the Serpens sub-regions.

\begin{table*}
\begin{center}
\caption{Summary of Anderson-Darling and Kolmogorov–Smirnov tests on regions of Orion A and Serpens that are divided along a selected $\delta$ and $\alpha$, respectively. The Table is formatted identically to Table \protect\ref{tb:vector_stats}.} \label{tb:vector_stats_add}
\begin{tabular}{ccccccc} 
\hline
\hline
 & \multicolumn{3}{c}{\textbf{Anderson-Darling Test}} & \multicolumn{3}{c}{\textbf{Kolmogorov–Smirnov Test}}\\
\hline
Region & $|\theta_\mathrm{C} - \theta_\mathcal{G}|$ & $|\theta_\mathrm{C} - \theta_{\mathrm{B}_\perp}|$ & $|\theta_\mathcal{G} - \theta_{\mathrm{B}_\perp}|$ & $|\theta_\mathrm{C} - \theta_\mathcal{G}|$ & $|\theta_\mathrm{C} - \theta_{\mathrm{B}_\perp}|$ & $|\theta_\mathcal{G} - \theta_{\mathrm{B}_\perp}|$\\
\hline
Orion A ($\delta < -5.5$~deg) & $-0.41$ ($\geq 0.25$) & $-0.82$ ($\geq 0.25$) & $-0.06$ ($\geq 0.25$) & $0.19$ ($0.51$) & $0.12$ ($0.87$) & $0.14$ ($0.79$)\\
Orion A ($\delta > -5.5$~deg) & $-0.14$ ($\geq 0.25$) & $-0.79$ ($\geq 0.25$) & $-0.62$ ($\geq 0.25$) & $0.19$ ($0.73$) & $0.13$ ($0.97$) & $0.16$ ($0.84$)\\
Serpens ($\alpha < 277.67$~deg) & $-0.55$ ($\geq 0.25$) & $0.61$ ($0.19$) & $-1.06$ ($\geq 0.25$) & $0.13$ ($0.84$) & $0.19$ ($0.31$) & $0.10$ ($0.97$)\\
Serpens ($\alpha > 277.67$~deg) & $-0.60$ ($\geq 0.25$) & $-1.04$ ($\geq 0.25$) & $-0.03$ ($\geq 0.25$) & $0.14$ ($0.95$) & $0.11$ ($0.98$) & $0.21$ ($0.45$)\\
\hline
\hline
\end{tabular}
\end{center}
\end{table*}

\subsection{Core kinematics and relative alignment by core type} \label{core_type_results}
For our cross-matched catalogue, we have the additional insight of knowing the core type of each dense core (i.e., whether they are starless, prestellar, or protostellar, see Section \ref{data_continuum} for details on the classification scheme). Of the 399 dense cores in our sample, 47 are identified as starless cores, 239 are prestellar cores, and 113 are protostellar cores. We analyze the derived velocity gradient magnitudes $\mathcal{G}$ and specific angular momenta $J/M$ distributions of the 329 cores with core type classifications that satisfy Equation \ref{eq:error_cond} and find no significant differences in the populations.

\begin{figure*}
    \centering
    \includegraphics[width=0.98\textwidth]{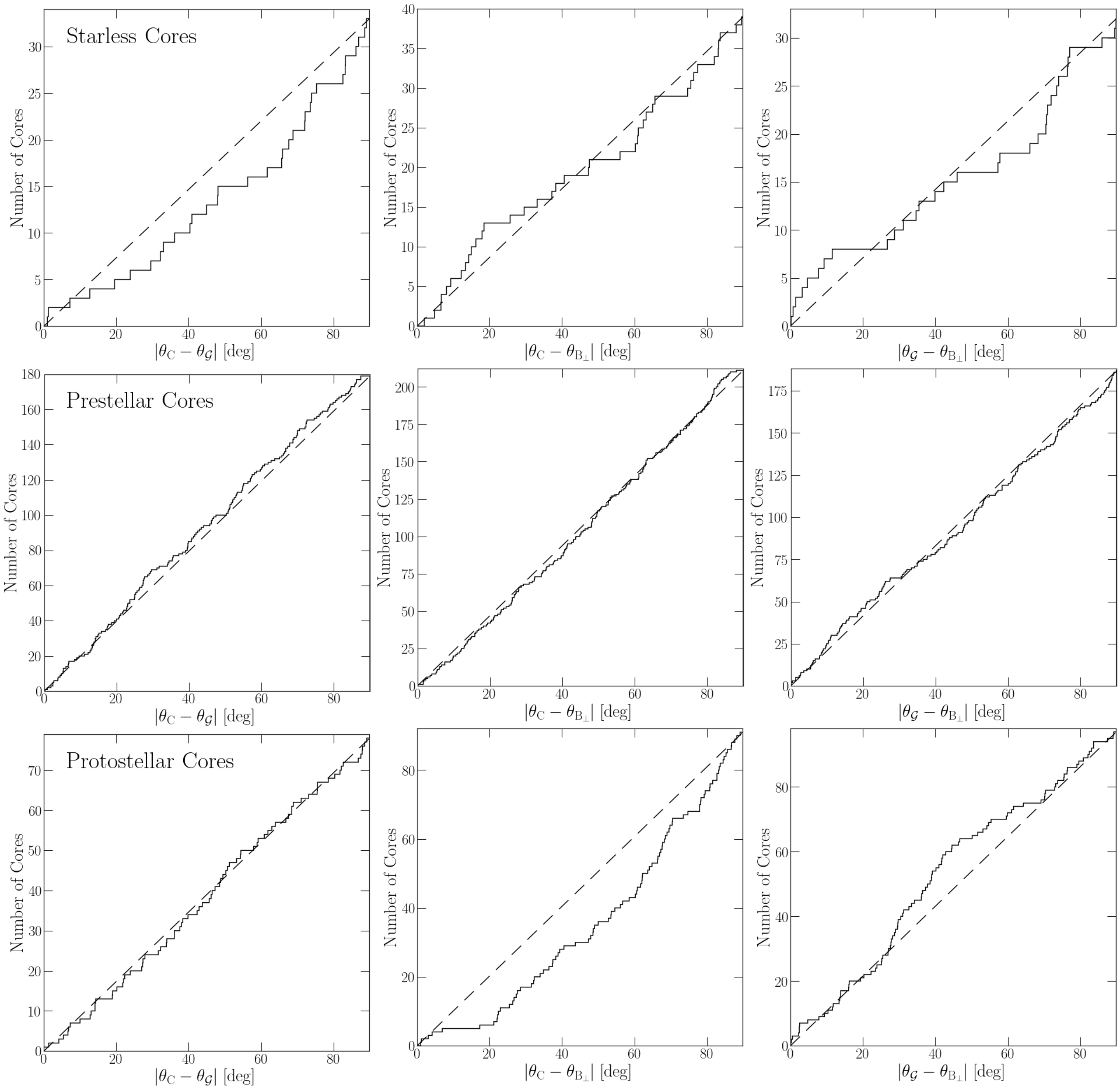}
    \caption{Cumulative distribution function comparing the relative alignment of $\theta_\mathrm{C}$, $\theta_\mathcal{G}$, and $\theta_{\mathrm{B}_\perp}$. The formatting for this Figure follows that of Figures \ref{fig:vectors_1} and \ref{fig:vectors_2}. The top row shows the cumulative distribution function for cores identified as starless, while the middle row is for all prestellar cores, and the bottom row is for protostellar cores.}
    \label{fig:vectors_core_type}
\end{figure*}

Similar to Figures \ref{fig:vectors_1} and \ref{fig:vectors_2}, we plot the CDFs for $|\theta_\mathrm{C} - \theta_\mathcal{G}|$, $|\theta_\mathrm{C} - \theta_{\mathrm{B}_\perp}|$, and $|\theta_\mathcal{G} - \theta_{\mathrm{B}_\perp}|$ in Figure \ref{fig:vectors_core_type}, this time separating the results based on core type. The top row of Figure \ref{fig:vectors_core_type} shows the relative alignment results for each pair of vectors for starless cores, followed by the results for prestellar cores in the middle row, and for protostellar cores in the bottom row. The summary of results for the Anderson-Darling and Kolmogorov-Smirnov tests on the distributions presented in Figure \ref{fig:vectors_core_type} is given in Table \ref{tb:vector_stats_coretype}. In most instances, the relative alignment between each of the pair of vectors is consistent with being randomly aligned, except in one case. There is strong evidence for a preferred anti-alignment between core orientation and ambient magnetic field direction for protostellar cores (with an Anderson-Darling test significance level of $0.01$ and a Kolmogorov-Smirnov test p-value of $0.05$). In this case, the $\geq 95\%$ significance criteria is met for both tests. This anti-alignment is unique to protostellar cores, as we do not see a similar preferential alignment or anti-alignment in starless or prestellar cores. We discuss the implications of the lack of alignment or alignment between these vectors in the following section.

\begin{table*}
\begin{center}
\caption{Summary of Anderson-Darling and Kolmogorov–Smirnov tests on starless, prestellar, and protostellar cores. The Table is formatted identically to Tables \protect\ref{tb:vector_stats} and \protect\ref{tb:vector_stats_add}.} \label{tb:vector_stats_coretype}
\begin{tabular}{ccccccc} 
\hline
\hline
 & \multicolumn{3}{c}{\textbf{Anderson-Darling Test}} & \multicolumn{3}{c}{\textbf{Kolmogorov–Smirnov Test}}\\
\hline
Core Type & $|\theta_\mathrm{C} - \theta_\mathcal{G}|$ & $|\theta_\mathrm{C} - \theta_{\mathrm{B}_\perp}|$ & $|\theta_\mathcal{G} - \theta_{\mathrm{B}_\perp}|$ & $|\theta_\mathrm{C} - \theta_\mathcal{G}|$ & $|\theta_\mathrm{C} - \theta_{\mathrm{B}_\perp}|$ & $|\theta_\mathcal{G} - \theta_{\mathrm{B}_\perp}|$\\
\hline
Starless & $-0.03$ ($\geq 0.25$) & $-1.00$ ($\geq 0.25$) & $-0.75$ ($\geq 0.25$) & $0.21$ ($0.45$) & $0.13$ ($0.91$) & $0.16$ ($0.84$)\\
Prestellar & $-0.71$ ($\geq 0.25$) & $-0.76$ ($\geq 0.25$) & $-0.93$ ($\geq 0.25$) & $0.06$ ($0.94$) & $0.04$ ($0.99$) & $0.04$ ($1.00$)\\
Protostellar & $-1.14$ ($\geq 0.25$) & $\boldsymbol{3.76}$ \textbf{($\boldsymbol{0.01}$)} & $-0.12$ ($\geq 0.25$) & $0.05$ ($1.00$) & $\boldsymbol{0.20}$ \textbf{($\boldsymbol{0.05}$)} & $0.14$ ($0.27$)\\
\hline
\hline
\end{tabular}
\end{center}
\end{table*}

\subsection{Hypothesis testing considerations} \label{stats_caveats}
\subsubsection{Accounting for multiple hypothesis tests applied on the same data}
For any individual instance in which we employ the Anderson-Darling or Kolmogorov-Smirnov hypothesis tests, we are determining whether a detection of alignment or anti-alignment is significant by rejecting the null hypothesis$-$that it is drawn from a uniform distribution$-$if the likelihood of observing the data is $\leq 0.05$. Every time we reorganize the data into a different configuration (e.g., demarcating cores by host cloud, dividing with respect to on-sky positions, or separating by core type) and repeat the hypothesis tests, we increase the likelihood of identifying a false positive detection. In our analysis, we conduct a total of $15$ hypothesis tests (one with the full catalogue, seven individual regions, four on-sky regions based on the cuts in Figures \ref{fig:vectors_ora} and \ref{fig:vectors_serpens}, and three core types), find two significant detections (anti-alignment in $|\theta_\mathcal{G} - \theta_{\mathrm{B}_\perp}|$ in Cepheus, and anti-alignment in $|\theta_\mathrm{C} - \theta_{\mathrm{B}_\perp}|$ for protostellar cores) satisfying our significance level of $0.05$.

The most conservative approach to address the concern of increasing false positive likelihood is by applying the Bonferroni correction \citep{bonferroni}. Rather than testing each hypothesis at the $0.05$ level, we would instead test them at the $0.05/15 \sim 0.003$ level to account for the total number of tests conducted. Following this approach, we cannot conclusively claim that either of our detections are significant to the desired level. This approach, however, does not definitively prove that our detections are false positives and, in addition, applying the Bonferroni correction also comes with the downside of increasing the likelihood of false negative events.

To get a better understanding of how likely it is that our detections are false positives, we check against the corresponding binomial distribution. For $n$ independent hypothesis tests with a probability of success of $p$ (and probability of failure $q = 1 - p$), the probability $P_k$ of getting $k$ successful trials is given by
\begin{equation}
P_k = \binom{n}{k} p^k q^{n - k}\,. \label{eq:binomial}
\end{equation}
In our case, $n = 15$, $k = 2$, $p = 0.05$, and $q = 0.95$, which gives $P_k \sim 0.13$. This result means that the probability that the two detections we see are false positives is only $\sim 13\%$. Since the likelihood of these detections being false positives is fairly low, we treat the anti-alignment seen in the $|\theta_\mathcal{G} - \theta_{\mathrm{B}_\perp}|$ distribution in Cepheus and the anti-alignment in $|\theta_\mathrm{C} - \theta_{\mathrm{B}_\perp}|$ for protostellar cores as significant and interpret the results accordingly.

\subsubsection{Distribution of measured p-values and its implications on the null hypothesis}
In our hypothesis tests, we test against three null hypotheses in a variety of different scenarios, namely: the core elongation axes and velocity gradient are randomly aligned, the core elongation axes and the magnetic field orientation are randomly aligned, and the velocity gradient and magnetic field orientation are randomly aligned. When the null hypothesis is true, we expect to see a uniform distribution of p-values from our hypothesis tests. This statement becomes immediately clear if we consider that the p-value, which itself is a random variable, is the probability integral transform of the associated test statistic \citep[a proof of this statement is provided in Chapter 2 by][]{rice2007}. The uniformity of p-values under the null hypothesis is better visualized by \cite{murdoch2008} who use Monte Carlo simulations to emphasize that p-values are random variables.

\begin{figure*}
    \centering
    \includegraphics[width=0.33\textwidth]{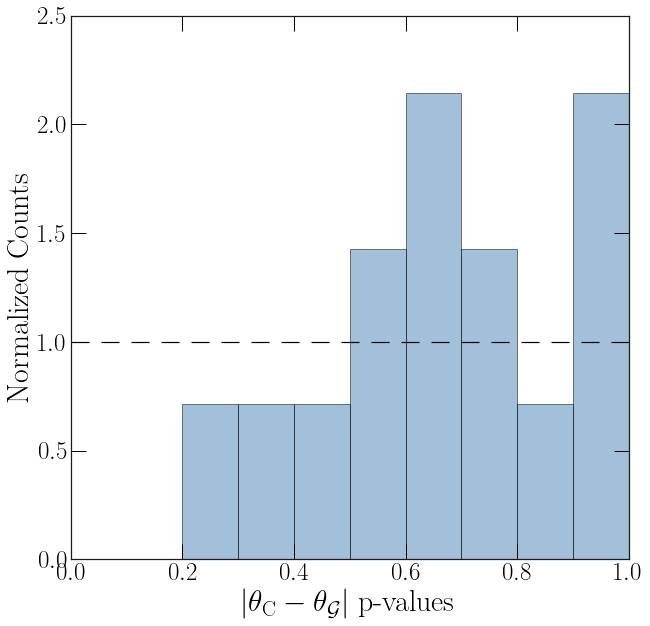}
    \includegraphics[width=0.33\textwidth]{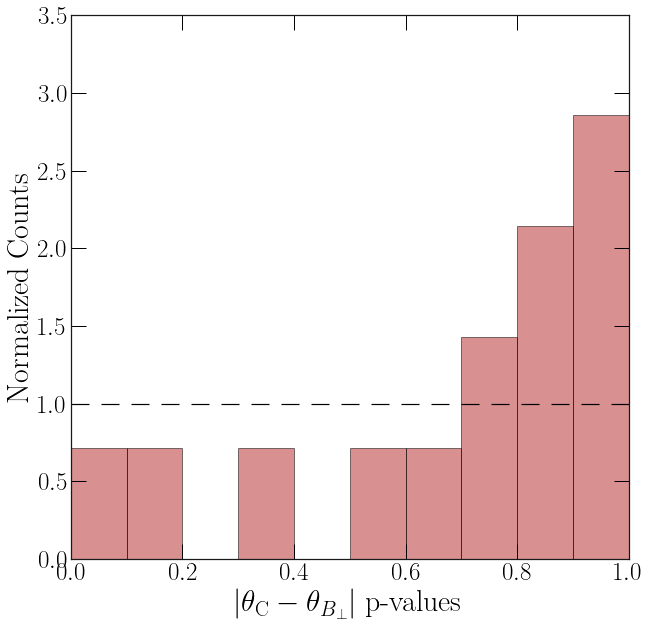}
    \includegraphics[width=0.3215\textwidth]{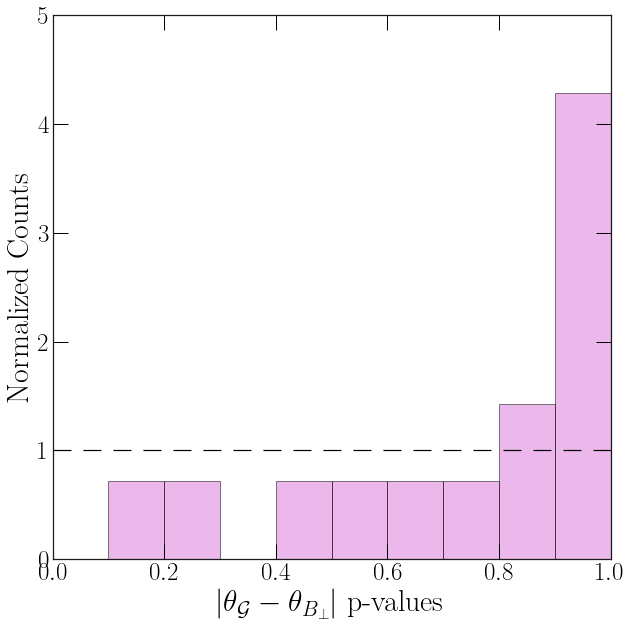}
    \caption{The distribution of Kolmogorov-Smirnov p-values for the three pairs of vectors (left: $|\theta_\mathrm{C} - \theta_\mathcal{G}|$, middle: $|\theta_\mathrm{C} - \theta_{\mathrm{B}_\perp}|$, and right: $|\theta_\mathcal{G} - \theta_{\mathrm{B}_\perp}|$) across the $15$ hypothesis tests we perform throughout this work. The counts have been normalized such that the area under the histogram sums to $1$. The dashed black line represents a uniform distribution over the same range.} 
    \label{fig:pvals}
\end{figure*}

In Figure \ref{fig:pvals}, we plot the normalized distribution of Kolmogorov-Smirnov test p-values for each of our $15$ null hypotheses and overlay a uniform distribution as a dashed black line. Note that due to the implementation of the Anderson-Darling test that we utilize ({\tt scipy.stats.anderson\_ksamp}), the p-values are floored at $0.001$ and capped at $0.25$, thus we only test for p-value uniformity with the Kolmogorov-Smirnov test. In all three cases, we see that the observed p-value distributions are not uniform and are instead skewed towards higher p-values. The skew suggests that the null hypotheses is not the best descriptor for the underlying correlation in the data. Yet, we also do not see many significant detections of preferred alignment or anti-alignment. Jointly, these two results may imply that $\theta_\mathrm{C}$, $\theta_\mathcal{G}$, and $\theta_{\mathrm{B}_\perp}$ are not truly randomly aligned with respect to one another and instead have some minor preferred orientation that does not always pass the significance criteria we chose.

\section{Discussion} \label{discussion}
\subsection{Velocity Gradient Magnitude and Angular Momenta} \label{sec4.1}
\subsubsection{Comparing with expectations of solid body rotation and turbulent cores} \label{sec4.1.1}
Of the 399 dense cores in our cross-matched sample, 329 satisfy Equation \ref{eq:error_cond} and have velocity distributions that are well fit with a 2D linear gradient, consistent with solid-body rotation projected on the plane of the sky. As we have previously noted, and will expand upon in Section \ref{sec4.1.2}, gas motions other than rotation can produce similar observed velocity gradients. In Section \ref{kinematics_results}, we computed the specific angular momentum as a function of core radius $J/M$ and determine a best-fit scaling relation of $J/M \propto R^{1.82 \pm 0.10}$ for the 329 cores. Our results agree to within uncertainties with the $J/M \propto R^{1.6 \pm 0.2}$ relation found by \cite{1993ApJ...406..528G} but is steeper than the $J/M \propto R^{1.5}$ scaling that \cite{2018ApJ...865...34C} found in their turbulent MHD simulation of bound cores. Our results agree with the late stage gravitationally bound cores in the decaying turbulence MC simulations by \cite{2010ApJ...723..425D} which scale as $J/M \propto R^{1.8}$ in 3D space. The spread in $J/M$ across our sample is quite large (spanning $\sim 10^{-5} - 10^{-1}~\mathrm{pc}~\mathrm{km}~\mathrm{s}^{-1}$) compared to that seen in hydrodynamic simulations of cores by \cite{2019ApJ...876...33K}, who find variations on only the $\lesssim 10\%$ level. It is possible that the additional spread in our measurements is introduced due to the limited 2D projection we observe instead of the full 3D core mass, radius, and velocity available in the \cite{2019ApJ...876...33K} simulation.

As noted in Section \ref{kinematics_results}, and in agreement with \cite{1993ApJ...406..528G}, the angular velocity gradient $\mathcal{G}$ only has a very marginal scaling with $R$. Thus, the expected $J/M$ and $R$ scaling relation for a sample of cores dominated by solid-body rotation would be $J/M \propto R^2$. On the other hand, we expect a $J/M \propto R^{1.5}$ scaling for turbulence dominated cores based on MHD simulation results \citep[e.g.,][]{2000ApJ...543..822B, 2018ApJ...865...34C}. \cite{2022ApJ...925...78A} are able to replicate the $J/M \propto R^{1.5}$ relation using a smoothed particle hydrodynamics simulation of the formation, collapse and fragmentation of giant MCs and conclude that the relationship emerges from a combination of gravitational contraction and angular momentum loss via turbulent viscosity. Our results are not in total agreement with either of these scenarios. Instead, our derived scaling relation lies in between the expected results of turbulence-dominated cores and rotating solid bodies. This result might suggest that while the observed velocity gradients have a significant contribution from solid body rotation, there is still a substantial turbulence component on core scales. Note that it is possible that our methodology is prone to observational bias, as we are only measuring the velocity gradients projected in two dimensions. In their simulation, \cite{2010ApJ...723..425D} see that $J/M$ is often overestimated when estimated using only a 2D velocity gradient. Further, the $J/M$ and $R$ scaling is shallower with only 2D information than the true scaling in three dimensions.

\subsubsection{Contributions to velocity gradients beyond rotation and the linearity of gradients} \label{sec4.1.2}
We probe core velocity gradients on $\sim 0.02 - 0.07$~pc scales and fit them as simple 2D linear velocity gradients. Here, we briefly discuss potential contributors to the bulk core motions other than rotation. Some of these mechanisms, such as outflows, would only affect a subset of our core sample. In general, determining the detailed velocity profile of an individual core would require focused, pixel-by-pixel analysis that is beyond the scope of this paper. Instead, our goal here is to apply a consistent method across a large sample of cores to analyze results on a population level. Despite the potential for individual core velocity distributions to be influenced by various bulk motions, the 2D linear velocity gradient analysis produces accurate measures of the core specific angular momenta and their orientations across a sample of cores.

As magnetized cores evolve and collapse into protostars, they may develop complex velocity structure that deviate from linear velocity gradients on smaller scales. For example, \cite{2012ApJ...761...40K} show that on $10^3$~AU scales ($\sim 5 \times 10^{-3}$~pc scales, smaller than probed by our data), simulated cores exhibit varying velocity gradients that depend on the viewing angle, magnetic field strength, and initial orientation of the rotation axis with respect to the magnetic field. Multiple velocity components along the line-of-sight have been identified in the GAS NH$_3$ data toward several regions, such as the NGC 1333 complex \citep{2020ApJ...891...84C} and L1688 \citep{choudhury_2020}, although as noted in Section 2.1 most spectra are well-characterized by a single velocity component. Toward the L1688 cores, the second velocity components tend to have low line brightness relative to the core component, and would not impact this analysis. 

Another consideration is that there may be multiple local gradients present across cores \cite[e.g.,][]{2002ApJ...572..238C, 2007A&A...470..221C} or contributions to the core velocity gradient from non-rotation gas motions \citep[e.g., mass flow along filaments and infalling gas][]{2020ApJ...891...84C, 2022ApJ...935...57C}. When detected, infall motions are generally subsonic on core scales \citep[e.g.][]{lee_1999,campbell_2016}. Infall motions, or the existence of multiple, significant local gradients contributing to the overall velocity gradient of a core, would manifest as broadening of the observed NH$_3$ line width in the GAS observations. We note, however, that only $48$ of the $329$ cores that satisfy Equation \ref{eq:error_cond} have a velocity dispersion that is larger than twice the sound speed at an assumed temperature of $15$~K. This suggests that most cores in our sample do not have significant contributions from multiple, distinct local gradients. 

The gravitationally bound, protostellar cores in our sample that satisfy Equation \ref{eq:error_cond} ($113$ of $329$) may host outflows, which can contribute to their observed velocity distributions \citep[e.g., in the case of G31.41+0.31;][we note, however, that this is a massive star-forming region where outflow impacts would be larger than in our low-mass protostellar cores]{2021A&A...648A.100B}. \cite{2011ApJ...740...45T} conduct the same $J/M$ analysis using 2D velocity gradients from NH$_3$ (1,1) and (2,2) and N$_2$H$^\mathrm{+}$ (1-0) observations for a sample of $17$ protostars. They find that, while some cores do contain outflows, they are highly collimated and do not significantly affect the large-scale, 2D velocity structure.

\cite{2000ApJ...543..822B} point out that linear line-of-sight velocity gradients can arise even when the underlying velocity profile is driven by turbulence, but note that the distribution of $J/M$ derived using the 2D linear fit method remains accurate across a sample of cores.

Finally, there may be some cores that contain a significant velocity gradient (i.e., they pass the significance criteria in Equation \ref{eq:error_cond}) but their velocity distribution is substantially non-linear. For a sample of 18 droplets in the GAS data set, \cite{2019ApJ...886..119C} determine the linearity of the velocity gradients by manually comparing the pixel-by-pixel CDFs of the local velocity gradient orientation, only finding $3$ that depict non-linear velocity gradients. The application of their methodology to a much larger sample of cores is beyond the scope of this work but merits a focused, independent study. The combination of a $3\sigma$ cut on the gradient (Equation \ref{eq:error_cond}) and small fraction of cores with non-linear gradients seen by \cite{2019ApJ...886..119C} in GAS data give us confidence that our sample is not dominated by non-linear gradients.

\subsection{Relative alignment of the core orientation, velocity gradient, and ambient magnetic field}
In the next section, we discuss the relative alignment, or lack thereof, between core orientations ($\theta_\mathrm{C}$), velocity gradient directions ($\theta_\mathcal{G}$), and the orientation of the ambient magnetic field ($\theta_{\mathrm{B}_\perp}$). Regarding $\theta_{\mathrm{B}_\perp}$, it is important to note that we trace magnetic field orientations on $\sim 0.25 - 0.8$~pc scales, which are larger than the core-scale morphology and velocity gradients we analyze. In Section \ref{b_cloud_to_core}, we explore the transition between cloud- and core-scale magnetic fields and argue that current observations support frequent alignment from \textit{Planck} to core scales. Higher resolution surveys tracking magnetic field orientations from clouds to cores are still needed to definitively resolve this question.

We see no preferred orientation of the velocity gradient direction within any region \citep[as seen for cores in Orion A by][]{2016PASJ...68...24T}. Further, we find no globally preferred alignment or anti-alignment between $\theta_\mathrm{C}$, $\theta_\mathcal{G}$, and $\theta_{\mathrm{B}_\perp}$ (final row of Figure \ref{fig:vectors_2}). The only region in which we find a statistically significant deviation from a randomly drawn population is in Cepheus (specifically in the L1228 and L1251 regions), which has an anti-alignment between $\theta_\mathcal{G}$ and $\theta_{\mathrm{B}_\perp}$ (see the last panel of Figure \ref{fig:vectors_1}). The $|\theta_\mathrm{C} - \theta_\mathcal{G}|$, $|\theta_\mathrm{C} - \theta_{\mathrm{B}_\perp}|$, and $|\theta_\mathcal{G} - \theta_{\mathrm{B}_\perp}|$ distributions in all other regions covered by our sample are consistent with being drawn from a random population. A significant anti-alignment exists between $\theta_\mathrm{C}$ and $\theta_{\mathrm{B}_\perp}$ specifically for protostellar cores (see the second panel in the last row of Figure \ref{fig:vectors_core_type}), but $\theta_\mathrm{C}$ and $\theta_{\mathrm{B}_\perp}$ are randomly aligned in all other situations.

\subsubsection{Comparing to the classical view of star forming cores} \label{sub:interpretation}
In the classical view of star formation the single, isolated dense cores can be characterized as an oblate spheroid with rotation about its minor axis. Here, the core is flattened along the magnetic field direction which is parallel to the minor axis \citep{1956MNRAS.116..503M, 1966MNRAS.132..359S, mouschovias_1976,1999ApJ...520..706C}. Following this framework, we expect to see the core elongated along the major axis $\theta_\mathrm{C}$, parallel to the velocity gradient direction $\theta_\mathcal{G}$, both of which are perpendicular to the magnetic field orientation $\theta_{\mathrm{B}_\perp}$. 

In the starless and prestellar stages, we observe random alignments between $\theta_\mathrm{C}$, $\theta_\mathcal{G}$, and $\theta_{\mathrm{B}_\perp}$, which is in disagreement with the classical view of star formation. Nevertheless, there is a strong preferential anti-alignment between $\theta_\mathrm{C}$ and $\theta_{\mathrm{B}_\perp}$ in the protostellar stage, which is in line with the classical picture. This difference suggests an evolution in the relative alignment between $\theta_\mathrm{C}$ and $\theta_{\mathrm{B}_\perp}$ from random to perpendicular, with more evolved, protostellar cores having their axes of elongation aligned perpendicular to the local magnetic field. 
Indeed, the protostellar cores in our sample have higher average densities than the starless and prestellar cores.  

In simulations of gravitational fragmentation in sheet-like layers, \citet{2009NewA...14..221B} find that initially magnetically supercritical runs produce elongated cores, where the shortest core axis is preferentially aligned with the magnetic field axis (in agreement with the anti-alignment we find between $\theta_\mathrm{C}$ and $\theta_{\mathrm{B}_\perp}$ in protostellar cores). They argue that dynamical, gravity-dominated fragmentation will accentuate anisotropies and produce the observed alignment in evolved cores. 
Furthermore, even in simulations where magnetic field strengths are relatively weak, cores initially formed without alignment between their minor axis and the local magnetic field direction will change orientation during gravitational collapse such that their minor axis becomes aligned with the local magnetic field direction \citep{matsumoto_2011}. Anisotropic accretion of mass onto cores may also drive a change in orientation as cores grow and evolve. \citet{2014ApJ...785...69C} find in simulations that in early stages of core formation in a post-shock medium, the magnetic field and velocity orientations are random and determined primarily by the local turbulence, while at later times cores gain mass anisotropically, becoming increasingly anti-aligned with the local B-field. The contraction on the core scale must be magnetically regulated if not magnetically dominated; cases of core contraction within relatively weak magnetic fields produce more spherically symmetric cores inconsistent with our observations \citep{2017ApJ...847..104O}. Given that the Planck observations trace the larger-scale magnetic field, a final possibility is that the observed anti-alignment at the protostellar stage results from a selection effect, whereby starless cores with this initial orientation relative to the larger-scale magnetic field are more likely to contract and form protostars.  

The lack of preferential alignment or anti-alignment in the $|\theta_\mathrm{C} - \theta_\mathcal{G}|$ distribution, even for protostellar cores, may be indicative of the cores being triaxial rather than oblate, which is in line with multiple turbulent simulations that find a predominately triaxial distribution of cores \citep[e.g., ][]{2000ApJS..128..287K, 2003ApJ...592..203G, 2004ApJ...607L..39B, 2004ApJ...605..800L, 2008ApJ...686.1174O, 2009ApJ...693..914O}, as well as observations \citep{basu_2000, jones_2001, lomax_2013}. 

\subsubsection{Comparing with previous simulations and observations of cores}
In most cases, we see that the core elongation and angular momentum direction is randomly oriented with respect to the ambient magnetic field. In general, this behaviour is more compatible with models of weakly magnetized cores. Specifically, the lack of correlation between the $\theta_\mathcal{G}$ and $\theta_{\mathrm{B}_\perp}$ vectors is difficult to reconcile with strongly magnetized cores. Instead, our results conform more closely with simulations of weakly magnetized cores reported by \cite{2017ApJ...834..201L} in which $\theta_\mathcal{G}$ and $\theta_{\mathrm{B}_\perp}$ are randomly aligned with respect to one another. \cite{2018ApJ...865...34C} find that their simulated cores are elongated most perpendicular to the ambient magnetic field and randomly aligned with respect to the velocity gradient, which is in agreement with our results for protostellar cores. Moreover, the cores in their simulated sample are triaxial rather than oblate and include considerable turbulence, further supporting the non-classical view discussed in Section \ref{sub:interpretation}. The random distribution of $|\theta_\mathcal{G} - \theta_{\mathrm{B}_\perp}|$ we see is also in agreement with highly episodic mass accretion onto cores resulting in changes in the angular momentum vector that lead over time to random alignment with $\theta_{\mathrm{B}_\perp}$, as described by \cite{2020ApJ...893...73K} in their MHD simulations.

The coverage of our cross-matched core sample overlaps with that of \cite{2020MNRAS.494.1971C} in the Perseus and Ophiuchus region, where they also see a largely random distribution of  $|\theta_\mathrm{C} - \theta_{\mathrm{B}_\perp}|$. The authors do find a significant anti-correlation between $|\theta_\mathrm{C} - \theta_{\mathrm{B}_\perp}|$ in the highly filamentary Taurus B211/213 region, not included here, parts of which are approximately magnetically critical \citep{li_2022}. Similarly using GAS data, \citet{hchen_2019} identify a subset of pressure-confined, velocity-coherent cores (`droplets') in Ophiuchus that may be in an early evolutionary state. The droplets show little correlation between their elongation and velocity gradient orientation, and additionally often appear disconnected in velocity from their surrounding environment. Furthermore, our starless and prestellar core results agree with that of \cite{2022MNRAS.517.1138S} who find a random distribution of relative angles between core orientation and the local magnetic field for an observed sample of 19 cores. The anti-alignment we see in protostellar cores, however, specifically diverges from their findings. 

For a sample of 200 protostellar outflows, \cite{2022ApJ...941...81X} find that the relative orientation angle of the outflow and the large-scale \textit{Planck} magnetic field shows a preference towards alignment$-$specifically, the distribution peaks around $30^\circ$, with a broad dispersion. If we assume the core orientation is perpendicular to the outflow direction, their results are in agreement with the $|\theta_\mathrm{C} - \theta_{\mathrm{B}_\perp}|$ anti-alignment seen in our protostellar sample. Several other studies have investigated the alignment of protostellar cores and outflows against magnetic field observations on still smaller scales. For instance, \cite{2021ApJ...907...33Y} look at core-scale (0.05–0.5 pc scales) magnetic fields relative to the outflow direction of a sample of 62 protostellar cores in nearby star-forming regions. They find a preferred orientation of $50^{\circ} \pm 15^{\circ}$ between outflow and magnetic field directions, which is in disagreement with our results for protostellar cores, again assuming the core orientation is perpendicular to the outflow direction. \cite{2021ApJ...907...33Y} note, however, that a random orientation in 3D space is less likely but cannot be ruled out. Similarly, \cite{2022ApJ...941...81X} find a decorrelation in outflow relative orientations with distance, which is suggestive of a random distribution in three dimensions. In addition, toward a sample of 29 protostellar cores within the high mass star-forming region W43-MM1, \cite{2020A&A...640A.111A} use high angular resolution polarimetry from ALMA and find that cores are oriented $20^{\circ} - 50^{\circ}$ relative to core-scale magnetic fields$-$in contradiction with our results which show a preferred anti-alignment for protostellar cores. The disagreement between our results and those of \cite{2021ApJ...907...33Y} and \cite{2020A&A...640A.111A} may suggest that on smaller scales, the core-scale magnetic field may be redirected and no longer aligns with the cloud-scale magnetic field seen at the \textit{Planck} resolution.

\subsection{Magnetic field orientation from cloud- to core-scales} \label{b_cloud_to_core}
It is not immediately clear whether the cloud-scale magnetic field orientation traced by \textit{Planck} observations is synonymous with the core-scale magnetic field. Certainly, the dissimilarity between results presented in this work and those by \cite{2020A&A...640A.111A} and \cite{2021ApJ...907...33Y} would suggest a disconnect in core- and cloud-scale magnetic field alignment. There is evidence, however, that polarized thermal millimeter/submillimeter emission on sub-pc scales is substantially correlated with optical starlight polarization, which is sensitive to $1-10$-pc scales in the intercloud medium \citep[as ascertained by][for a sample of 25 cores in the Orion molecular cloud]{2009ApJ...704..891L}. More recently, \cite{2022ApJ...941..122C} compare \textit{Planck} and JCMT BISTRO dust polarization observations towards the DR21 filament and find that the two are well aligned, suggesting a smooth transition of magnetic field orientation from $\sim 0.1 - 10$~pc scales. In addition, \cite{2022MNRAS.510.6085L} compare polarimetric observations of the B211 region in Taurus, taken with the High-resolution Airborne Wideband Camera Plus (HAWC+) onboard the Stratospheric Observatory for Infrared Astronomy (SOFIA), with lower resolution \textit{Planck} dust polarization. They find that the peaks of the magnetic field position angle distribution in both \textit{Planck} and SOFIA HAWC+ results are in agreement but that the SOFIA HAWC+ position angle distribution is more dispersed because it resolves the small scale variations in the chaotic, interacting region in the northwestern part of the main filament. In MHD simulations, \cite{2020ApJ...893...73K} find that core-scale magnetic fields are not isotropic and tend to be aligned perpendicular to the host filament, inheriting their orientation from larger, cloud-scale magnetic fields. Simulations suggest that the correlation between the local and cloud-scale magnetic field orientation will depend, however, on the magnetic field strength and Mach number of the region \citep{matsumoto_2011}. 

In general, some scatter in the cloud- to core-scale magnetic field direction correlation is evident and could alter the relative alignment results for individual cores, particularly those in small, inhomogeneous regions such as the one detailed by \cite{2022MNRAS.510.6085L}. Nonetheless, the prevailing concordance between the two scales \citep[as shown by][]{2009ApJ...704..891L, 2020ApJ...893...73K, 2022ApJ...941..122C, 2022MNRAS.510.6085L} implies that our large sample should be sensitive to statistical trends in any alignment or anti-alignment of $\theta_\mathrm{C}$ and $\theta_\mathcal{G}$ with respect to the core scale magnetic field, even with \textit{Planck} observations of the cloud scale $\theta_{\mathrm{B}_\perp}$. To truly determine the extent of potential alignment of core morphologies and velocity gradients with magnetic fields at comparable, core scales, we will need high resolution polarimetry (e.g., using BISTRO data) across a large sample of cores in the future.

\subsection{Magnetization by region} \label{mag_by_region}
As touched on by \cite{2022MNRAS.510.6085L}, there are scenarios in which the magnetic field orientation becomes disordered, especially on smaller scales within a MC. As such, a useful way to characterize the disorder present in $\theta_{\mathrm{B}_\perp}$ measurements for a given region is to use the local polarization angle dispersion function $\mathcal{S}$ \citep[for details, see][]{2015A&A...576A.104P, 2016A&A...586A.138P, 2016ApJ...824..134F}. \cite{2016ApJ...824..134F} find that the polarization fraction $p = P/I$, where $P$ is the total linearly polarized intensity and $I$ is the total intensity, is negatively correlated with $\mathcal{S}$ on cloud scales. Thus, regions in which we observe high $p$ and low $\mathcal{S}$ are indicative of an ordered magnetic field, while regions with low $p$ and high $\mathcal{S}$ likely have disordered magnetic fields or the magnetic field in that region is orientated primarily along the line of sight rather than in the plane of the sky \citep[e.g.,][]{2001ApJ...546..980O, 2016ApJ...824..134F, 2021MNRAS.503.5006S}. The overall order or disorder of magnetic fields in a region may impact the relative alignment we see between $\theta_\mathrm{C}$, $\theta_\mathcal{G}$, and $\theta_{\mathrm{B}_\perp}$. For example, the relative alignment between $\theta_{\mathrm{B}_\perp}$ and the two other vectors may be more random in a region with a disordered magnetic field compared to a region with an ordered field.

\cite{2021MNRAS.503.5006S} measure the median $p$ and $\mathcal{S}$ across multiple MCs, three of which overlap with this work: Perseus, Ophiuchus, and Cepheus. Their respective median values are $p = 3.8\%$ and $\mathcal{S} = 10.93^\circ$ for Perseus, $p = 5.1\%$ and $\mathcal{S} = 7.18^\circ$ for Ophiuchus, and $p = 4.7\%$ and $\mathcal{S} = 5.62^\circ$ for Cepheus \citep[see Table 2 by][]{2021MNRAS.503.5006S}. Their findings agree with the level of dispersion we measure in the $\theta_{\mathrm{B}_\perp}$ distributions for our sample (see Figure \ref{fig:b_dist} and Table \ref{tb:b_dist}), namely that Cepheus has the smallest $\sigma_{B\perp}$ while Perseus and Ophiuchus span a significantly larger range of $\theta_{\mathrm{B}_\perp}$. This behaviour suggests that Cepheus hosts the most ordered magnetic fields, with magnetic fields in Ophiuchus being less ordered than those in Cepheus, and Perseus containing the most disordered magnetic fields. In our analysis, we find that the L1228 and L1251 regions of Cepheus are the only ones to show a preferential anti-alignment between $\theta_\mathcal{G}$ and $\theta_{\mathrm{B}_\perp}$, while the $|\theta_\mathrm{C} - \theta_{\mathrm{B}_\perp}|$ and $|\theta_\mathcal{G} - \theta_{\mathrm{B}_\perp}|$ distributions in Perseus and Ophiuchus were consistent with being drawn from a random population. In conjunction with the results by \cite{2021MNRAS.503.5006S}, our findings could imply that preferential orientation of core velocity gradients with the local magnetic field is more prevalent in regions with ordered magnetic fields (such as L1228 and L1251 in Cepheus) and random alignment is more common for regions with disordered fields (such as Perseus and Ophiuchus). Note, however, that this may not always be the case$-$Orion A South also has a small $\sigma_{B\perp}$ but does not show any strong alignment or anti-alignment of core elongation or velocity gradient with the magnetic field orientation. Unlike with Cepheus, we do not have the $p$ or $\mathcal{S}$ across Orion A South and therefore cannot be sure that the region hosts an ordered magnetic field.

\section{Conclusions} \label{conclusions}
We perform a systematic analysis of core kinematics, specific angular momenta $J/M$, and the relative alignment of core elongation $\theta_\mathrm{C}$, velocity gradient $\theta_\mathcal{G}$, and ambient magnetic field orientation $\theta_{\mathrm{B}_\perp}$ for a sample of 399 dense cores identified in the Green Bank Ammonia Survey (GAS) and cross-matched with a continuum source in the \textit{Herschel} Gould Belt Survey (HGBS) or the James Clerk Maxwell Telescope Gould Belt Survey (JCMT GBS). The ambient magnetic field orientation is derived from Planck maps of dust polarization at 353~GHz. Of the 399 dense cores, 329 exhibit velocity gradients that are well fit by solid body rotation curves projected on the sky, providing us the largest sample of cores with line of sight velocity information to date. The key conclusions of this work are summarized below:
\begin{enumerate}
    \item The specific angular momentum $J/M$ of cores in our sample ranges from $\sim 10^{-5} - 10^{-1}~\mathrm{pc}~\mathrm{km}~\mathrm{s}^{-1}$ and scales with the core radius $R$ as $J/M \propto R^{1.82 \pm 0.10}$. The derived scaling relation falls in between the expected scaling for an ideal solid body rotating core ($J/M \propto R^2$) and a turbulence dominated core ($J/M \propto R^{1.5}$). Our result is not in complete agreement with either scenario and suggests that the velocity gradients across cores have significant contributions from both solid body rotation and turbulent motions.
    
    \item We find no globally preferred orientation between the core elongation, core velocity gradient, and ambient magnetic field across our cross-matched sample. In most regions, the $|\theta_\mathrm{C} - \theta_\mathcal{G}|$, $|\theta_\mathrm{C} - \theta_{\mathrm{B}_\perp}|$, and $|\theta_\mathcal{G} - \theta_{\mathrm{B}_\perp}|$ distributions are all consistent with being drawn from a random distribution. In general, our results disagree with the classical view of star forming cores as magnetized spheres, instead favouring a triaxial, low-magnetization model of cores.
    
    \item There is a preferred anti-alignment between the angular momentum axis and the ambient magnetic field in the L1228 and L1251 regions of Cepheus, which also have arguably the most ordered magnetic fields from all the regions we consider. Most other regions have significantly larger dispersions in $\theta_{\mathrm{B}_\perp}$ angles, indicating that they may host disordered magnetic fields. Our findings could indicate that a preferential orientation between the core velocity gradient and magnetic field direction is more prevalent in regions with ordered magnetic fields.

    \item The elongation axis for protostellar cores has a unique preference for anti-alignment with the ambient magnetic field that is not observed in prestellar or starless cores. This result suggests the relative alignment between $\theta_\mathrm{C}$ and $\theta_{\mathrm{B}_\perp}$ evolves from randomly oriented in the case of starless and prestellar cores to anti-aligned for protostellar cores. These results are in agreement with simulations of core contraction in magnetically-regulated (but not dominant) environments, or where core growth occurs through anisotropic accretion.
\end{enumerate}

While there is some evidence for a smooth transition from cloud- to core-scale magnetic fields in molecular clouds, higher resolution polarimetry overlapping with the GAS observations is needed to understand fully the role of magnetic fields on core scales. Whether the relative alignment between core elongation, velocity gradient, and magnetic field orientation of our sample remains congruent between cloud- and core-scale fields will further elucidate the role of magnetic fields in transition to smaller scales and their relative impact on the star formation process.

\section*{Acknowledgements} \label{ack}
The authors thank Joshua S. Speagle for useful feedback on the manuscript. 
We also thank the anonymous reviewer for careful reading of the manuscript and for providing constructive comments.
AP is funded by an Ontario Graduate Scholarship.
LMF acknowledges the support of the Natural Sciences and Engineering Research Council of Canada (NSERC) through Discovery Grant RGPIN/06266-2020, and funding through the Queen’s University Research Initiation Grant.
AG acknowledges support from the NSF under grants CAREER 2142300 and AST 2008101.
HK acknowledges support from an NSERC Discovery Grant.
SO acknowledges support from NSF CAREER 1748571.
ER acknowledges the support of the Natural Sciences and Engineering Research Council of Canada (NSERC), funding reference number RGPIN-2022-03499.
The University of Toronto operates on the traditional land of the Huron-Wendat, the Seneca, and most recently, the Mississaugas of the Credit River; we are grateful to have the opportunity to work on this land. 
The Green Bank Observatory is a facility of the National Science Foundation operated under cooperative agreement by Associated Universities, Inc. 
This research made use of {\tt astrodendro}, a Python package to compute dendrograms of Astronomical data (\href{http://www.dendrograms.org}{http://www.dendrograms.org/}). 
This research has made use of data from the \textit{Herschel} Gould Belt survey (HGBS) project (\href{http://gouldbelt-herschel.cea.fr}{http://gouldbelt-herschel.cea.fr}). The HGBS is a \textit{Herschel} Key Programme jointly carried out by SPIRE Specialist Astronomy Group 3 (SAG 3), scientists of several institutes in the PACS Consortium (CEA Saclay, INAF-IFSI Rome and INAF-Arcetri, KU Leuven, MPIA Heidelberg), and scientists of the \textit{Herschel} Science Center (HSC).

\section*{Data availability} \label{data_avail}
The full cross-matched catalogue generated as a part of this work has been published online with the Canadian Advanced Network for Astronomical Research (\href{https://doi.org/10.11570/23.0008}{doi:10.11570/23.0008)}. The catalogue is formatted as a {\tt txt} file and a description of each column of the catalogue is provided as a header within the file.

\bibliographystyle{mnras}
\bibliography{ref}

\begin{thebibliography}{}
\makeatletter
\relax
\def\mn@urlcharsother{\let\do\@makeother \do\$\do\&\do\#\do\^\do\_\do\%\do\~}
\def\mn@doi{\begingroup\mn@urlcharsother \@ifnextchar [ {\mn@doi@}
  {\mn@doi@[]}}
\def\mn@doi@[#1]#2{\def\@tempa{#1}\ifx\@tempa\@empty \href
  {http://dx.doi.org/#2} {doi:#2}\else \href {http://dx.doi.org/#2} {#1}\fi
  \endgroup}
\def\mn@eprint#1#2{\mn@eprint@#1:#2::\@nil}
\def\mn@eprint@arXiv#1{\href {http://arxiv.org/abs/#1} {{\tt arXiv:#1}}}
\def\mn@eprint@dblp#1{\href {http://dblp.uni-trier.de/rec/bibtex/#1.xml}
  {dblp:#1}}
\def\mn@eprint@#1:#2:#3:#4\@nil{\def\@tempa {#1}\def\@tempb {#2}\def\@tempc
  {#3}\ifx \@tempc \@empty \let \@tempc \@tempb \let \@tempb \@tempa \fi \ifx
  \@tempb \@empty \def\@tempb {arXiv}\fi \@ifundefined
  {mn@eprint@\@tempb}{\@tempb:\@tempc}{\expandafter \expandafter \csname
  mn@eprint@\@tempb\endcsname \expandafter{\@tempc}}}

\bibitem[\protect\citeauthoryear{Anderson \& Darling}{Anderson \&
  Darling}{1954}]{ADtest}
Anderson T.,  Darling D.,  1954, Journal of the American Statistical
  Association, 49, 765

\bibitem[\protect\citeauthoryear{{Andre}, {Ward-Thompson}  \&
  {Barsony}}{{Andre} et~al.}{2000}]{2000prpl.conf...59A}
{Andre} P.,  {Ward-Thompson} D.,   {Barsony} M.,  2000, in {Mannings} V.,
  {Boss} A.~P.,   {Russell} S.~S.,  eds, Protostars and Planets IV. p.~59
  (\mn@eprint {arXiv} {astro-ph/9903284}),
  \mn@doi{10.48550/arXiv.astro-ph/9903284}

\bibitem[\protect\citeauthoryear{{Andr{\'e}} et~al.,}{{Andr{\'e}}
  et~al.}{2010}]{2010A&A...518L.102A}
{Andr{\'e}} P.,  et~al., 2010, \mn@doi [\aap] {10.1051/0004-6361/201014666},
  \href {https://ui.adsabs.harvard.edu/abs/2010A&A...518L.102A} {518, L102}

\bibitem[\protect\citeauthoryear{{Andr{\'e}}, {Di Francesco}, {Ward-Thompson},
  {Inutsuka}, {Pudritz}  \& {Pineda}}{{Andr{\'e}}
  et~al.}{2014}]{2014prpl.conf...27A}
{Andr{\'e}} P.,  {Di Francesco} J.,  {Ward-Thompson} D.,  {Inutsuka} S.~I.,
  {Pudritz} R.~E.,   {Pineda} J.~E.,  2014, in {Beuther} H.,  {Klessen} R.~S.,
  {Dullemond} C.~P.,   {Henning} T.,  eds, Protostars and Planets VI. p.~27
  (\mn@eprint {arXiv} {1312.6232}),
  \mn@doi{10.2458/azu\_uapress\_9780816531240-ch002}

\bibitem[\protect\citeauthoryear{{Arce-Tord} et~al.,}{{Arce-Tord}
  et~al.}{2020}]{2020A&A...640A.111A}
{Arce-Tord} C.,  et~al., 2020, \mn@doi [\aap] {10.1051/0004-6361/202038024},
  \href {https://ui.adsabs.harvard.edu/abs/2020A&A...640A.111A} {640, A111}

\bibitem[\protect\citeauthoryear{{Arroyo-Ch{\'a}vez} \&
  {V{\'a}zquez-Semadeni}}{{Arroyo-Ch{\'a}vez} \&
  {V{\'a}zquez-Semadeni}}{2022}]{2022ApJ...925...78A}
{Arroyo-Ch{\'a}vez} G.,  {V{\'a}zquez-Semadeni} E.,  2022, \mn@doi [\apj]
  {10.3847/1538-4357/ac3915}, \href
  {https://ui.adsabs.harvard.edu/abs/2022ApJ...925...78A} {925, 78}

\bibitem[\protect\citeauthoryear{{Ballesteros-Paredes}, {Klessen}, {Mac Low}
  \& {Vazquez-Semadeni}}{{Ballesteros-Paredes}
  et~al.}{2007}]{2007prpl.conf...63B}
{Ballesteros-Paredes} J.,  {Klessen} R.~S.,  {Mac Low} M.~M.,
  {Vazquez-Semadeni} E.,  2007, in {Reipurth} B.,  {Jewitt} D.,   {Keil} K.,
  eds, Protostars and Planets V. p.~63 (\mn@eprint {arXiv} {astro-ph/0603357})

\bibitem[\protect\citeauthoryear{{Basu}}{{Basu}}{2000}]{basu_2000}
{Basu} S.,  2000, \mn@doi [\apjl] {10.1086/312885}, \href
  {https://ui.adsabs.harvard.edu/abs/2000ApJ...540L.103B} {540, L103}

\bibitem[\protect\citeauthoryear{{Basu} \& {Ciolek}}{{Basu} \&
  {Ciolek}}{2004}]{2004ApJ...607L..39B}
{Basu} S.,  {Ciolek} G.~E.,  2004, \mn@doi [\apjl] {10.1086/421464}, \href
  {https://ui.adsabs.harvard.edu/abs/2004ApJ...607L..39B} {607, L39}

\bibitem[\protect\citeauthoryear{{Basu}, {Ciolek}  \& {Wurster}}{{Basu}
  et~al.}{2009}]{2009NewA...14..221B}
{Basu} S.,  {Ciolek} G.~E.,   {Wurster} J.,  2009, \mn@doi [\na]
  {10.1016/j.newast.2008.07.006}, \href
  {https://ui.adsabs.harvard.edu/abs/2009NewA...14..221B} {14, 221}

\bibitem[\protect\citeauthoryear{{Beltr{\'a}n} et~al.,}{{Beltr{\'a}n}
  et~al.}{2021}]{2021A&A...648A.100B}
{Beltr{\'a}n} M.~T.,  et~al., 2021, \mn@doi [\aap]
  {10.1051/0004-6361/202040121}, \href
  {https://ui.adsabs.harvard.edu/abs/2021A&A...648A.100B} {648, A100}

\bibitem[\protect\citeauthoryear{{Bland} \& {Altman}}{{Bland} \&
  {Altman}}{1995}]{bonferroni}
{Bland} J.~M.,  {Altman} D.~G.,  1995, \mn@doi [BMJ]
  {10.1136/bmj.310.6973.170}, 310

\bibitem[\protect\citeauthoryear{{Burkert} \& {Bodenheimer}}{{Burkert} \&
  {Bodenheimer}}{2000}]{2000ApJ...543..822B}
{Burkert} A.,  {Bodenheimer} P.,  2000, \mn@doi [\apj] {10.1086/317122}, \href
  {https://ui.adsabs.harvard.edu/abs/2000ApJ...543..822B} {543, 822}

\bibitem[\protect\citeauthoryear{{Campbell}, {Friesen}, {Martin}, {Caselli},
  {Kauffmann}  \& {Pineda}}{{Campbell} et~al.}{2016}]{campbell_2016}
{Campbell} J.~L.,  {Friesen} R.~K.,  {Martin} P.~G.,  {Caselli} P.,
  {Kauffmann} J.,   {Pineda} J.~E.,  2016, \mn@doi [\apj]
  {10.3847/0004-637X/819/2/143}, \href
  {https://ui.adsabs.harvard.edu/abs/2016ApJ...819..143C} {819, 143}

\bibitem[\protect\citeauthoryear{{Caselli}, {Benson}, {Myers}  \&
  {Tafalla}}{{Caselli} et~al.}{2002}]{2002ApJ...572..238C}
{Caselli} P.,  {Benson} P.~J.,  {Myers} P.~C.,   {Tafalla} M.,  2002, \mn@doi
  [\apj] {10.1086/340195}, \href
  {https://ui.adsabs.harvard.edu/abs/2002ApJ...572..238C} {572, 238}

\bibitem[\protect\citeauthoryear{{Chen} \& {Ostriker}}{{Chen} \&
  {Ostriker}}{2014}]{2014ApJ...785...69C}
{Chen} C.-Y.,  {Ostriker} E.~C.,  2014, \mn@doi [\apj]
  {10.1088/0004-637X/785/1/69}, \href
  {https://ui.adsabs.harvard.edu/abs/2014ApJ...785...69C} {785, 69}

\bibitem[\protect\citeauthoryear{{Chen} \& {Ostriker}}{{Chen} \&
  {Ostriker}}{2018}]{2018ApJ...865...34C}
{Chen} C.-Y.,  {Ostriker} E.~C.,  2018, \mn@doi [\apj]
  {10.3847/1538-4357/aad905}, \href
  {https://ui.adsabs.harvard.edu/abs/2018ApJ...865...34C} {865, 34}

\bibitem[\protect\citeauthoryear{{Chen}, {Launhardt}  \& {Henning}}{{Chen}
  et~al.}{2007}]{2007ApJ...669.1058C}
{Chen} X.,  {Launhardt} R.,   {Henning} T.,  2007, \mn@doi [\apj]
  {10.1086/521868}, \href
  {https://ui.adsabs.harvard.edu/abs/2007ApJ...669.1058C} {669, 1058}

\bibitem[\protect\citeauthoryear{{Chen} et~al.,}{{Chen}
  et~al.}{2019a}]{2019MNRAS.490..527C}
{Chen} C.-Y.,  et~al., 2019a, \mn@doi [\mnras] {10.1093/mnras/stz2633}, \href
  {https://ui.adsabs.harvard.edu/abs/2019MNRAS.490..527C} {490, 527}

\bibitem[\protect\citeauthoryear{{Chen} et~al.,}{{Chen}
  et~al.}{2019b}]{2019ApJ...886..119C}
{Chen} H. H.-H.,  et~al., 2019b, \mn@doi [\apj] {10.3847/1538-4357/ab4ce9},
  \href {https://ui.adsabs.harvard.edu/abs/2019ApJ...886..119C} {886, 119}

\bibitem[\protect\citeauthoryear{{Chen} et~al.,}{{Chen}
  et~al.}{2019c}]{hchen_2019}
{Chen} H. H.-H.,  et~al., 2019c, \mn@doi [\apj] {10.3847/1538-4357/ab4ce9},
  \href {https://ui.adsabs.harvard.edu/abs/2019ApJ...886..119C} {886, 119}

\bibitem[\protect\citeauthoryear{{Chen} et~al.,}{{Chen}
  et~al.}{2020a}]{2020MNRAS.494.1971C}
{Chen} C.-Y.,  et~al., 2020a, \mn@doi [\mnras] {10.1093/mnras/staa835}, \href
  {https://ui.adsabs.harvard.edu/abs/2020MNRAS.494.1971C} {494, 1971}

\bibitem[\protect\citeauthoryear{{Chen} et~al.,}{{Chen}
  et~al.}{2020b}]{2020ApJ...891...84C}
{Chen} M. C.-Y.,  et~al., 2020b, \mn@doi [\apj] {10.3847/1538-4357/ab7378},
  \href {https://ui.adsabs.harvard.edu/abs/2020ApJ...891...84C} {891, 84}

\bibitem[\protect\citeauthoryear{{Chen}, {Di Francesco}, {Pineda}, {Offner}  \&
  {Friesen}}{{Chen} et~al.}{2022}]{2022ApJ...935...57C}
{Chen} M. C.-Y.,  {Di Francesco} J.,  {Pineda} J.~E.,  {Offner} S. S.~R.,
  {Friesen} R.~K.,  2022, \mn@doi [\apj] {10.3847/1538-4357/ac7d4a}, \href
  {https://ui.adsabs.harvard.edu/abs/2022ApJ...935...57C} {935, 57}

\bibitem[\protect\citeauthoryear{{Ching} et~al.,}{{Ching}
  et~al.}{2022}]{2022ApJ...941..122C}
{Ching} T.-C.,  et~al., 2022, \mn@doi [\apj] {10.3847/1538-4357/ac9dfb}, \href
  {https://ui.adsabs.harvard.edu/abs/2022ApJ...941..122C} {941, 122}

\bibitem[\protect\citeauthoryear{{Choudhury} et~al.,}{{Choudhury}
  et~al.}{2020}]{choudhury_2020}
{Choudhury} S.,  et~al., 2020, \mn@doi [\aap] {10.1051/0004-6361/202037955},
  \href {https://ui.adsabs.harvard.edu/abs/2020A&A...640L...6C} {640, L6}

\bibitem[\protect\citeauthoryear{{Crapsi}, {Caselli}, {Walmsley}  \&
  {Tafalla}}{{Crapsi} et~al.}{2007}]{2007A&A...470..221C}
{Crapsi} A.,  {Caselli} P.,  {Walmsley} M.~C.,   {Tafalla} M.,  2007, \mn@doi
  [\aap] {10.1051/0004-6361:20077613}, \href
  {https://ui.adsabs.harvard.edu/abs/2007A&A...470..221C} {470, 221}

\bibitem[\protect\citeauthoryear{{Crutcher}}{{Crutcher}}{1999}]{1999ApJ...520..706C}
{Crutcher} R.~M.,  1999, \mn@doi [\apj] {10.1086/307483}, \href
  {https://ui.adsabs.harvard.edu/abs/1999ApJ...520..706C} {520, 706}

\bibitem[\protect\citeauthoryear{{Crutcher}}{{Crutcher}}{2012}]{2012ARA&A..50...29C}
{Crutcher} R.~M.,  2012, \mn@doi [\araa] {10.1146/annurev-astro-081811-125514},
  \href {https://ui.adsabs.harvard.edu/abs/2012ARA&A..50...29C} {50, 29}

\bibitem[\protect\citeauthoryear{{Crutcher}, {Troland}, {Goodman}, {Heiles},
  {Kazes}  \& {Myers}}{{Crutcher} et~al.}{1993}]{1993ApJ...407..175C}
{Crutcher} R.~M.,  {Troland} T.~H.,  {Goodman} A.~A.,  {Heiles} C.,  {Kazes}
  I.,   {Myers} P.~C.,  1993, \mn@doi [\apj] {10.1086/172503}, \href
  {https://ui.adsabs.harvard.edu/abs/1993ApJ...407..175C} {407, 175}

\bibitem[\protect\citeauthoryear{{Davis} \& {Greenstein}}{{Davis} \&
  {Greenstein}}{1951}]{1951ApJ...114..206D}
{Davis} Leverett J.,  {Greenstein} J.~L.,  1951, \mn@doi [\apj]
  {10.1086/145464}, \href
  {https://ui.adsabs.harvard.edu/abs/1951ApJ...114..206D} {114, 206}

\bibitem[\protect\citeauthoryear{{Dempsey} et~al.,}{{Dempsey}
  et~al.}{2013}]{2013MNRAS.430.2534D}
{Dempsey} J.~T.,  et~al., 2013, \mn@doi [\mnras] {10.1093/mnras/stt090}, \href
  {https://ui.adsabs.harvard.edu/abs/2013MNRAS.430.2534D} {430, 2534}

\bibitem[\protect\citeauthoryear{{Di Francesco}, {Evans}, {Caselli}, {Myers},
  {Shirley}, {Aikawa}  \& {Tafalla}}{{Di Francesco}
  et~al.}{2007}]{2007prpl.conf...17D}
{Di Francesco} J.,  {Evans} N.~J. I.,  {Caselli} P.,  {Myers} P.~C.,  {Shirley}
  Y.,  {Aikawa} Y.,   {Tafalla} M.,  2007, in {Reipurth} B.,  {Jewitt} D.,
  {Keil} K.,  eds, Protostars and Planets V. p.~17 (\mn@eprint {arXiv}
  {astro-ph/0602379}), \mn@doi{10.48550/arXiv.astro-ph/0602379}

\bibitem[\protect\citeauthoryear{{Di Francesco} et~al.,}{{Di Francesco}
  et~al.}{2020}]{2020ApJ...904..172D}
{Di Francesco} J.,  et~al., 2020, \mn@doi [\apj] {10.3847/1538-4357/abc016},
  \href {https://ui.adsabs.harvard.edu/abs/2020ApJ...904..172D} {904, 172}

\bibitem[\protect\citeauthoryear{{Dib}, {Hennebelle}, {Pineda}, {Csengeri},
  {Bontemps}, {Audit}  \& {Goodman}}{{Dib} et~al.}{2010}]{2010ApJ...723..425D}
{Dib} S.,  {Hennebelle} P.,  {Pineda} J.~E.,  {Csengeri} T.,  {Bontemps} S.,
  {Audit} E.,   {Goodman} A.~A.,  2010, \mn@doi [\apj]
  {10.1088/0004-637X/723/1/425}, \href
  {https://ui.adsabs.harvard.edu/abs/2010ApJ...723..425D} {723, 425}

\bibitem[\protect\citeauthoryear{{Fissel} et~al.,}{{Fissel}
  et~al.}{2016}]{2016ApJ...824..134F}
{Fissel} L.~M.,  et~al., 2016, \mn@doi [\apj] {10.3847/0004-637X/824/2/134},
  \href {https://ui.adsabs.harvard.edu/abs/2016ApJ...824..134F} {824, 134}

\bibitem[\protect\citeauthoryear{{Friesen} et~al.,}{{Friesen}
  et~al.}{2017}]{2017ApJ...843...63F}
{Friesen} R.~K.,  et~al., 2017, \mn@doi [\apj] {10.3847/1538-4357/aa6d58},
  \href {https://ui.adsabs.harvard.edu/abs/2017ApJ...843...63F} {843, 63}

\bibitem[\protect\citeauthoryear{{Fromang}, {Hennebelle}  \&
  {Teyssier}}{{Fromang} et~al.}{2006}]{2006A&A...457..371F}
{Fromang} S.,  {Hennebelle} P.,   {Teyssier} R.,  2006, \mn@doi [\aap]
  {10.1051/0004-6361:20065371}, \href
  {https://ui.adsabs.harvard.edu/abs/2006A&A...457..371F} {457, 371}

\bibitem[\protect\citeauthoryear{{Gammie}, {Lin}, {Stone}  \&
  {Ostriker}}{{Gammie} et~al.}{2003}]{2003ApJ...592..203G}
{Gammie} C.~F.,  {Lin} Y.-T.,  {Stone} J.~M.,   {Ostriker} E.~C.,  2003,
  \mn@doi [\apj] {10.1086/375635}, \href
  {https://ui.adsabs.harvard.edu/abs/2003ApJ...592..203G} {592, 203}

\bibitem[\protect\citeauthoryear{{G{\'o}mez} \&
  {V{\'a}zquez-Semadeni}}{{G{\'o}mez} \&
  {V{\'a}zquez-Semadeni}}{2014}]{2014ApJ...791..124G}
{G{\'o}mez} G.~C.,  {V{\'a}zquez-Semadeni} E.,  2014, \mn@doi [\apj]
  {10.1088/0004-637X/791/2/124}, \href
  {https://ui.adsabs.harvard.edu/abs/2014ApJ...791..124G} {791, 124}

\bibitem[\protect\citeauthoryear{{Gong} \& {Ostriker}}{{Gong} \&
  {Ostriker}}{2015}]{2015ApJ...806...31G}
{Gong} M.,  {Ostriker} E.~C.,  2015, \mn@doi [\apj]
  {10.1088/0004-637X/806/1/31}, \href
  {https://ui.adsabs.harvard.edu/abs/2015ApJ...806...31G} {806, 31}

\bibitem[\protect\citeauthoryear{{Goodman}, {Benson}, {Fuller}  \&
  {Myers}}{{Goodman} et~al.}{1993}]{1993ApJ...406..528G}
{Goodman} A.~A.,  {Benson} P.~J.,  {Fuller} G.~A.,   {Myers} P.~C.,  1993,
  \mn@doi [\apj] {10.1086/172465}, \href
  {https://ui.adsabs.harvard.edu/abs/1993ApJ...406..528G} {406, 528}

\bibitem[\protect\citeauthoryear{{Heiles}, {Goodman}, {McKee}  \&
  {Zweibel}}{{Heiles} et~al.}{1993}]{1993prpl.conf..279H}
{Heiles} C.,  {Goodman} A.~A.,  {McKee} C.~F.,   {Zweibel} E.~G.,  1993, in
  {Levy} E.~H.,  {Lunine} J.~I.,  eds, Protostars and Planets III. p.~279

\bibitem[\protect\citeauthoryear{{Hirota} et~al.,}{{Hirota}
  et~al.}{2008}]{2008PASJ...60...37H}
{Hirota} T.,  et~al., 2008, \mn@doi [\pasj] {10.1093/pasj/60.1.37}, \href
  {https://ui.adsabs.harvard.edu/abs/2008PASJ...60...37H} {60, 37}

\bibitem[\protect\citeauthoryear{{Hosking} \& {Whitworth}}{{Hosking} \&
  {Whitworth}}{2004}]{2004MNRAS.347.1001H}
{Hosking} J.~G.,  {Whitworth} A.~P.,  2004, \mn@doi [\mnras]
  {10.1111/j.1365-2966.2004.07274.x}, \href
  {https://ui.adsabs.harvard.edu/abs/2004MNRAS.347.1001H} {347, 1001}

\bibitem[\protect\citeauthoryear{{Jones}, {Basu}  \& {Dubinski}}{{Jones}
  et~al.}{2001}]{jones_2001}
{Jones} C.~E.,  {Basu} S.,   {Dubinski} J.,  2001, \mn@doi [\apj]
  {10.1086/320093}, \href
  {https://ui.adsabs.harvard.edu/abs/2001ApJ...551..387J} {551, 387}

\bibitem[\protect\citeauthoryear{{Kataoka}, {Machida}  \& {Tomisaka}}{{Kataoka}
  et~al.}{2012}]{2012ApJ...761...40K}
{Kataoka} A.,  {Machida} M.~N.,   {Tomisaka} K.,  2012, \mn@doi [\apj]
  {10.1088/0004-637X/761/1/40}, \href
  {https://ui.adsabs.harvard.edu/abs/2012ApJ...761...40K} {761, 40}

\bibitem[\protect\citeauthoryear{{Kauffmann}, {Bertoldi}, {Bourke}, {Evans}  \&
  {Lee}}{{Kauffmann} et~al.}{2008}]{2008A&A...487..993K}
{Kauffmann} J.,  {Bertoldi} F.,  {Bourke} T.~L.,  {Evans} N.~J. I.,   {Lee}
  C.~W.,  2008, \mn@doi [\aap] {10.1051/0004-6361:200809481}, \href
  {https://ui.adsabs.harvard.edu/abs/2008A&A...487..993K} {487, 993}

\bibitem[\protect\citeauthoryear{{Klessen} \& {Burkert}}{{Klessen} \&
  {Burkert}}{2000}]{2000ApJS..128..287K}
{Klessen} R.~S.,  {Burkert} A.,  2000, \mn@doi [\apjs] {10.1086/313371}, \href
  {https://ui.adsabs.harvard.edu/abs/2000ApJS..128..287K} {128, 287}

\bibitem[\protect\citeauthoryear{Klotz}{Klotz}{1967}]{KStest}
Klotz J.,  1967, Journal of the American Statistical Association, 62, 932

\bibitem[\protect\citeauthoryear{{K{\"o}nyves} et~al.,}{{K{\"o}nyves}
  et~al.}{2015}]{2015A&A...584A..91K}
{K{\"o}nyves} V.,  et~al., 2015, \mn@doi [\aap] {10.1051/0004-6361/201525861},
  \href {https://ui.adsabs.harvard.edu/abs/2015A&A...584A..91K} {584, A91}

\bibitem[\protect\citeauthoryear{{K{\"o}nyves} et~al.,}{{K{\"o}nyves}
  et~al.}{2020}]{2020A&A...635A..34K}
{K{\"o}nyves} V.,  et~al., 2020, \mn@doi [\aap] {10.1051/0004-6361/201834753},
  \href {https://ui.adsabs.harvard.edu/abs/2020A&A...635A..34K} {635, A34}

\bibitem[\protect\citeauthoryear{{Kounkel} et~al.,}{{Kounkel}
  et~al.}{2017}]{2017ApJ...834..142K}
{Kounkel} M.,  et~al., 2017, \mn@doi [\apj] {10.3847/1538-4357/834/2/142},
  \href {https://ui.adsabs.harvard.edu/abs/2017ApJ...834..142K} {834, 142}

\bibitem[\protect\citeauthoryear{{Kusune} et~al.,}{{Kusune}
  et~al.}{2019}]{kusune_2019}
{Kusune} T.,  et~al., 2019, \mn@doi [\pasj] {10.1093/pasj/psz040}, \href
  {https://ui.adsabs.harvard.edu/abs/2019PASJ...71S...5K} {71, S5}

\bibitem[\protect\citeauthoryear{{Kuznetsova}, {Hartmann}  \&
  {Heitsch}}{{Kuznetsova} et~al.}{2019}]{2019ApJ...876...33K}
{Kuznetsova} A.,  {Hartmann} L.,   {Heitsch} F.,  2019, \mn@doi [\apj]
  {10.3847/1538-4357/ab12ce}, \href
  {https://ui.adsabs.harvard.edu/abs/2019ApJ...876...33K} {876, 33}

\bibitem[\protect\citeauthoryear{{Kuznetsova}, {Hartmann}  \&
  {Heitsch}}{{Kuznetsova} et~al.}{2020}]{2020ApJ...893...73K}
{Kuznetsova} A.,  {Hartmann} L.,   {Heitsch} F.,  2020, \mn@doi [\apj]
  {10.3847/1538-4357/ab7eac}, \href
  {https://ui.adsabs.harvard.edu/abs/2020ApJ...893...73K} {893, 73}

\bibitem[\protect\citeauthoryear{{Ladjelate} et~al.,}{{Ladjelate}
  et~al.}{2020}]{2020A&A...638A..74L}
{Ladjelate} B.,  et~al., 2020, \mn@doi [\aap] {10.1051/0004-6361/201936442},
  \href {https://ui.adsabs.harvard.edu/abs/2020A&A...638A..74L} {638, A74}

\bibitem[\protect\citeauthoryear{{Lazarian}}{{Lazarian}}{2007}]{2007JQSRT.106..225L}
{Lazarian} A.,  2007, \mn@doi [\jqsrt] {10.1016/j.jqsrt.2007.01.038}, \href
  {https://ui.adsabs.harvard.edu/abs/2007JQSRT.106..225L} {106, 225}

\bibitem[\protect\citeauthoryear{{Lee}, {Myers}  \& {Tafalla}}{{Lee}
  et~al.}{1999}]{lee_1999}
{Lee} C.~W.,  {Myers} P.~C.,   {Tafalla} M.,  1999, \mn@doi [\apj]
  {10.1086/308027}, \href
  {https://ui.adsabs.harvard.edu/abs/1999ApJ...526..788L} {526, 788}

\bibitem[\protect\citeauthoryear{{Lee}, {Hull}  \& {Offner}}{{Lee}
  et~al.}{2017}]{2017ApJ...834..201L}
{Lee} J. W.~Y.,  {Hull} C. L.~H.,   {Offner} S. S.~R.,  2017, \mn@doi [\apj]
  {10.3847/1538-4357/834/2/201}, \href
  {https://ui.adsabs.harvard.edu/abs/2017ApJ...834..201L} {834, 201}

\bibitem[\protect\citeauthoryear{{Li}, {Norman}, {Mac Low}  \& {Heitsch}}{{Li}
  et~al.}{2004}]{2004ApJ...605..800L}
{Li} P.~S.,  {Norman} M.~L.,  {Mac Low} M.-M.,   {Heitsch} F.,  2004, \mn@doi
  [\apj] {10.1086/382652}, \href
  {https://ui.adsabs.harvard.edu/abs/2004ApJ...605..800L} {605, 800}

\bibitem[\protect\citeauthoryear{{Li}, {Dowell}, {Goodman}, {Hildebrand}  \&
  {Novak}}{{Li} et~al.}{2009}]{2009ApJ...704..891L}
{Li} H.-b.,  {Dowell} C.~D.,  {Goodman} A.,  {Hildebrand} R.,   {Novak} G.,
  2009, \mn@doi [\apj] {10.1088/0004-637X/704/2/891}, \href
  {https://ui.adsabs.harvard.edu/abs/2009ApJ...704..891L} {704, 891}

\bibitem[\protect\citeauthoryear{{Li}, {Lopez-Rodriguez}, {Ajeddig},
  {Andr{\'e}}, {McKee}, {Rho}  \& {Klein}}{{Li} et~al.}{2022a}]{li_2022}
{Li} P.~S.,  {Lopez-Rodriguez} E.,  {Ajeddig} H.,  {Andr{\'e}} P.,  {McKee}
  C.~F.,  {Rho} J.,   {Klein} R.~I.,  2022a, \mn@doi [\mnras]
  {10.1093/mnras/stab3448}, \href
  {https://ui.adsabs.harvard.edu/abs/2022MNRAS.510.6085L} {510, 6085}

\bibitem[\protect\citeauthoryear{{Li}, {Lopez-Rodriguez}, {Ajeddig},
  {Andr{\'e}}, {McKee}, {Rho}  \& {Klein}}{{Li}
  et~al.}{2022b}]{2022MNRAS.510.6085L}
{Li} P.~S.,  {Lopez-Rodriguez} E.,  {Ajeddig} H.,  {Andr{\'e}} P.,  {McKee}
  C.~F.,  {Rho} J.,   {Klein} R.~I.,  2022b, \mn@doi [\mnras]
  {10.1093/mnras/stab3448}, \href
  {https://ui.adsabs.harvard.edu/abs/2022MNRAS.510.6085L} {510, 6085}

\bibitem[\protect\citeauthoryear{{Lomax}, {Whitworth}  \& {Cartwright}}{{Lomax}
  et~al.}{2013}]{lomax_2013}
{Lomax} O.,  {Whitworth} A.~P.,   {Cartwright} A.,  2013, \mn@doi [\mnras]
  {10.1093/mnras/stt1764}, \href
  {https://ui.adsabs.harvard.edu/abs/2013MNRAS.436.2680L} {436, 2680}

\bibitem[\protect\citeauthoryear{{Mairs} et~al.,}{{Mairs}
  et~al.}{2016}]{2016MNRAS.461.4022M}
{Mairs} S.,  et~al., 2016, \mn@doi [\mnras] {10.1093/mnras/stw1550}, \href
  {https://ui.adsabs.harvard.edu/abs/2016MNRAS.461.4022M} {461, 4022}

\bibitem[\protect\citeauthoryear{{Matsumoto} \& {Hanawa}}{{Matsumoto} \&
  {Hanawa}}{2011}]{matsumoto_2011}
{Matsumoto} T.,  {Hanawa} T.,  2011, \mn@doi [\apj]
  {10.1088/0004-637X/728/1/47}, \href
  {https://ui.adsabs.harvard.edu/abs/2011ApJ...728...47M} {728, 47}

\bibitem[\protect\citeauthoryear{{McKee} \& {Ostriker}}{{McKee} \&
  {Ostriker}}{2007}]{2007ARA&A..45..565M}
{McKee} C.~F.,  {Ostriker} E.~C.,  2007, \mn@doi [\araa]
  {10.1146/annurev.astro.45.051806.110602}, \href
  {https://ui.adsabs.harvard.edu/abs/2007ARA&A..45..565M} {45, 565}

\bibitem[\protect\citeauthoryear{{Megeath} et~al.,}{{Megeath}
  et~al.}{2012}]{megeath_2012}
{Megeath} S.~T.,  et~al., 2012, \mn@doi [\aj] {10.1088/0004-6256/144/6/192},
  \href {https://ui.adsabs.harvard.edu/abs/2012AJ....144..192M} {144, 192}

\bibitem[\protect\citeauthoryear{{Mellon} \& {Li}}{{Mellon} \&
  {Li}}{2008}]{2008ApJ...681.1356M}
{Mellon} R.~R.,  {Li} Z.-Y.,  2008, \mn@doi [\apj] {10.1086/587542}, \href
  {https://ui.adsabs.harvard.edu/abs/2008ApJ...681.1356M} {681, 1356}

\bibitem[\protect\citeauthoryear{{Men'shchikov}, {Andr{\'e}}, {Didelon},
  {Motte}, {Hennemann}  \& {Schneider}}{{Men'shchikov}
  et~al.}{2012}]{2012A&A...542A..81M}
{Men'shchikov} A.,  {Andr{\'e}} P.,  {Didelon} P.,  {Motte} F.,  {Hennemann}
  M.,   {Schneider} N.,  2012, \mn@doi [\aap] {10.1051/0004-6361/201218797},
  \href {https://ui.adsabs.harvard.edu/abs/2012A&A...542A..81M} {542, A81}

\bibitem[\protect\citeauthoryear{{Mestel} \& {Spitzer}}{{Mestel} \&
  {Spitzer}}{1956}]{1956MNRAS.116..503M}
{Mestel} L.,  {Spitzer} L. J.,  1956, \mn@doi [\mnras]
  {10.1093/mnras/116.5.503}, \href
  {https://ui.adsabs.harvard.edu/abs/1956MNRAS.116..503M} {116, 503}

\bibitem[\protect\citeauthoryear{{Moscadelli}, {Li}, {Cesaroni}, {Sanna}, {Xu}
  \& {Zhang}}{{Moscadelli} et~al.}{2013}]{2013A&A...549A.122M}
{Moscadelli} L.,  {Li} J.~J.,  {Cesaroni} R.,  {Sanna} A.,  {Xu} Y.,   {Zhang}
  Q.,  2013, \mn@doi [\aap] {10.1051/0004-6361/201220497}, \href
  {https://ui.adsabs.harvard.edu/abs/2013A&A...549A.122M} {549, A122}

\bibitem[\protect\citeauthoryear{{Mouschovias}}{{Mouschovias}}{1976}]{mouschovias_1976}
{Mouschovias} T.~C.,  1976, \mn@doi [\apj] {10.1086/154478}, \href
  {https://ui.adsabs.harvard.edu/abs/1976ApJ...207..141M} {207, 141}

\bibitem[\protect\citeauthoryear{{Muench}, {Getman}, {Hillenbrand}  \&
  {Preibisch}}{{Muench} et~al.}{2008}]{2008hsf1.book..483M}
{Muench} A.,  {Getman} K.,  {Hillenbrand} L.,   {Preibisch} T.,  2008, in
  {Reipurth} B.,  ed., , Vol.~4, Handbook of Star Forming Regions, Volume I.
Astronomical Society of the Pacific Conference Series, p.~483,
  \mn@doi{10.48550/arXiv.0812.1323}

\bibitem[\protect\citeauthoryear{{Murdoch}, {Tsai}  \& {Adcock}}{{Murdoch}
  et~al.}{2008}]{murdoch2008}
{Murdoch} D.~J.,  {Tsai} Y.-L.,   {Adcock} J.,  2008, The American
  Statistician, 62, 242

\bibitem[\protect\citeauthoryear{{Ng} et~al.,}{{Ng}
  et~al.}{2020}]{2020MNRAS.496.2836N}
{Ng} C.,  et~al., 2020, \mn@doi [\mnras] {10.1093/mnras/staa1658}, \href
  {https://ui.adsabs.harvard.edu/abs/2020MNRAS.496.2836N} {496, 2836}

\bibitem[\protect\citeauthoryear{{Offner} \& {Chaban}}{{Offner} \&
  {Chaban}}{2017}]{2017ApJ...847..104O}
{Offner} S. S.~R.,  {Chaban} J.,  2017, \mn@doi [\apj]
  {10.3847/1538-4357/aa8996}, \href
  {https://ui.adsabs.harvard.edu/abs/2017ApJ...847..104O} {847, 104}

\bibitem[\protect\citeauthoryear{{Offner} \& {Krumholz}}{{Offner} \&
  {Krumholz}}{2009}]{2009ApJ...693..914O}
{Offner} S. S.~R.,  {Krumholz} M.~R.,  2009, \mn@doi [\apj]
  {10.1088/0004-637X/693/1/914}, \href
  {https://ui.adsabs.harvard.edu/abs/2009ApJ...693..914O} {693, 914}

\bibitem[\protect\citeauthoryear{{Offner}, {Klein}  \& {McKee}}{{Offner}
  et~al.}{2008}]{2008ApJ...686.1174O}
{Offner} S. S.~R.,  {Klein} R.~I.,   {McKee} C.~F.,  2008, \mn@doi [\apj]
  {10.1086/590238}, \href
  {https://ui.adsabs.harvard.edu/abs/2008ApJ...686.1174O} {686, 1174}

\bibitem[\protect\citeauthoryear{{Ortiz-Le{\'o}n} et~al.,}{{Ortiz-Le{\'o}n}
  et~al.}{2018a}]{2018ApJ...865...73O}
{Ortiz-Le{\'o}n} G.~N.,  et~al., 2018a, \mn@doi [\apj]
  {10.3847/1538-4357/aada49}, \href
  {https://ui.adsabs.harvard.edu/abs/2018ApJ...865...73O} {865, 73}

\bibitem[\protect\citeauthoryear{{Ortiz-Le{\'o}n} et~al.,}{{Ortiz-Le{\'o}n}
  et~al.}{2018b}]{2018ApJ...869L..33O}
{Ortiz-Le{\'o}n} G.~N.,  et~al., 2018b, \mn@doi [\apjl]
  {10.3847/2041-8213/aaf6ad}, \href
  {https://ui.adsabs.harvard.edu/abs/2018ApJ...869L..33O} {869, L33}

\bibitem[\protect\citeauthoryear{{Ostriker}, {Stone}  \& {Gammie}}{{Ostriker}
  et~al.}{2001}]{2001ApJ...546..980O}
{Ostriker} E.~C.,  {Stone} J.~M.,   {Gammie} C.~F.,  2001, \mn@doi [\apj]
  {10.1086/318290}, \href
  {https://ui.adsabs.harvard.edu/abs/2001ApJ...546..980O} {546, 980}

\bibitem[\protect\citeauthoryear{{Palmeirim} et~al.,}{{Palmeirim}
  et~al.}{2013}]{2013A&A...550A..38P}
{Palmeirim} P.,  et~al., 2013, \mn@doi [\aap] {10.1051/0004-6361/201220500},
  \href {https://ui.adsabs.harvard.edu/abs/2013A&A...550A..38P} {550, A38}

\bibitem[\protect\citeauthoryear{{Pandhi}, {Hutschenreuter}, {West}, {Gaensler}
   \& {Stock}}{{Pandhi} et~al.}{2022}]{2022MNRAS.516.4739P}
{Pandhi} A.,  {Hutschenreuter} S.,  {West} J.~L.,  {Gaensler} B.~M.,   {Stock}
  A.,  2022, \mn@doi [\mnras] {10.1093/mnras/stac2314}, \href
  {https://ui.adsabs.harvard.edu/abs/2022MNRAS.516.4739P} {516, 4739}

\bibitem[\protect\citeauthoryear{{Pattle}, {Fissel}, {Tahani}, {Liu}  \&
  {Ntormousi}}{{Pattle} et~al.}{2022}]{2022arXiv220311179P}
{Pattle} K.,  {Fissel} L.,  {Tahani} M.,  {Liu} T.,   {Ntormousi} E.,  2022,
  \mn@doi [arXiv e-prints] {10.48550/arXiv.2203.11179}, \href
  {https://ui.adsabs.harvard.edu/abs/2022arXiv220311179P} {p. arXiv:2203.11179}

\bibitem[\protect\citeauthoryear{{Pezzuto} et~al.,}{{Pezzuto}
  et~al.}{2021}]{2021A&A...645A..55P}
{Pezzuto} S.,  et~al., 2021, \mn@doi [\aap] {10.1051/0004-6361/201936534},
  \href {https://ui.adsabs.harvard.edu/abs/2021A&A...645A..55P} {645, A55}

\bibitem[\protect\citeauthoryear{{Pillai} et~al.,}{{Pillai}
  et~al.}{2020}]{pillai_2020}
{Pillai} T. G.~S.,  et~al., 2020, \mn@doi [Nature Astronomy]
  {10.1038/s41550-020-1172-6}, \href
  {https://ui.adsabs.harvard.edu/abs/2020NatAs...4.1195P} {4, 1195}

\bibitem[\protect\citeauthoryear{{Pineda}, {Goodman}, {Arce}, {Caselli},
  {Foster}, {Myers}  \& {Rosolowsky}}{{Pineda} et~al.}{2010}]{pineda_2010}
{Pineda} J.~E.,  {Goodman} A.~A.,  {Arce} H.~G.,  {Caselli} P.,  {Foster}
  J.~B.,  {Myers} P.~C.,   {Rosolowsky} E.~W.,  2010, \mn@doi [\apjl]
  {10.1088/2041-8205/712/1/L116}, \href
  {https://ui.adsabs.harvard.edu/abs/2010ApJ...712L.116P} {712, L116}

\bibitem[\protect\citeauthoryear{{Pineda}, {Zhao}, {Schmiedeke}, {Segura-Cox},
  {Caselli}, {Myers}, {Tobin}  \& {Dunham}}{{Pineda}
  et~al.}{2019}]{2019ApJ...882..103P}
{Pineda} J.~E.,  {Zhao} B.,  {Schmiedeke} A.,  {Segura-Cox} D.~M.,  {Caselli}
  P.,  {Myers} P.~C.,  {Tobin} J.~J.,   {Dunham} M.,  2019, \mn@doi [\apj]
  {10.3847/1538-4357/ab2cd1}, \href
  {https://ui.adsabs.harvard.edu/abs/2019ApJ...882..103P} {882, 103}

\bibitem[\protect\citeauthoryear{{Pineda} et~al.,}{{Pineda}
  et~al.}{2022}]{2022arXiv220503935P}
{Pineda} J.~E.,  et~al., 2022, arXiv e-prints, \href
  {https://ui.adsabs.harvard.edu/abs/2022arXiv220503935P} {p. arXiv:2205.03935}

\bibitem[\protect\citeauthoryear{{Pirogov}, {Zinchenko}, {Caselli}, {Johansson}
   \& {Myers}}{{Pirogov} et~al.}{2003}]{2003A&A...405..639P}
{Pirogov} L.,  {Zinchenko} I.,  {Caselli} P.,  {Johansson} L.~E.~B.,   {Myers}
  P.~C.,  2003, \mn@doi [\aap] {10.1051/0004-6361:20030659}, \href
  {https://ui.adsabs.harvard.edu/abs/2003A&A...405..639P} {405, 639}

\bibitem[\protect\citeauthoryear{{Planck Collaboration Int. XIX}}{{Planck
  Collaboration Int. XIX}}{2015}]{2015A&A...576A.104P}
{Planck Collaboration Int. XIX} 2015, \mn@doi [\aap]
  {10.1051/0004-6361/201424082}, \href
  {https://ui.adsabs.harvard.edu/abs/2015A&A...576A.104P} {576, A104}

\bibitem[\protect\citeauthoryear{{Planck Collaboration Int. XXXV}}{{Planck
  Collaboration Int. XXXV}}{2016}]{2016A&A...586A.138P}
{Planck Collaboration Int. XXXV} 2016, \mn@doi [\aap]
  {10.1051/0004-6361/201525896}, \href
  {https://ui.adsabs.harvard.edu/abs/2016A&A...586A.138P} {586, A138}

\bibitem[\protect\citeauthoryear{{Price} \& {Bate}}{{Price} \&
  {Bate}}{2007}]{2007MNRAS.377...77P}
{Price} D.~J.,  {Bate} M.~R.,  2007, \mn@doi [\mnras]
  {10.1111/j.1365-2966.2007.11621.x}, \href
  {https://ui.adsabs.harvard.edu/abs/2007MNRAS.377...77P} {377, 77}

\bibitem[\protect\citeauthoryear{{Punanova}, {Caselli}, {Pineda}, {Pon},
  {Tafalla}, {Hacar}  \& {Bizzocchi}}{{Punanova}
  et~al.}{2018}]{2018A&A...617A..27P}
{Punanova} A.,  {Caselli} P.,  {Pineda} J.~E.,  {Pon} A.,  {Tafalla} M.,
  {Hacar} A.,   {Bizzocchi} L.,  2018, \mn@doi [\aap]
  {10.1051/0004-6361/201731159}, \href
  {https://ui.adsabs.harvard.edu/abs/2018A&A...617A..27P} {617, A27}

\bibitem[\protect\citeauthoryear{{Rice}}{{Rice}}{2007}]{rice2007}
{Rice} J.~A.,  2007, {Mathematical Statistics and Data Analysis}.
Duxbury

\bibitem[\protect\citeauthoryear{{Rosolowsky}, {Pineda}, {Kauffmann}  \&
  {Goodman}}{{Rosolowsky} et~al.}{2008}]{2008ApJ...679.1338R}
{Rosolowsky} E.~W.,  {Pineda} J.~E.,  {Kauffmann} J.,   {Goodman} A.~A.,  2008,
  \mn@doi [\apj] {10.1086/587685}, \href
  {https://ui.adsabs.harvard.edu/abs/2008ApJ...679.1338R} {679, 1338}

\bibitem[\protect\citeauthoryear{{Sharma}, {Gopinathan}, {Soam}, {Lee}  \&
  {Seshadri}}{{Sharma} et~al.}{2022}]{2022MNRAS.517.1138S}
{Sharma} E.,  {Gopinathan} M.,  {Soam} A.,  {Lee} C.~W.,   {Seshadri} T.~R.,
  2022, \mn@doi [\mnras] {10.1093/mnras/stac2487}, \href
  {https://ui.adsabs.harvard.edu/abs/2022MNRAS.517.1138S} {517, 1138}

\bibitem[\protect\citeauthoryear{{Shirley}}{{Shirley}}{2015}]{2015PASP..127..299S}
{Shirley} Y.~L.,  2015, \mn@doi [\pasp]
  {10.1086/68034210.48550/arXiv.1501.01629}, \href
  {https://ui.adsabs.harvard.edu/abs/2015PASP..127..299S} {127, 299}

\bibitem[\protect\citeauthoryear{{Shu}, {Adams}  \& {Lizano}}{{Shu}
  et~al.}{1987}]{1987ARA&A..25...23S}
{Shu} F.~H.,  {Adams} F.~C.,   {Lizano} S.,  1987, \mn@doi [\araa]
  {10.1146/annurev.aa.25.090187.000323}, \href
  {https://ui.adsabs.harvard.edu/abs/1987ARA&A..25...23S} {25, 23}

\bibitem[\protect\citeauthoryear{{Silsbee}, {Ivlev}, {Sipil{\"a}}, {Caselli}
  \& {Zhao}}{{Silsbee} et~al.}{2020}]{2020A&A...641A..39S}
{Silsbee} K.,  {Ivlev} A.~V.,  {Sipil{\"a}} O.,  {Caselli} P.,   {Zhao} B.,
  2020, \mn@doi [\aap] {10.1051/0004-6361/202038063}, \href
  {https://ui.adsabs.harvard.edu/abs/2020A&A...641A..39S} {641, A39}

\bibitem[\protect\citeauthoryear{{Sokolov}, {Pineda}, {Buchner}  \&
  {Caselli}}{{Sokolov} et~al.}{2020}]{sokolov_2020}
{Sokolov} V.,  {Pineda} J.~E.,  {Buchner} J.,   {Caselli} P.,  2020, \mn@doi
  [\apjl] {10.3847/2041-8213/ab8018}, \href
  {https://ui.adsabs.harvard.edu/abs/2020ApJ...892L..32S} {892, L32}

\bibitem[\protect\citeauthoryear{{Soler}}{{Soler}}{2019}]{2019A&A...629A..96S}
{Soler} J.~D.,  2019, \mn@doi [\aap] {10.1051/0004-6361/201935779}, \href
  {https://ui.adsabs.harvard.edu/abs/2019A&A...629A..96S} {629, A96}

\bibitem[\protect\citeauthoryear{{Strittmatter}}{{Strittmatter}}{1966}]{1966MNRAS.132..359S}
{Strittmatter} P.~A.,  1966, \mn@doi [\mnras] {10.1093/mnras/132.2.359}, \href
  {https://ui.adsabs.harvard.edu/abs/1966MNRAS.132..359S} {132, 359}

\bibitem[\protect\citeauthoryear{{Stutz} \& {Kainulainen}}{{Stutz} \&
  {Kainulainen}}{2015}]{2015A&A...577L...6S}
{Stutz} A.~M.,  {Kainulainen} J.,  2015, \mn@doi [\aap]
  {10.1051/0004-6361/201526243}, \href
  {https://ui.adsabs.harvard.edu/abs/2015A&A...577L...6S} {577, L6}

\bibitem[\protect\citeauthoryear{{Sugitani} et~al.,}{{Sugitani}
  et~al.}{2011}]{sugitani_2011}
{Sugitani} K.,  et~al., 2011, \mn@doi [\apj] {10.1088/0004-637X/734/1/63},
  \href {https://ui.adsabs.harvard.edu/abs/2011ApJ...734...63S} {734, 63}

\bibitem[\protect\citeauthoryear{{Sullivan}, {Fissel}, {King}, {Chen}, {Li}  \&
  {Soler}}{{Sullivan} et~al.}{2021}]{2021MNRAS.503.5006S}
{Sullivan} C.~H.,  {Fissel} L.~M.,  {King} P.~K.,  {Chen} C.~Y.,  {Li} Z.~Y.,
  {Soler} J.~D.,  2021, \mn@doi [\mnras] {10.1093/mnras/stab596}, \href
  {https://ui.adsabs.harvard.edu/abs/2021MNRAS.503.5006S} {503, 5006}

\bibitem[\protect\citeauthoryear{{Tahani} et~al.,}{{Tahani}
  et~al.}{2022}]{tahani_2022}
{Tahani} M.,  et~al., 2022, \mn@doi [arXiv e-prints]
  {10.48550/arXiv.2212.10884}, \href
  {https://ui.adsabs.harvard.edu/abs/2022arXiv221210884T} {p. arXiv:2212.10884}

\bibitem[\protect\citeauthoryear{{Tatematsu}, {Ohashi}, {Sanhueza}, {Nguyen
  Luong}, {Umemoto}  \& {Mizuno}}{{Tatematsu}
  et~al.}{2016}]{2016PASJ...68...24T}
{Tatematsu} K.,  {Ohashi} S.,  {Sanhueza} P.,  {Nguyen Luong} Q.,  {Umemoto}
  T.,   {Mizuno} N.,  2016, \mn@doi [\pasj] {10.1093/pasj/psw002}, \href
  {https://ui.adsabs.harvard.edu/abs/2016PASJ...68...24T} {68, 24}

\bibitem[\protect\citeauthoryear{{Tobin} et~al.,}{{Tobin}
  et~al.}{2011}]{2011ApJ...740...45T}
{Tobin} J.~J.,  et~al., 2011, \mn@doi [\apj] {10.1088/0004-637X/740/1/45},
  \href {https://ui.adsabs.harvard.edu/abs/2011ApJ...740...45T} {740, 45}

\bibitem[\protect\citeauthoryear{{Tsukamoto} et~al.,}{{Tsukamoto}
  et~al.}{2022}]{2022arXiv220913765T}
{Tsukamoto} Y.,  et~al., 2022, \mn@doi [arXiv e-prints]
  {10.48550/arXiv.2209.13765}, \href
  {https://ui.adsabs.harvard.edu/abs/2022arXiv220913765T} {p. arXiv:2209.13765}

\bibitem[\protect\citeauthoryear{Virtanen et~al.,}{Virtanen
  et~al.}{2020}]{2020SciPy-NMeth}
Virtanen P.,  et~al., 2020, \mn@doi [Nature Methods]
  {10.1038/s41592-019-0686-2}, \href {https://rdcu.be/b08Wh} {17, 261}

\bibitem[\protect\citeauthoryear{{Ward-Thompson} et~al.,}{{Ward-Thompson}
  et~al.}{2007}]{2007PASP..119..855W}
{Ward-Thompson} D.,  et~al., 2007, \mn@doi [\pasp] {10.1086/521277}, \href
  {https://ui.adsabs.harvard.edu/abs/2007PASP..119..855W} {119, 855}

\bibitem[\protect\citeauthoryear{{Xu}, {Offner}, {Gutermuth}  \& {Tan}}{{Xu}
  et~al.}{2022}]{2022ApJ...941...81X}
{Xu} D.,  {Offner} S. S.~R.,  {Gutermuth} R.,   {Tan} J.~C.,  2022, \mn@doi
  [\apj] {10.3847/1538-4357/aca153}, \href
  {https://ui.adsabs.harvard.edu/abs/2022ApJ...941...81X} {941, 81}

\bibitem[\protect\citeauthoryear{{Yan}, {Zhang}, {Xu}, {Guo}, {Macquart},
  {Tang}  \& {Walsh}}{{Yan} et~al.}{2019}]{2019A&A...624A...6Y}
{Yan} Q.-Z.,  {Zhang} B.,  {Xu} Y.,  {Guo} S.,  {Macquart} J.-P.,  {Tang}
  Z.-H.,   {Walsh} A.~J.,  2019, \mn@doi [\aap] {10.1051/0004-6361/201834337},
  \href {https://ui.adsabs.harvard.edu/abs/2019A&A...624A...6Y} {624, A6}

\bibitem[\protect\citeauthoryear{{Yen}, {Koch}, {Takakuwa}, {Ho}, {Ohashi}  \&
  {Tang}}{{Yen} et~al.}{2015}]{2015ApJ...799..193Y}
{Yen} H.-W.,  {Koch} P.~M.,  {Takakuwa} S.,  {Ho} P. T.~P.,  {Ohashi} N.,
  {Tang} Y.-W.,  2015, \mn@doi [\apj] {10.1088/0004-637X/799/2/193}, \href
  {https://ui.adsabs.harvard.edu/abs/2015ApJ...799..193Y} {799, 193}

\bibitem[\protect\citeauthoryear{{Yen} et~al.,}{{Yen}
  et~al.}{2021}]{2021ApJ...907...33Y}
{Yen} H.-W.,  et~al., 2021, \mn@doi [\apj] {10.3847/1538-4357/abca99}, \href
  {https://ui.adsabs.harvard.edu/abs/2021ApJ...907...33Y} {907, 33}

\bibitem[\protect\citeauthoryear{{Zhao}, {Caselli}, {Li}, {Krasnopolsky},
  {Shang}  \& {Nakamura}}{{Zhao} et~al.}{2016}]{2016MNRAS.460.2050Z}
{Zhao} B.,  {Caselli} P.,  {Li} Z.-Y.,  {Krasnopolsky} R.,  {Shang} H.,
  {Nakamura} F.,  2016, \mn@doi [\mnras] {10.1093/mnras/stw1124}, \href
  {https://ui.adsabs.harvard.edu/abs/2016MNRAS.460.2050Z} {460, 2050}

\bibitem[\protect\citeauthoryear{{Zhao}, {Caselli}, {Li}  \&
  {Krasnopolsky}}{{Zhao} et~al.}{2018}]{2018MNRAS.473.4868Z}
{Zhao} B.,  {Caselli} P.,  {Li} Z.-Y.,   {Krasnopolsky} R.,  2018, \mn@doi
  [\mnras] {10.1093/mnras/stx2617}, \href
  {https://ui.adsabs.harvard.edu/abs/2018MNRAS.473.4868Z} {473, 4868}

\bibitem[\protect\citeauthoryear{{Ziegler}}{{Ziegler}}{2005}]{2005A&A...435..385Z}
{Ziegler} U.,  2005, \mn@doi [\aap] {10.1051/0004-6361:20042451}, \href
  {https://ui.adsabs.harvard.edu/abs/2005A&A...435..385Z} {435, 385}

\bibitem[\protect\citeauthoryear{{Zucker}, {Schlafly}, {Speagle}, {Green},
  {Portillo}, {Finkbeiner}  \& {Goodman}}{{Zucker}
  et~al.}{2018}]{2018ApJ...869...83Z}
{Zucker} C.,  {Schlafly} E.~F.,  {Speagle} J.~S.,  {Green} G.~M.,  {Portillo}
  S. K.~N.,  {Finkbeiner} D.~P.,   {Goodman} A.~A.,  2018, \mn@doi [\apj]
  {10.3847/1538-4357/aae97c}, \href
  {https://ui.adsabs.harvard.edu/abs/2018ApJ...869...83Z} {869, 83}

\makeatother
\end{thebibliography}

\appendix

\section{Summary of cross-matched catalogue} \label{app:catalog}
We present summary statistics of the core sizes and masses for our cross-matched sample in Table \ref{tb:mean_vals}. The results are divided into the $17$ regions introduced in Table \ref{tb:cross_match} and the name of each region and the number of cross-matched cores it encompasses is given in columns 1 and 2, respectively. Column 3, 4, and 5 list the median, mean, and standard deviation of the core size $R_\mathrm{core}$ in pc, while columns 6, 7, and 8 provide the median, mean, and standard deviation of the core mass $M_\mathrm{core}$ in $\mathrm{M}_\odot$. In regions with only one cross-matched core, the standard deviation is not presented. The final row of Table \ref{tb:mean_vals} provides the same statistics described above but for the entire cross-matched sample of $399$ cores.

Table \ref{tb:catalog} presents an excerpt of ten cores from the full cross-matched catalogue used in this work. For these ten cores, the continuum data is obtained from \cite{2021A&A...645A..55P}. Table \ref{tb:catalog} presents a summary of properties used in this work for all 399 cross-matched cores. The host cloud and region of each core is given in columns 1 and 2, respectively, with the GAS dendrogram index for the region presented in column 3. We note that the dendrogram index starts from 0 for each region and therefore the indices are only unique within a given region. The on-sky position (as measured in the GAS data) in J2000 Right Ascension and declination is shown in columns 4 and 5. Columns 6 and 7 give the core radius $R_\mathrm{core}$ and mass $M_\mathrm{core}$, which is obtained via cross-matching with continuum catalogues. Column 8 shows the best fit 2D linear velocity gradient magnitude. An upper limit of $3\sigma_\mathcal{G}$ is reported when Equation \ref{eq:error_cond} is not met. The core elongation, velocity gradient, and ambient magnetic field orientation vectors are presented in columns 9$-$11. The core type, namely ``starless,'' ``prestellar,'' or ``protostellar,'' is given in column 12. Column 13 lists which significance criteria each core passes with E for the core elongation cut (Equation \ref{eq:elong_cond}), $\mathcal{G}$ for the velocity gradient cut (Equation \ref{eq:error_cond}), and P for the polarized intensity cut (Equation \ref{eq:PI_cond}). The full catalogue is available online and also contains the corresponding uncertainties, where available, for parameters presented in Table \ref{tb:catalog} (see Section \ref{data_avail} for details).

\begin{table*}
\begin{center}
\caption{Summary statistics (median $\mathrm{Med}$, mean $\mu$, and standard deviation $\sigma$) of the core sizes and masses by region for the cross-matched sample used in this work. For regions with only one cross-matched core, the core size and mass standard deviations are not provided.} \label{tb:mean_vals}
\begin{tabular}{ccccccccc} 
\hline
\hline
Region & Cross-matched& $\mathrm{Med}_{R_\mathrm{core}}$ & $\mu_{R_\mathrm{core}}$ & $\sigma_{R_\mathrm{core}}$ & $\mathrm{Med}_{M_\mathrm{core}}$ & $\mu_{M_\mathrm{core}}$ & $\sigma_{M_\mathrm{core}}$\\
 & cores & (pc) & (pc) & (pc) & (M$_\odot$) & (M$_\odot$) & (M$_\odot$)\\
\hline
Perseus B1 & $15$ & $0.02$ & $0.03$ & $0.01$ & $0.89$ & $1.43$ & $1.53$\\
Perseus B1E & $1$ & $0.06$ & $0.06$ & $-$ & $2.31$ & $2.31$ & $-$\\
Perseus NGC1333 & $26$ & $0.02$ & $0.02$ & $0.01$ & $1.69$ & $2.68$ & $3.27$\\
Perseus IC348 & $17$ & $0.03$ & $0.03$ & $0.01$ & $1.29$ & $1.41$ & $1.00$\\
Perseus L1448 & $1$ & $0.02$ & $0.02$ & $-$ & $1.55$ & $1.55$ & $-$\\
Perseus L1451 & $4$ & $0.05$ & $0.04$ & $0.02$ & $2.89$ & $2.86$ & $0.79$\\
Perseus L1455 & $11$ & $0.04$ & $0.03$ & $0.02$ & $1.78$ & $1.70$ & $1.53$\\
Ophiuchus L1688 & $38$ & $0.01$ & $0.02$ & $0.01$ & $0.16$ & $0.66$ & $1.38$\\
Ophiuchus L1689 & $12$ & $0.01$ & $0.01$ & $0.01$ & $0.29$ & $1.01$ & $1.52$\\
Serpens W40 & $103$ & $0.03$ & $0.03$ & $0.01$ & $0.73$ & $1.23$ & $1.40$\\
Serpens MWC297 & $2$ & $0.03$ & $0.03$ & $0.01$ & $1.93$ & $1.93$ & $1.13$\\
Cepheus L1228 & $1$ & $0.06$ & $0.06$ & $-$ & $1.66$ & $1.66$ & $-$\\
Cepheus L1251 & $9$ & $0.02$ & $0.03$ & $0.02$ & $1.31$ & $2.30$ & $2.28$\\
Orion B NGC2023 & $19$ & $0.03$ & $0.03$ & $0.01$ & $3.59$ & $5.50$ & $8.03$\\
Orion B NGC2068 & $10$ & $0.03$ & $0.03$ & $0.01$ & $6.43$ & $6.29$ & $3.59$\\
Orion A & $97$ & $0.05$ & $0.04$ & $0.02$ & $1.31$ & $4.67$ & $17.25$\\
Orion A South & $33$ & $0.06$ & $0.06$ & $0.03$ & $1.03$ & $1.71$ & $2.69$\\
\hline
\textbf{Total} & \textbf{399} & \textbf{0.03} & \textbf{0.03} & \textbf{0.02} & \textbf{1.03} & \textbf{2.55} & \textbf{9.00}\\
\hline
\hline
\end{tabular}
\end{center}
\end{table*}

\setlength{\tabcolsep}{3.73pt}

\onecolumn
\begin{center}
\begin{longtable}{ccccccccccccc}
\caption{An excerpt of the full cross-matched catalogue generated in this work, presenting the columns relevant to the results presented in Section \ref{results}. The continuum data for the ten cores shown in this excerpt is from \protect\cite{2021A&A...645A..55P}. The full catalogue is available for download online (see Section \ref{data_avail} for details).} \label{tb:catalog}\\
\hline 
\hline
Cloud & Region & Index & $\alpha$ & $\delta$ & $R_\mathrm{core}$ & $M_\mathrm{core}$ & $|\mathcal{G}|$ & $\theta_\mathrm{C}$ & $\theta_\mathcal{G}$ & $\theta_{\mathrm{B}_\perp}$ & Core Type & Flags\\
 &  &  & (deg) & (deg) & (pc) & ($M_\odot$) & (km s$^{-1}$ pc$^{-1}$) & (deg) & (deg) & (deg) &  & \\
\hline
\endfirsthead

\multicolumn{6}{c}
{{\bfseries \tablename\ \thetable{} -- continued from previous page}} \\
\hline
Cloud & Region & Index & $\alpha$ & $\delta$ & $R_\mathrm{core}$ & $M_\mathrm{core}$ & $|\mathcal{G}|$ & $\theta_\mathrm{C}$ & $\theta_\mathcal{G}$ & $\theta_{\mathrm{B}_\perp}$ & Core Type & Flags\\
 &  &  & (deg) & (deg) & (pc) & ($M_\odot$) & (km s$^{-1}$ pc$^{-1}$) & (deg) & (deg) & (deg) &  & \\
\hline
\endhead

\hline \multicolumn{2}{c}{{Continued on next page}} \\ \hline
\endfoot
\hline \hline
\endlastfoot

Perseus & B1 & 3 & 53.068816 & 30.825568 & 0.01 & 1.54 & 1.00 & 96.0 & 41.64 & 136.81 & protostellar & E, $\mathcal{G}$, P\\
Perseus & B1 & 5 & 53.081083 & 30.859995 & 0.02 & 0.14 & 1.94 & 54.0 & 155.14 & 126.91 & prestellar & E, $\mathcal{G}$, P\\
Perseus & B1 & 4 & 53.135829 & 30.840802 & 0.03 & 0.89 & 2.97 & 115.0 & -99.13 & 133.32 & prestellar & E, $\mathcal{G}$, P\\
Perseus & B1 & 11 & 53.141501 & 30.941490 & 0.04 & 0.19 & $<$4.53 & 84.0 & 82.89 & 133.09 & prestellar & E, P\\
Perseus & B1 & 6 & 53.148311 & 30.882467 & 0.04 & 0.35 & 5.00 & 86.0 & 99.32 & 136.61 & prestellar & E, $\mathcal{G}$, P\\
Perseus & B1 & 27 & 53.184508 & 30.999739 & 0.05 & 3.13 & 1.18 & 169.0 & 127.29 & 135.73 & prestellar & E, $\mathcal{G}$, P\\
Perseus & B1 & 58 & 53.257974 & 31.348152 & 0.05 & 1.42 & 0.70 & 23.0 & 70.55 & 107.45 & prestellar & E, $\mathcal{G}$, P\\
Perseus & B1 & 34 & 53.263066 & 31.075834 & 0.02 & 0.84 & 0.95 & 4.0 & 148.56 & 117.77 & prestellar & E, $\mathcal{G}$, P\\
Perseus & B1 & 40 & 53.272518 & 31.108471 & 0.02 & 1.56 & $<$0.40 & 48.0 & -118.98 & 116.34 & prestellar & E, P\\
Perseus & B1 & 43 & 53.318785 & 31.116537 & 0.01 & 0.75 & 5.51 & 71.0 & -157.60 & 118.39 & protostellar & E, $\mathcal{G}$, P\\

\end{longtable}
\end{center}
\twocolumn

\section{Core identification and cross-matching results} \label{app:cross_match}
\begin{figure*}
    \centering
    \includegraphics[width=0.48\textwidth]{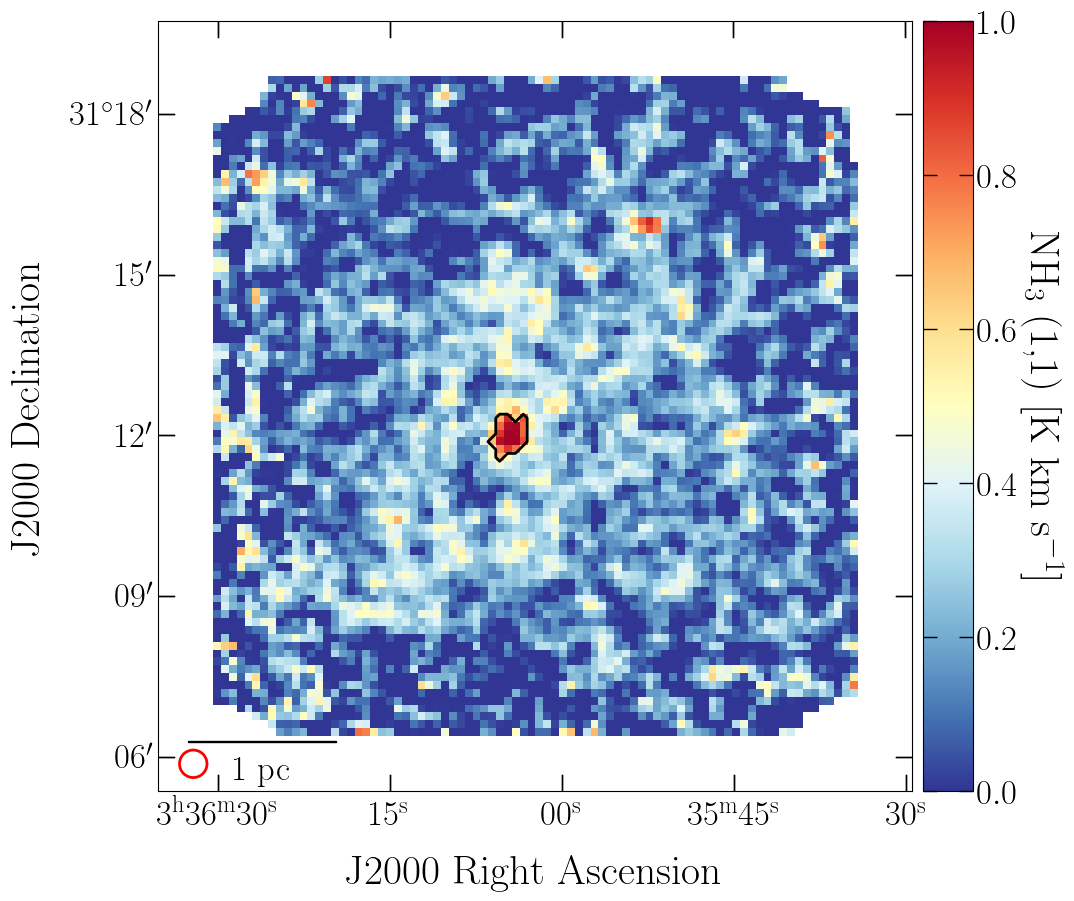}
    \includegraphics[width=0.48\textwidth]{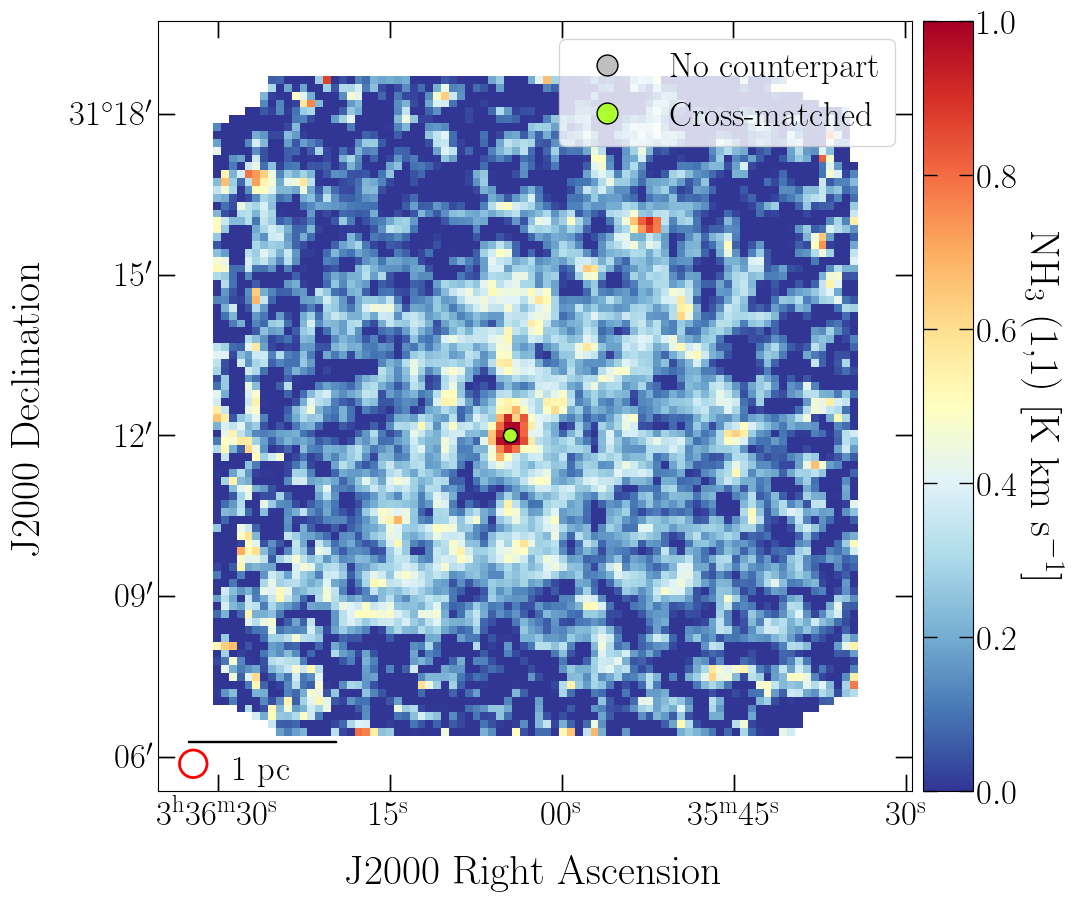}
    \caption{Analogous to Figure \ref{fig:b1_cores} but for the Perseus B1E region. In this region, the single identified core in GAS was uniquely cross-matched with a continuum counterpart in the HGBS data.}
    \label{fig:b1e_cores}
\end{figure*}

\begin{figure*}
    \centering
    \includegraphics[width=0.48\textwidth]{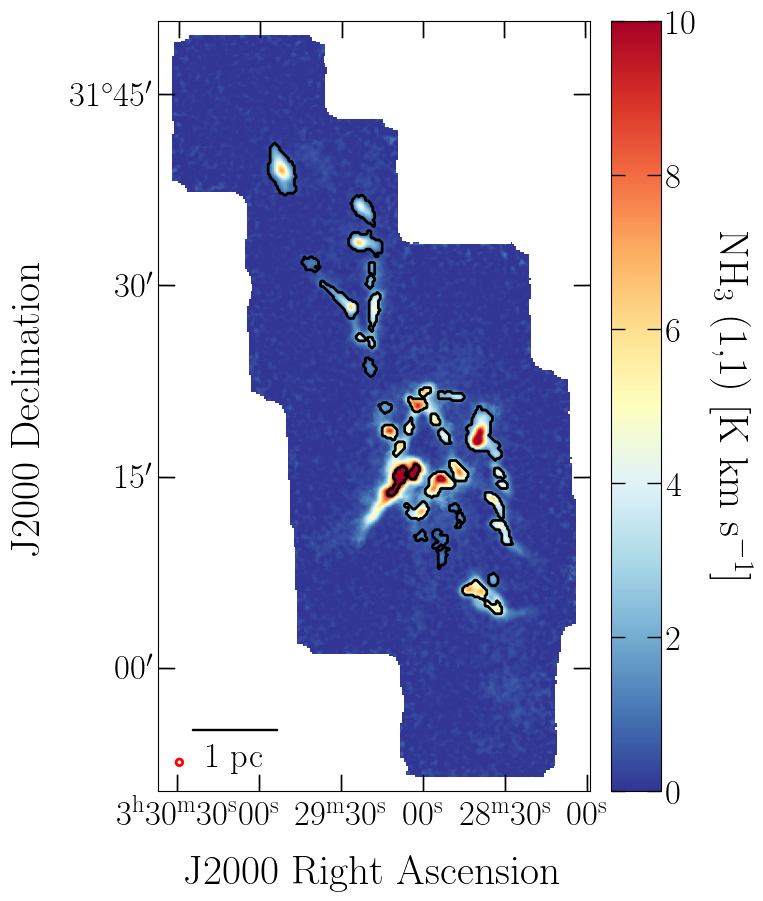}
    \includegraphics[width=0.48\textwidth]{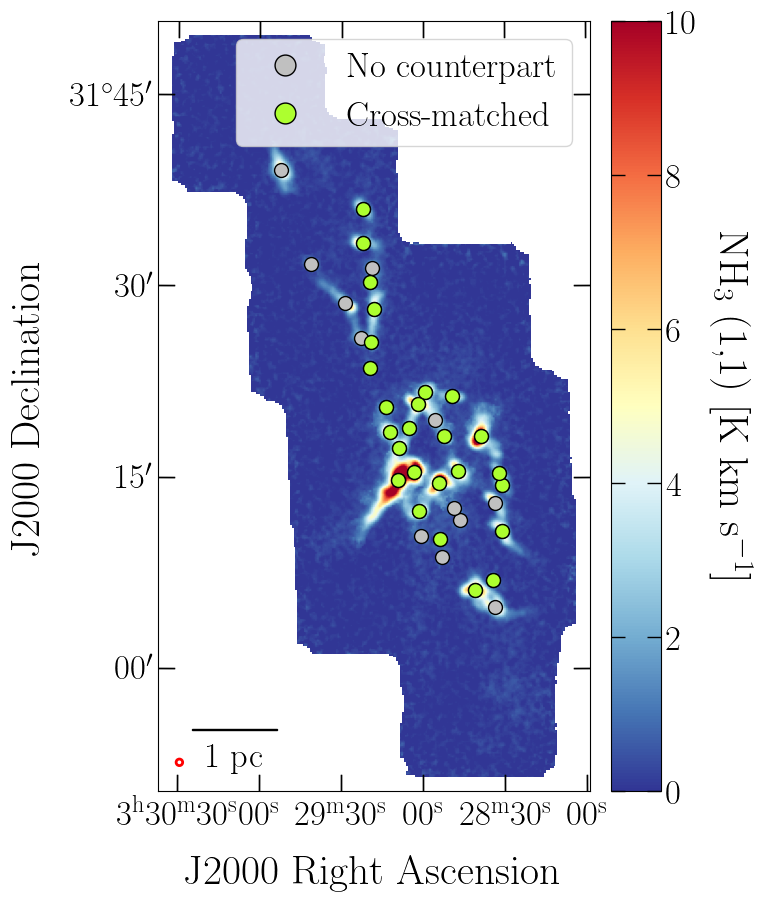}
    \caption{Analogous to Figure \ref{fig:b1_cores} but for the Perseus NGC1333 region. In this region, $26$ of the $38$ cores identified in GAS were uniquely cross-matched with a continuum counterpart in the HGBS data.}
    \label{fig:ngc1333_cores}
\end{figure*}

\begin{figure*}
    \centering
    \includegraphics[width=0.48\textwidth]{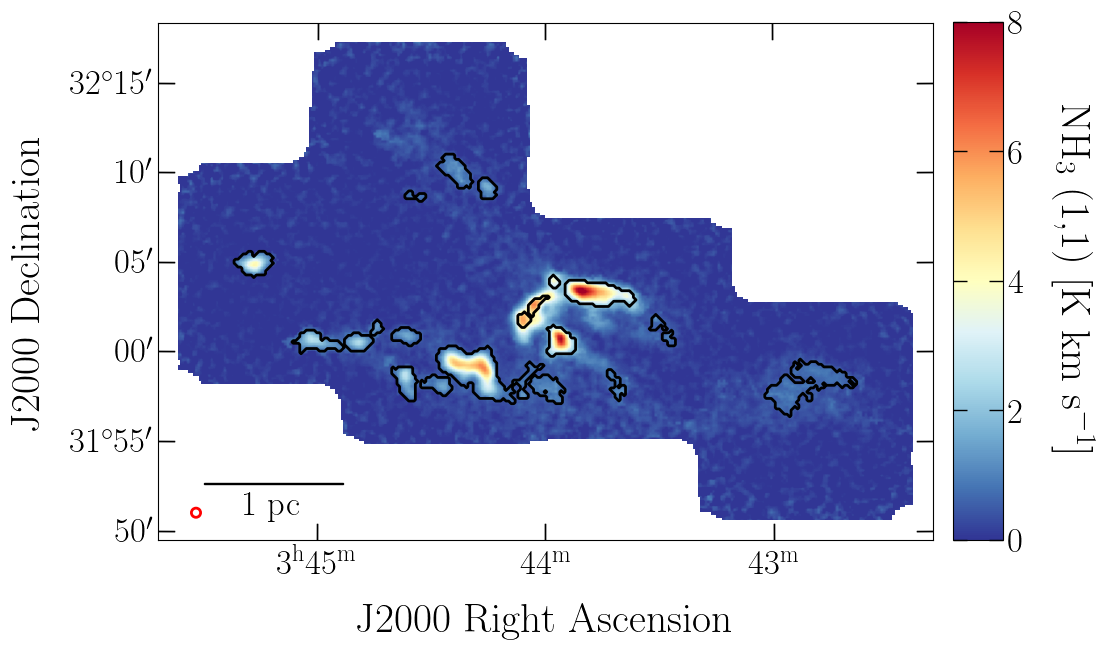}
    \includegraphics[width=0.48\textwidth]{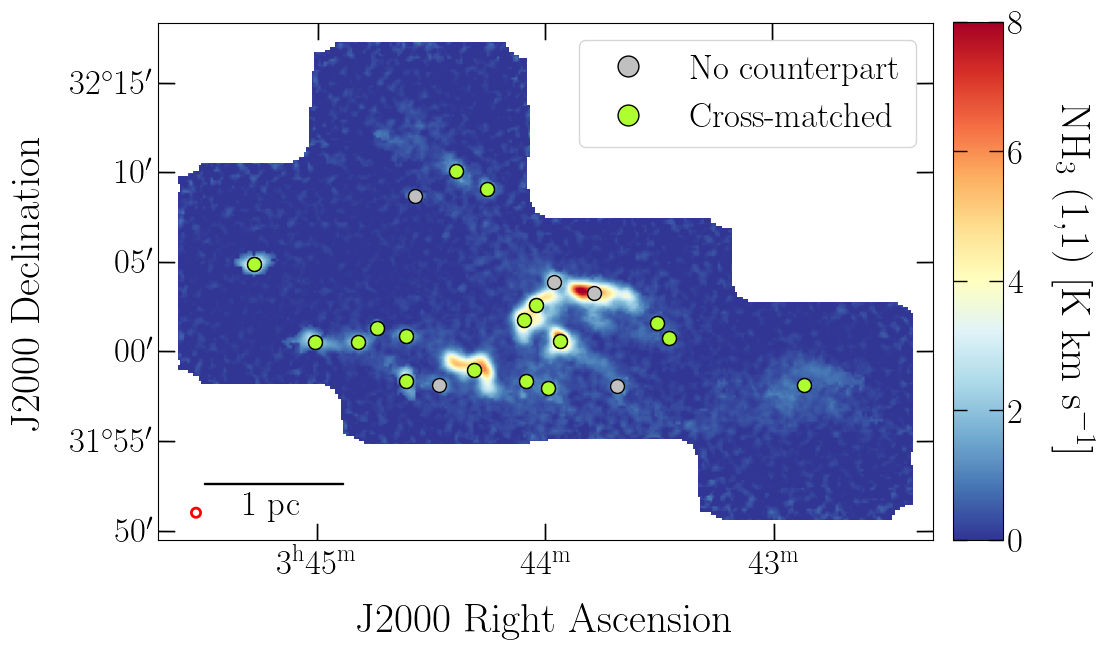}
    \caption{Analogous to Figure \ref{fig:b1_cores} but for the Perseus IC348 region. In this region, $17$ of the $22$ cores identified in GAS were uniquely cross-matched with a continuum counterpart in the HGBS data.}
    \label{fig:ic348_cores}
\end{figure*}

\begin{figure*}
    \centering
    \includegraphics[width=0.48\textwidth]{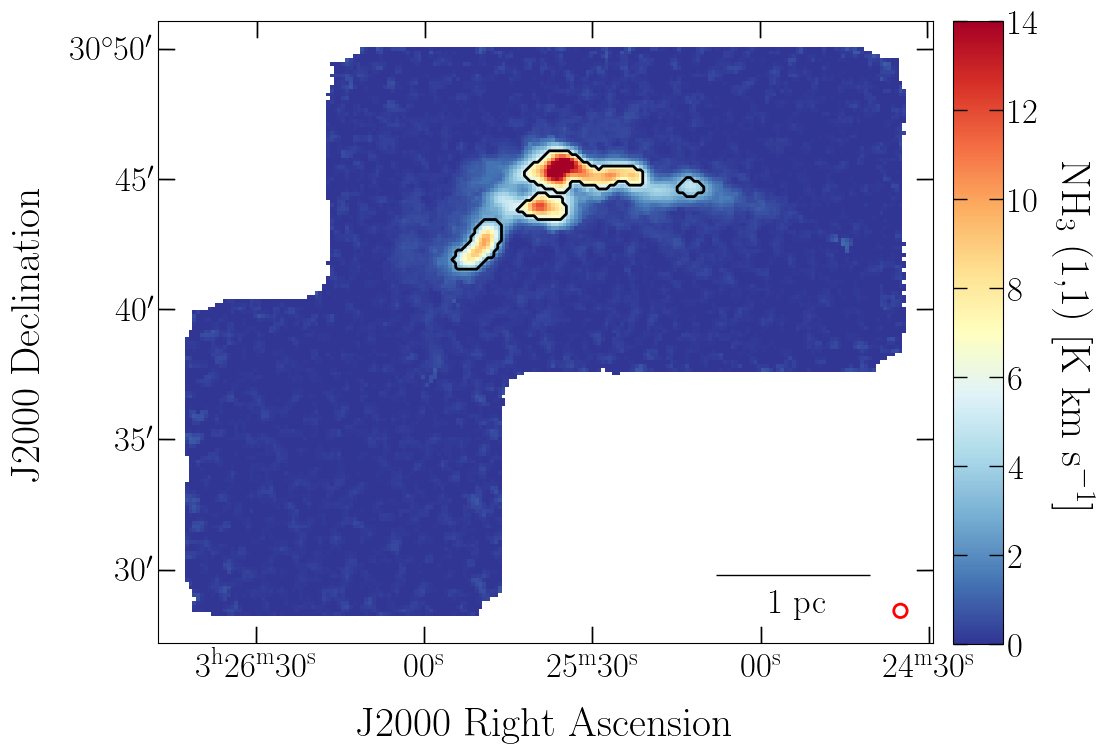}
    \includegraphics[width=0.48\textwidth]{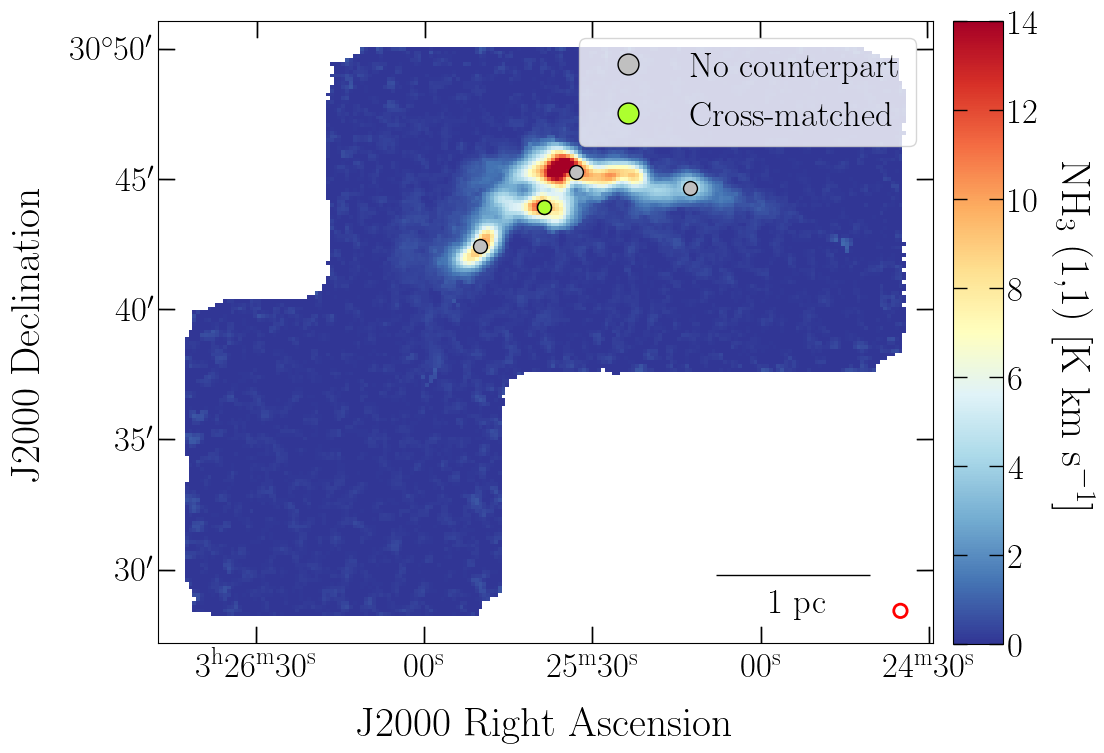}
    \caption{Analogous to Figure \ref{fig:b1_cores} but for the Perseus L1448 region. In this region, $1$ of the $4$ cores identified in GAS were uniquely cross-matched with a continuum counterpart in the HGBS data.}
    \label{fig:l1448_cores}
\end{figure*}

\begin{figure*}
    \centering
    \includegraphics[width=0.48\textwidth]{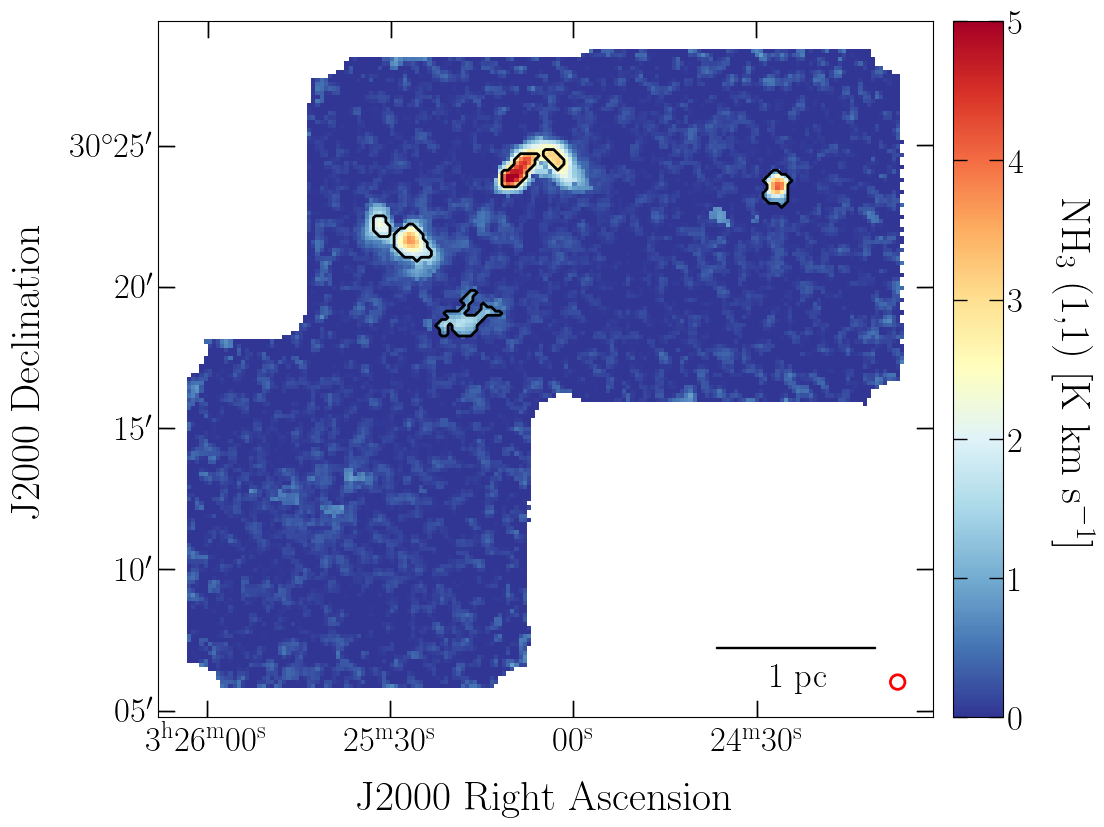}
    \includegraphics[width=0.48\textwidth]{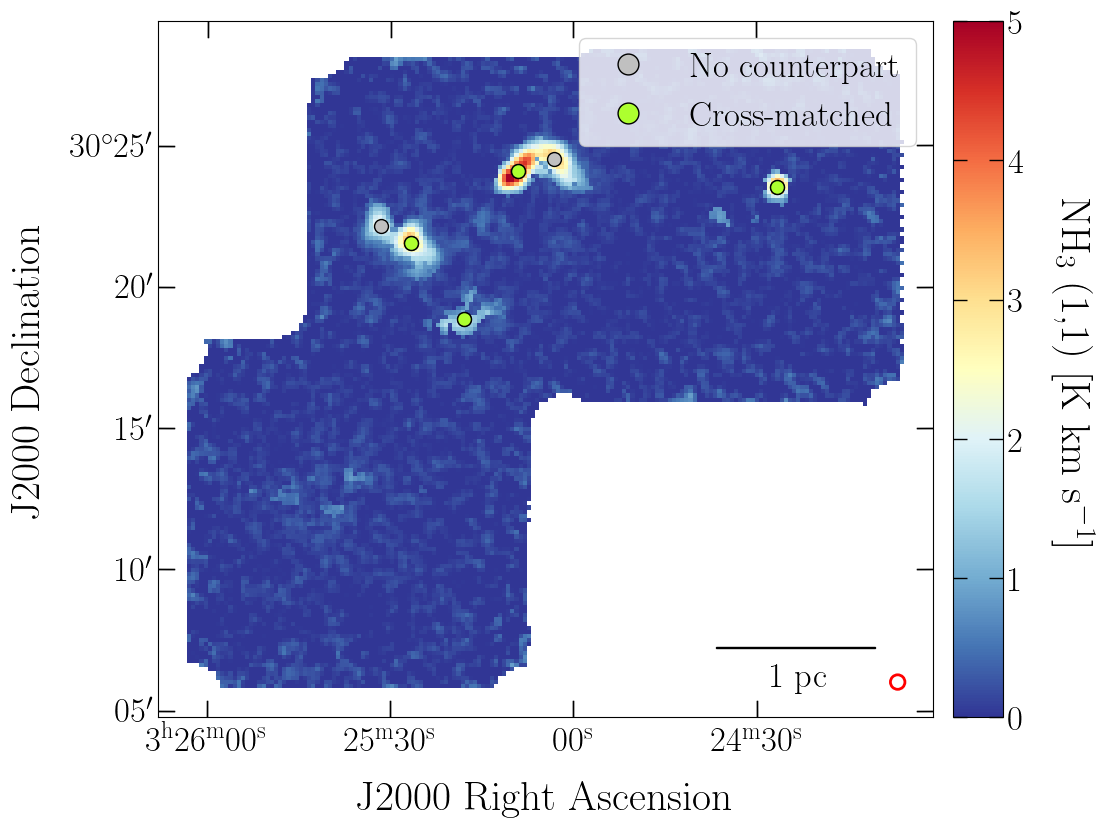}
    \caption{Analogous to Figure \ref{fig:b1_cores} but for the Perseus L1451 region. In this region, $4$ of the $6$ cores identified in GAS were uniquely cross-matched with a continuum counterpart in the HGBS data.}
    \label{fig:l1451_cores}
\end{figure*}

\begin{figure*}
    \centering
    \includegraphics[width=0.48\textwidth]{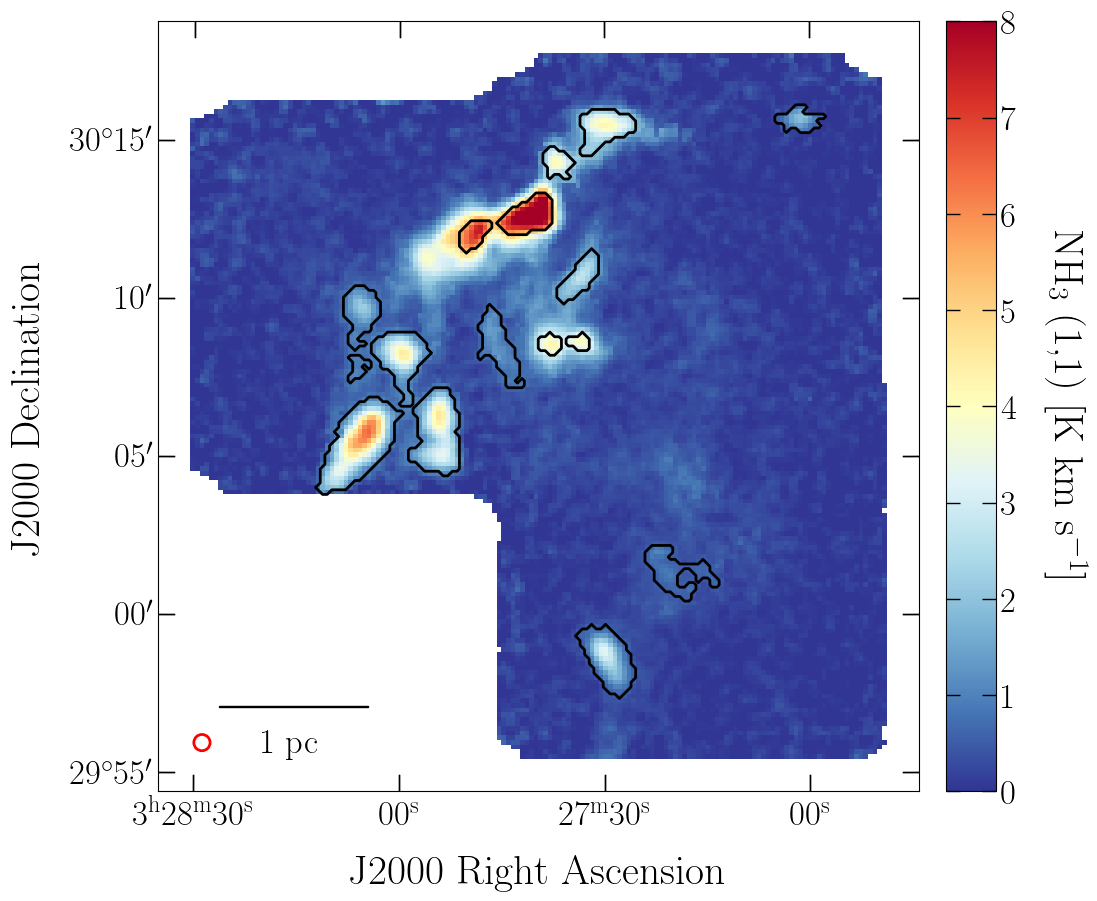}
    \includegraphics[width=0.48\textwidth]{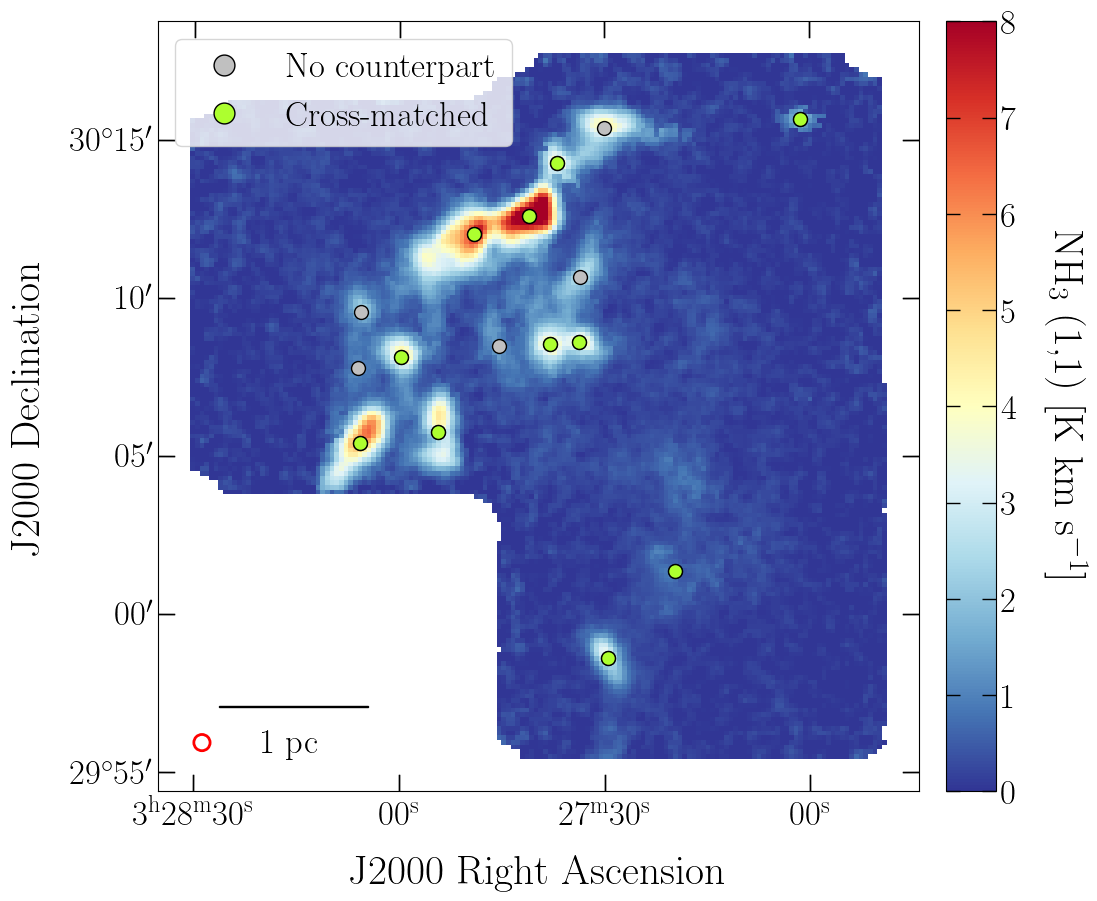}
    \caption{Analogous to Figure \ref{fig:b1_cores} but for the Perseus L1455 region. In this region, $11$ of the $16$ cores identified in GAS were uniquely cross-matched with a continuum counterpart in the HGBS data.}
    \label{fig:l1455_cores}
\end{figure*}

\begin{figure*}
    \centering
    \includegraphics[width=0.48\textwidth]{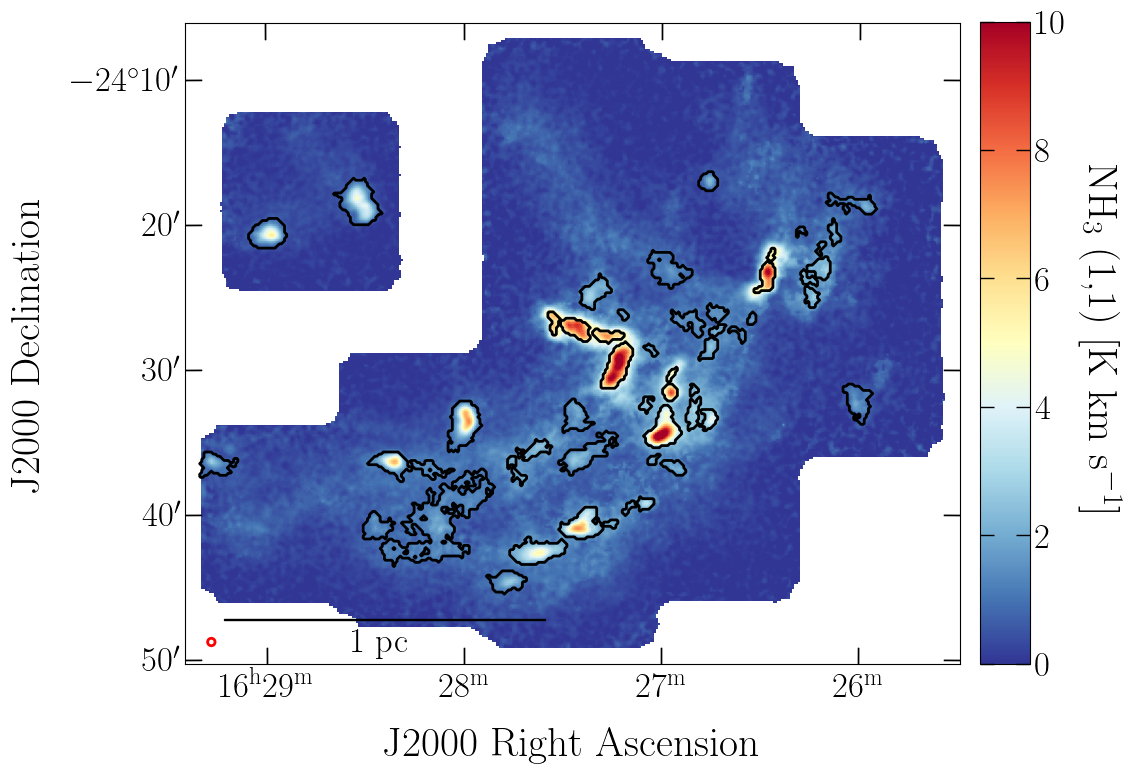}
    \includegraphics[width=0.48\textwidth]{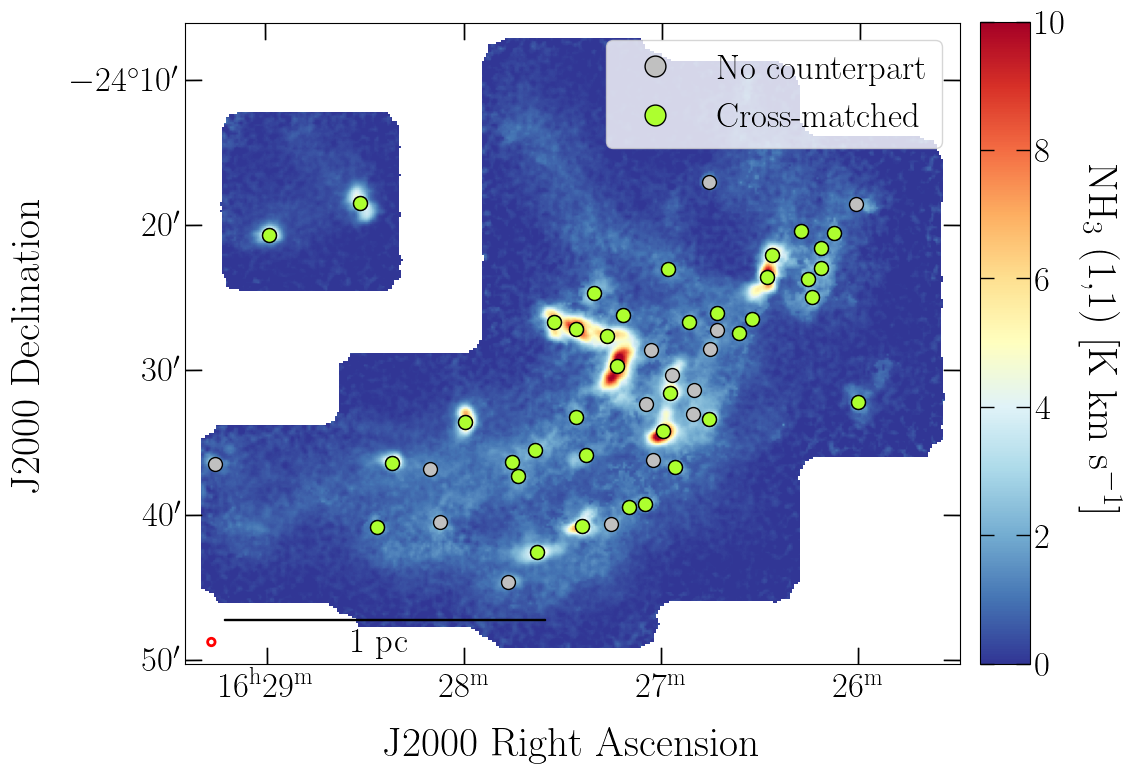}
    \caption{Analogous to Figure \ref{fig:b1_cores} but for the Ophiuchus L1688 region. In this region, $38$ of the $53$ cores identified in GAS were uniquely cross-matched with a continuum counterpart in the HGBS data.}
    \label{fig:l1688_cores}
\end{figure*}

\begin{figure*}
    \centering
    \includegraphics[width=0.48\textwidth]{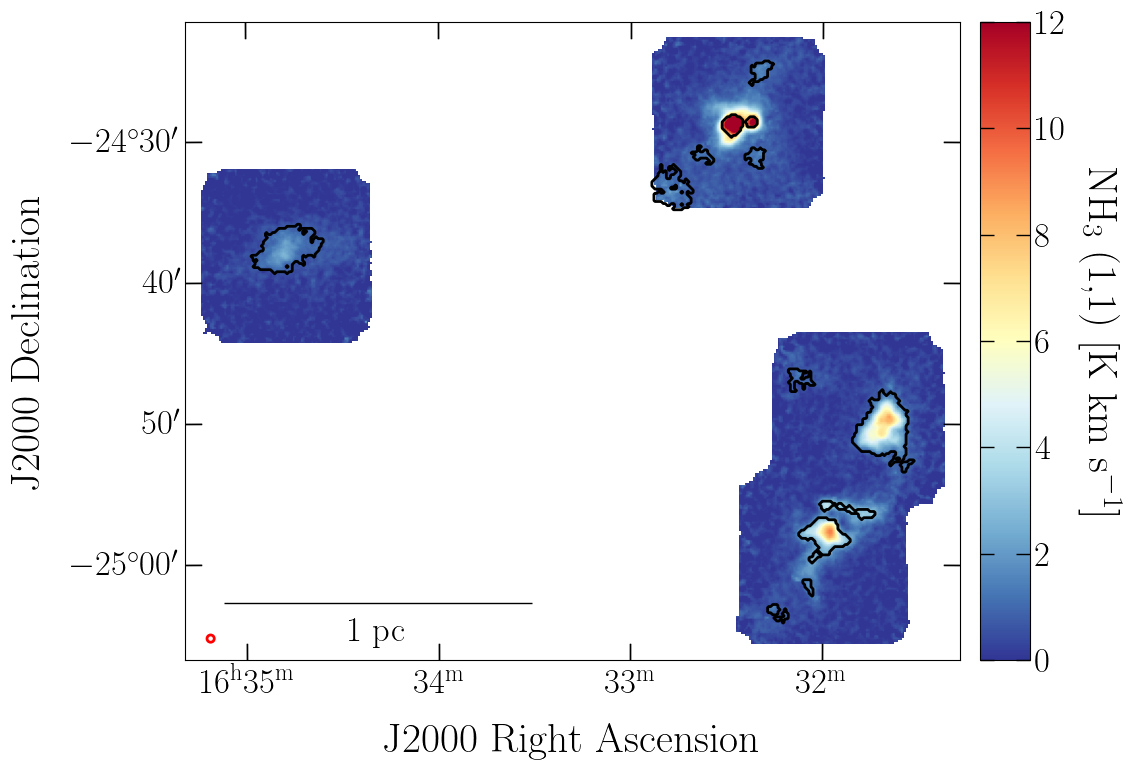}
    \includegraphics[width=0.48\textwidth]{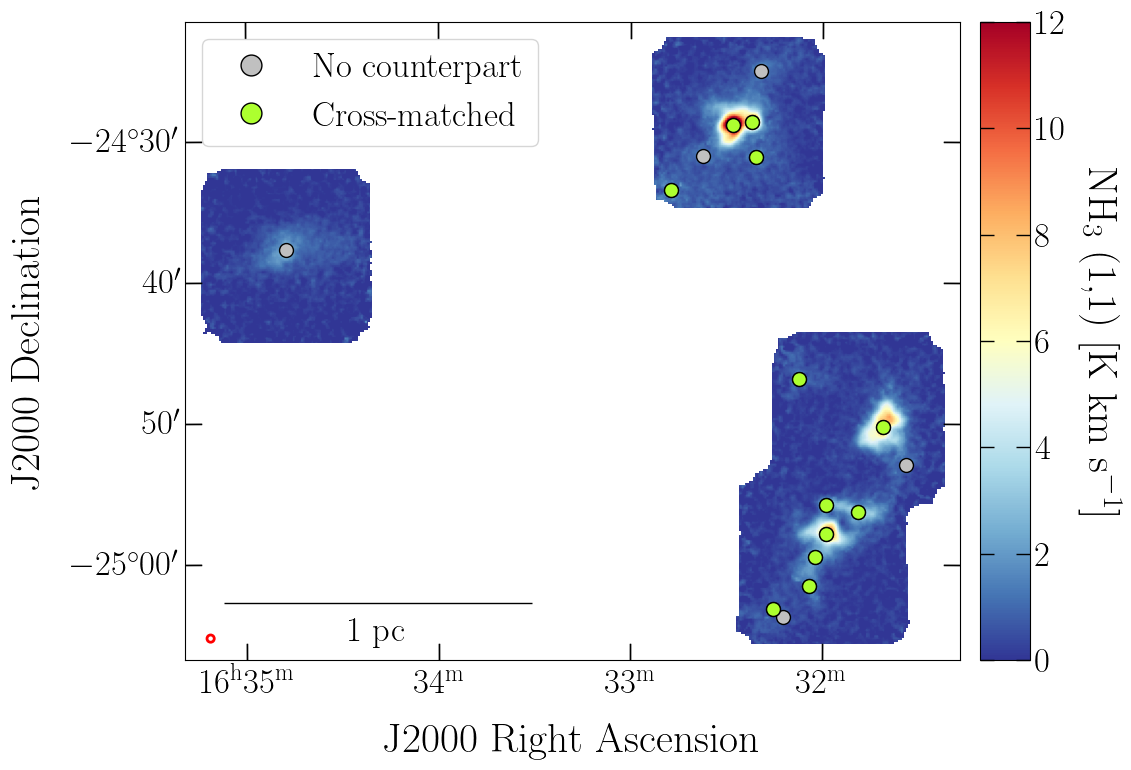}
    \caption{Analogous to Figure \ref{fig:b1_cores} but for the Ophiuchus L1689 region. In this region, $12$ of the $17$ cores identified in GAS were uniquely cross-matched with a continuum counterpart in the HGBS data.}
    \label{fig:l1689_cores}
\end{figure*}

\begin{figure*}
    \centering
    \includegraphics[width=0.48\textwidth]{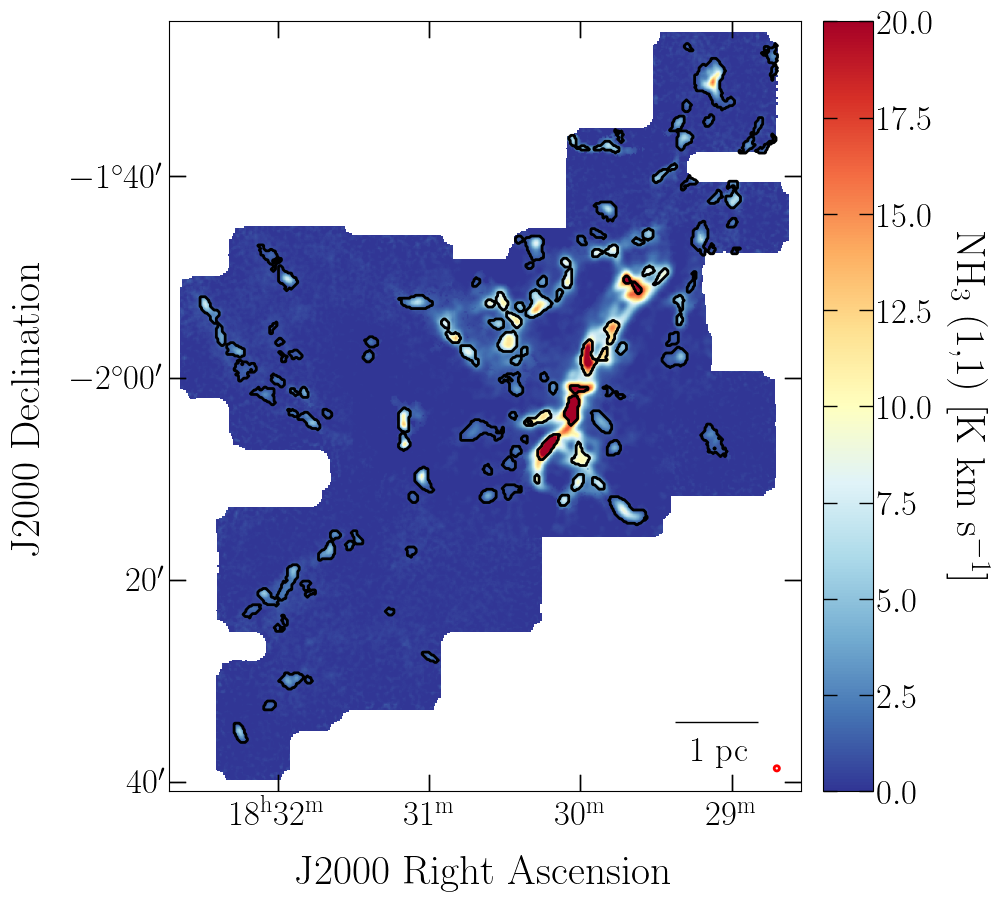}
    \includegraphics[width=0.48\textwidth]{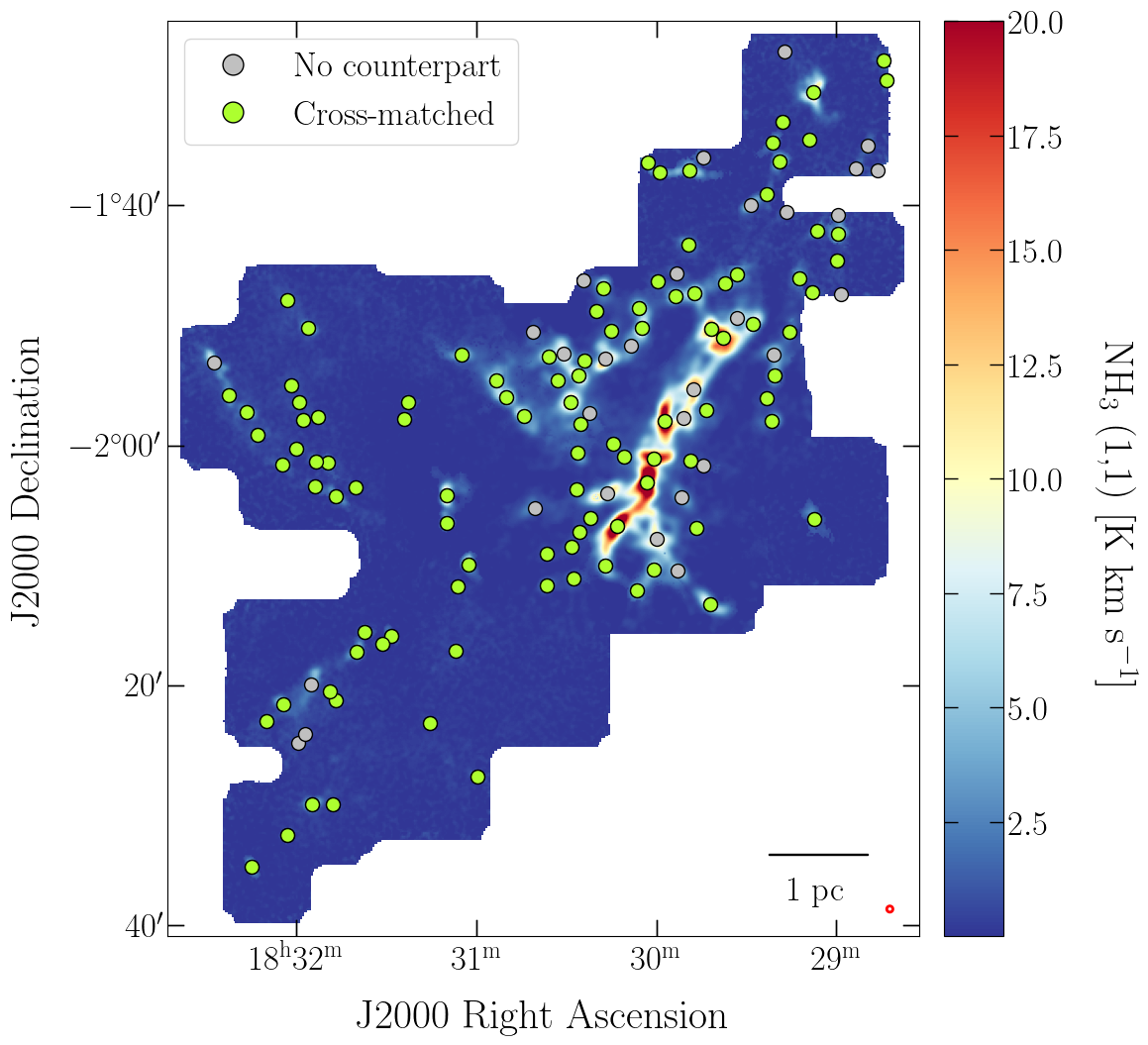}
    \caption{Analogous to Figure \ref{fig:b1_cores} but for the Serpens W40 region. In this region, $103$ of the $133$ cores identified in GAS were uniquely cross-matched with a continuum counterpart in the HGBS data.}
    \label{fig:w40_cores}
\end{figure*}

\begin{figure*}
    \centering
    \includegraphics[width=0.48\textwidth]{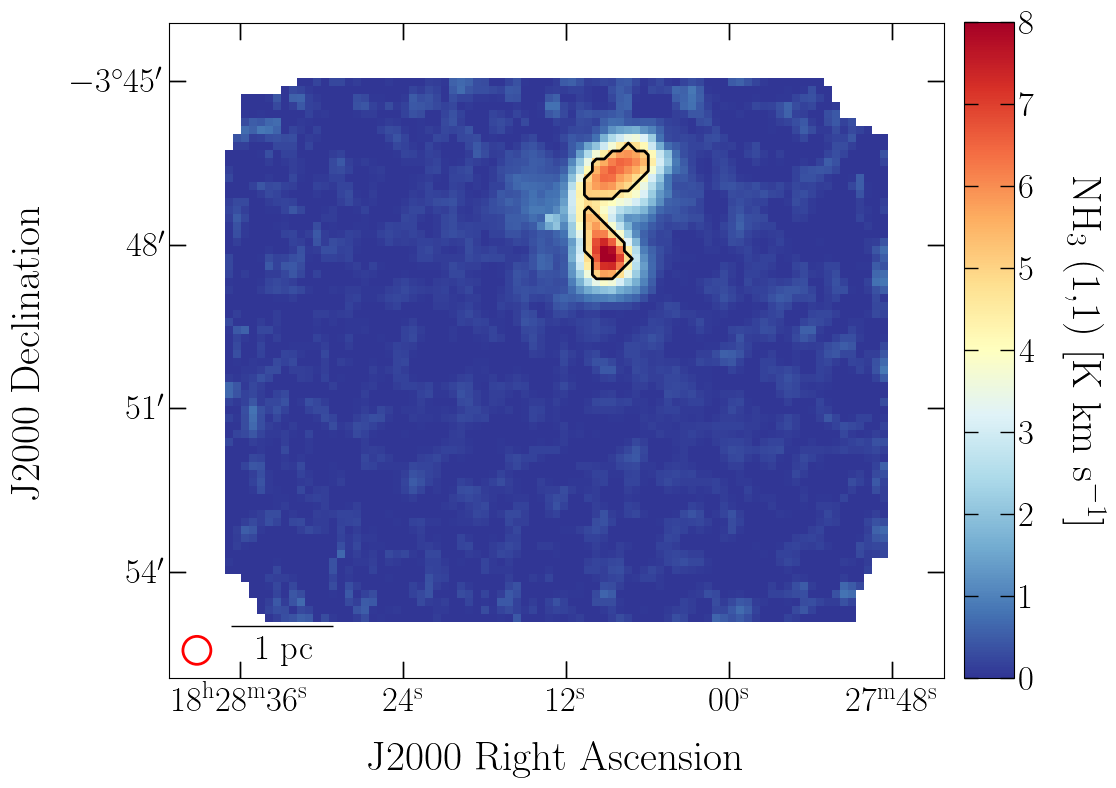}
    \includegraphics[width=0.48\textwidth]{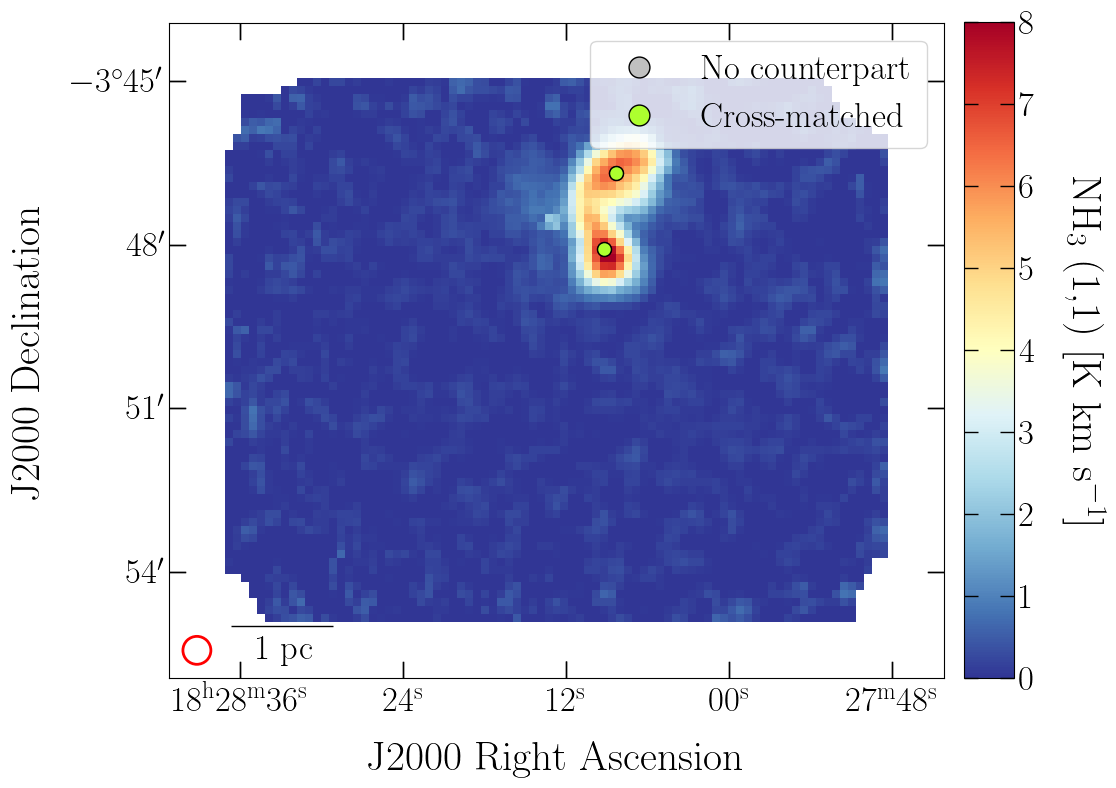}
    \caption{Analogous to Figure \ref{fig:b1_cores} but for the Serpens MWC297 region. In this region, both of the cores identified in GAS were uniquely cross-matched with a continuum counterpart in the HGBS data.}
    \label{fig:mwc297_cores}
\end{figure*}

\begin{figure*}
    \centering
    \includegraphics[width=0.48\textwidth]{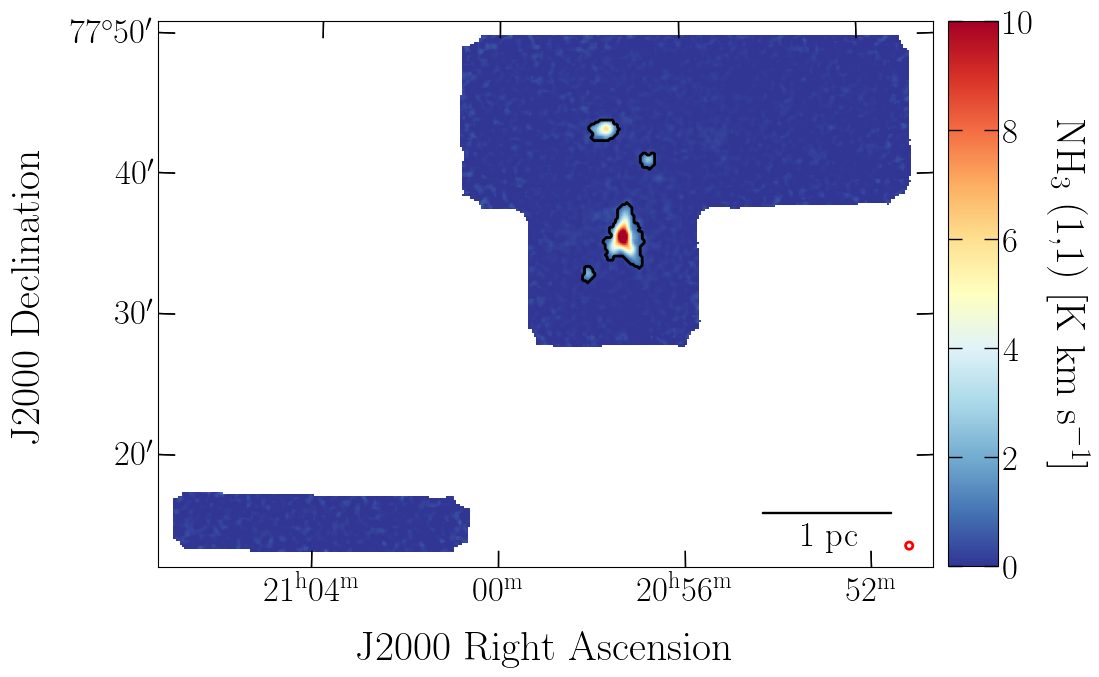}
    \includegraphics[width=0.48\textwidth]{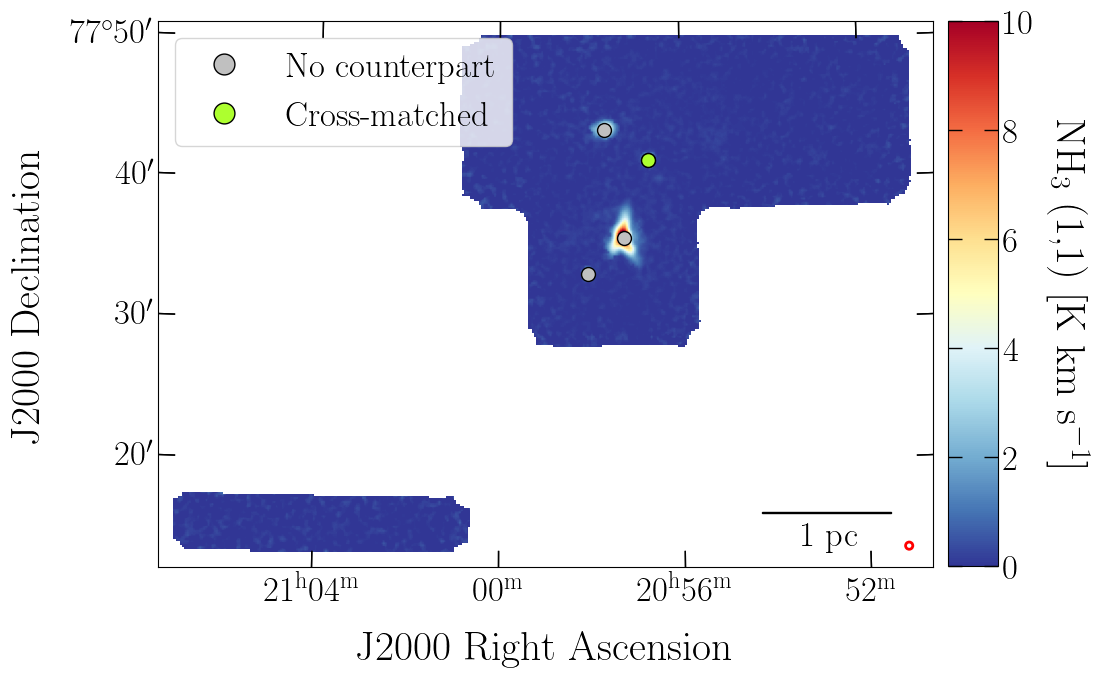}
    \caption{Analogous to Figure \ref{fig:b1_cores} but for the Cepheus L1228 region. In this region, $1$ of the $4$ cores identified in GAS were uniquely cross-matched with a continuum counterpart in the HGBS data.}
    \label{fig:l1228_cores}
\end{figure*}

\begin{figure*}
    \centering
    \includegraphics[width=0.48\textwidth]{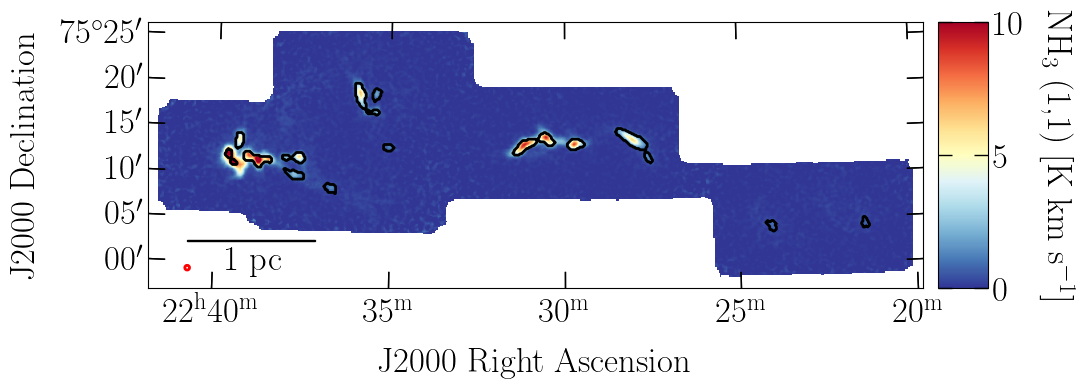}
    \includegraphics[width=0.48\textwidth]{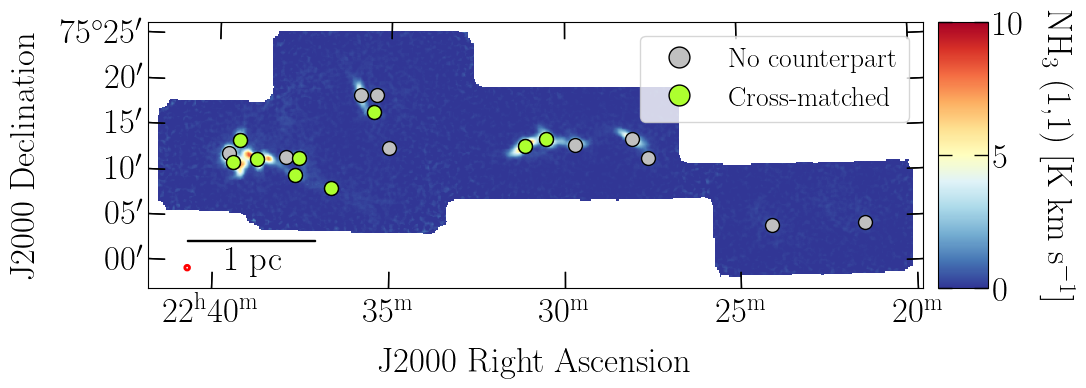}
    \caption{Analogous to Figure \ref{fig:b1_cores} but for the Cepheus L1251 region. In this region, $9$ of the $19$ cores identified in GAS were uniquely cross-matched with a continuum counterpart in the HGBS data.}
    \label{fig:l1251_cores}
\end{figure*}

\begin{figure*}
    \centering
    \includegraphics[width=0.48\textwidth]{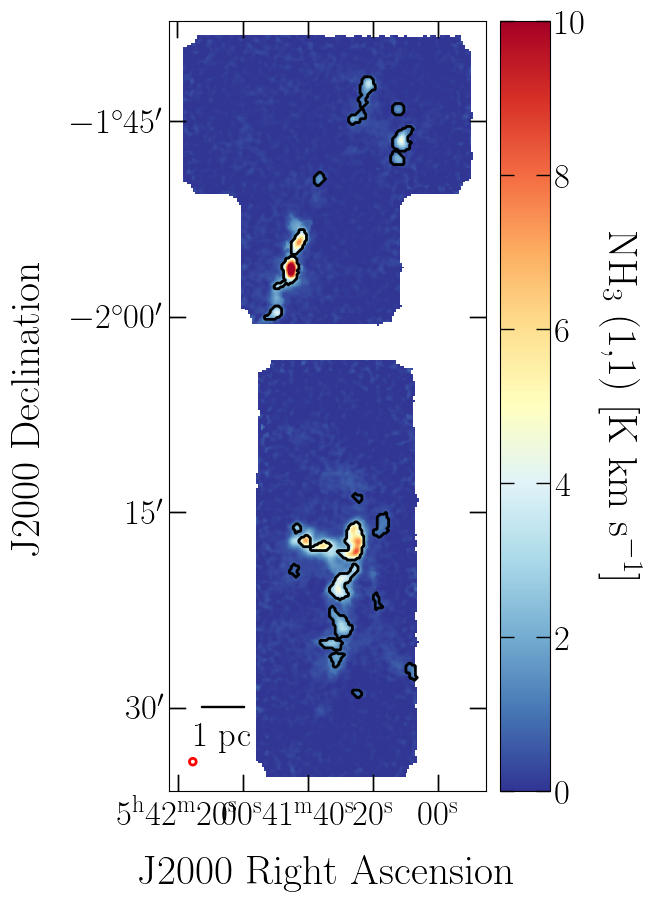}
    \includegraphics[width=0.48\textwidth]{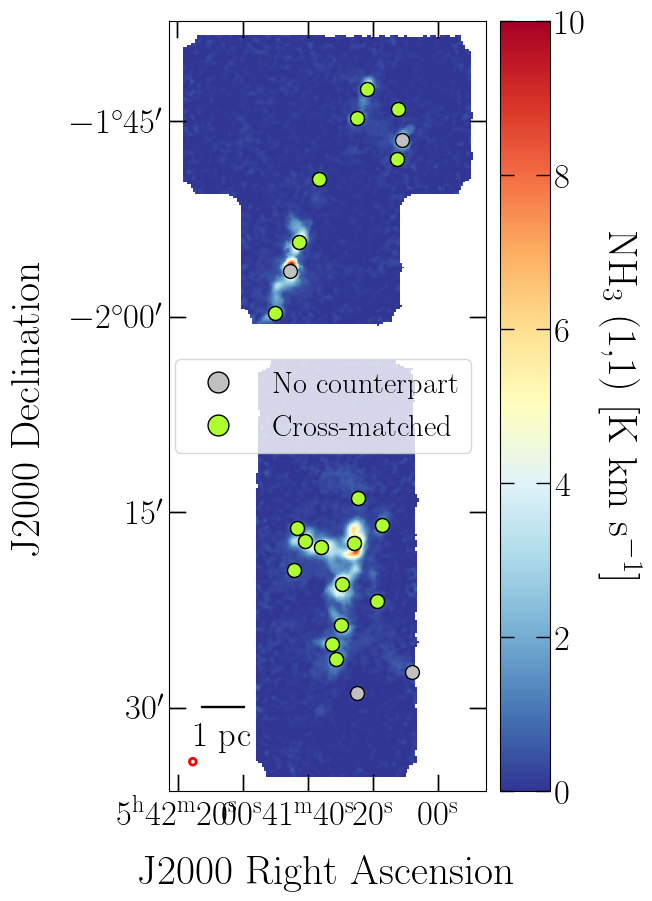}
    \caption{Analogous to Figure \ref{fig:b1_cores} but for the Orion B NGC2023 region. In this region, $19$ of the $23$ cores identified in GAS were uniquely cross-matched with a continuum counterpart in the HGBS data.}
    \label{fig:ngc2023_cores}
\end{figure*}

\begin{figure*}
    \centering
    \includegraphics[width=0.48\textwidth]{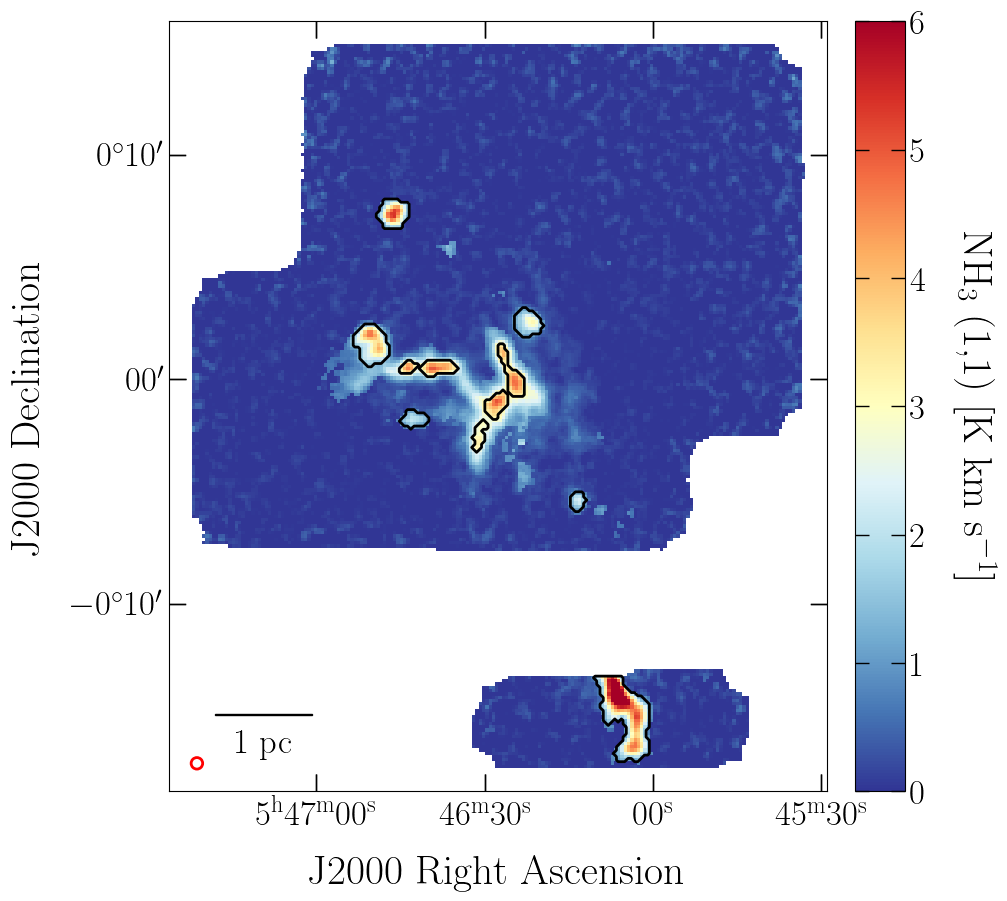}
    \includegraphics[width=0.48\textwidth]{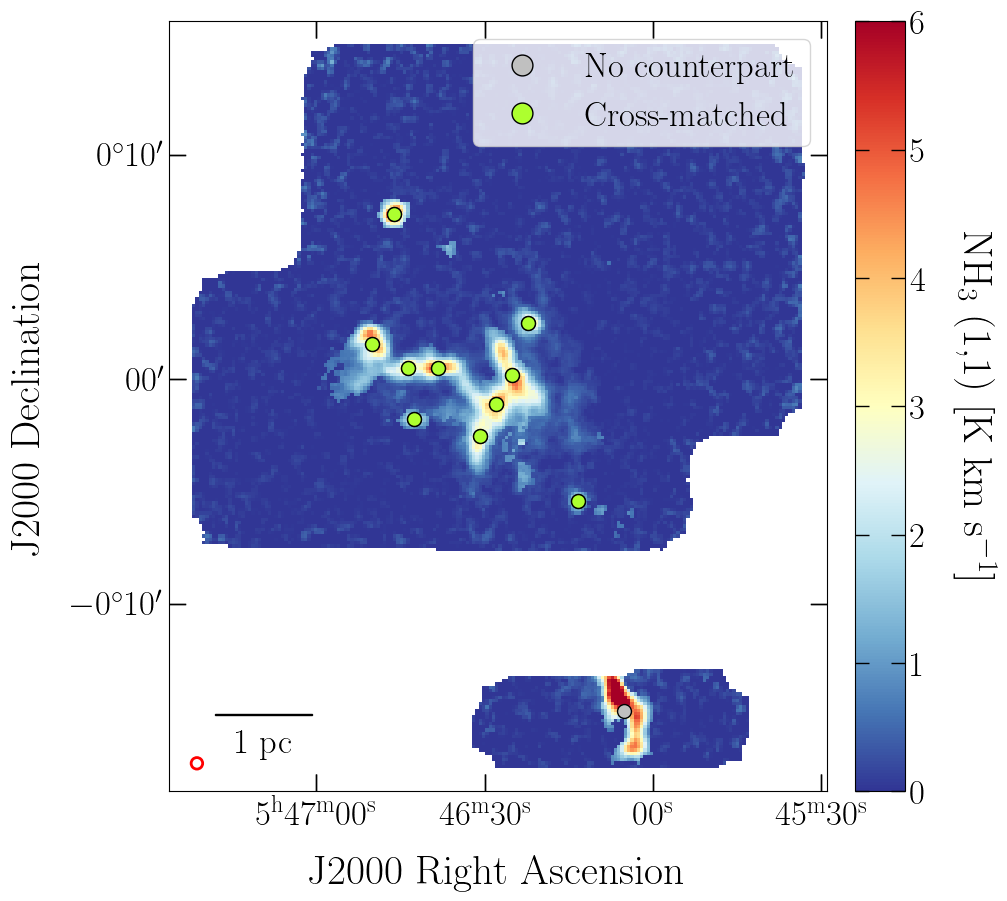}
    \caption{Analogous to Figure \ref{fig:b1_cores} but for the Orion B NGC2068 region. In this region, $10$ of the $11$ cores identified in GAS were uniquely cross-matched with a continuum counterpart in the HGBS data.}
    \label{fig:ngc2068_cores}
\end{figure*}

\begin{figure*}
    \centering
    \includegraphics[width=0.48\textwidth]{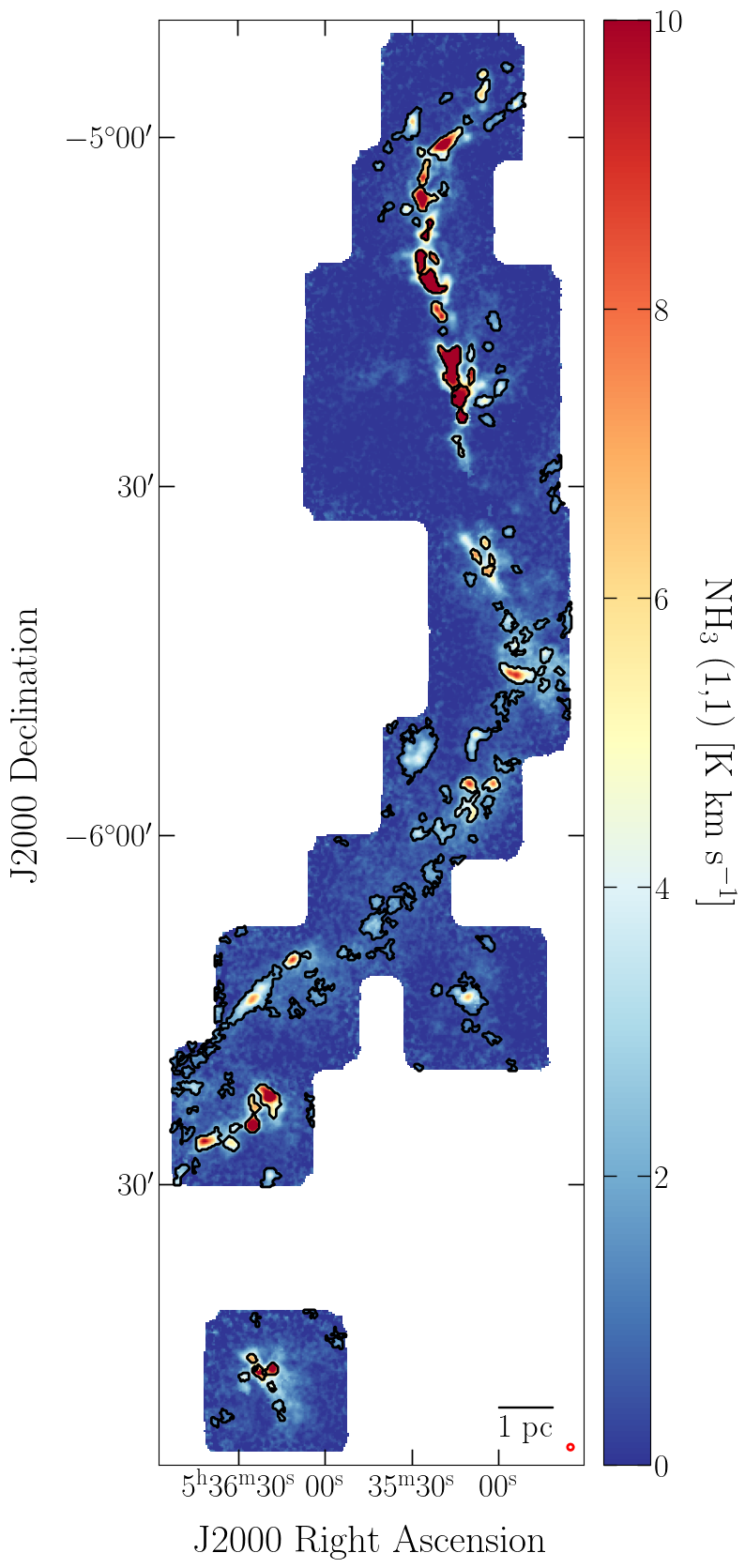}
    \includegraphics[width=0.48\textwidth]{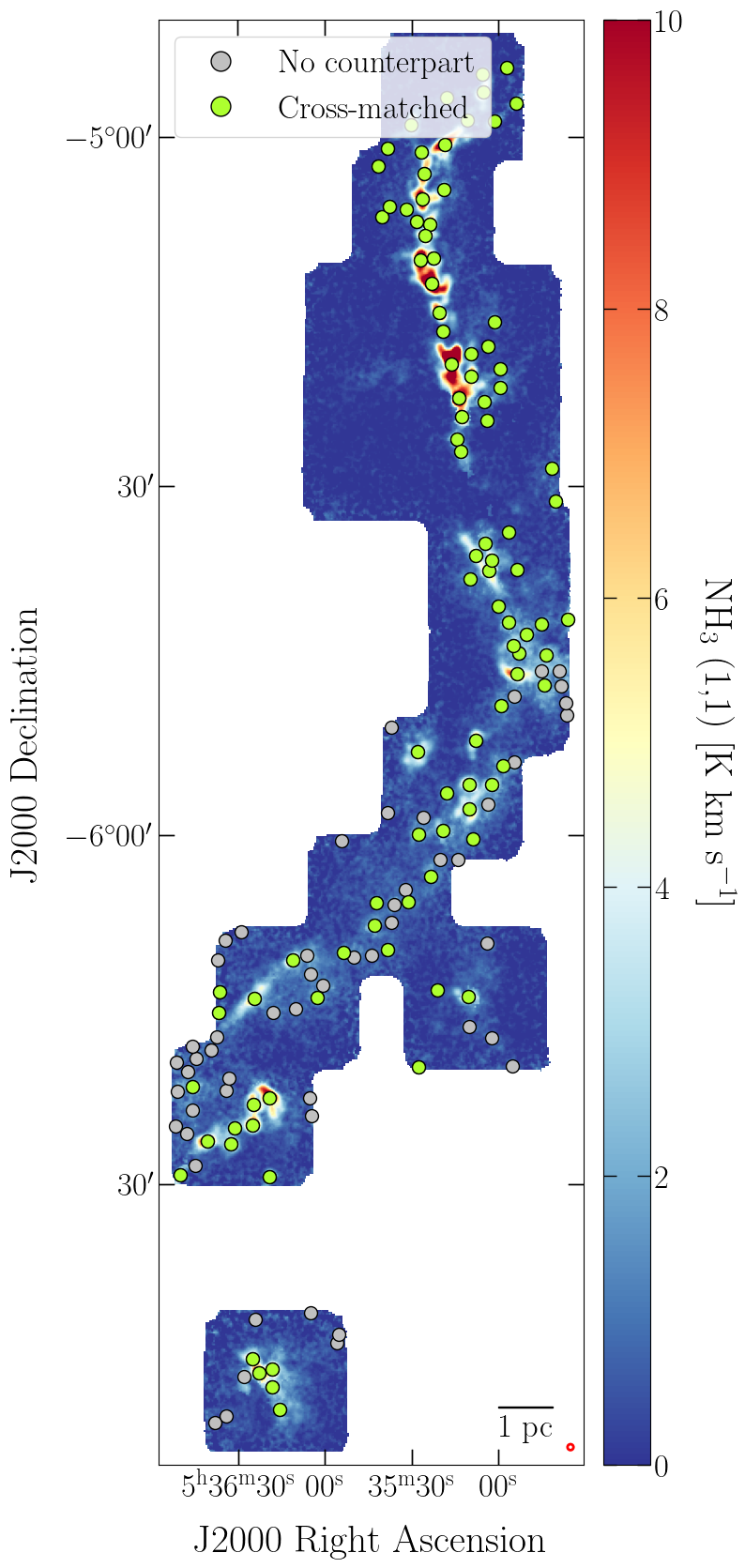}
    \caption{Analogous to Figure \ref{fig:b1_cores} but for the Orion A region. In this region, $97$ of the $150$ cores identified in GAS were uniquely cross-matched with a continuum counterpart in the JCMT GBS data.}
    \label{fig:oriona_cores}
\end{figure*}

\begin{figure*}
    \centering
    \includegraphics[width=0.48\textwidth]{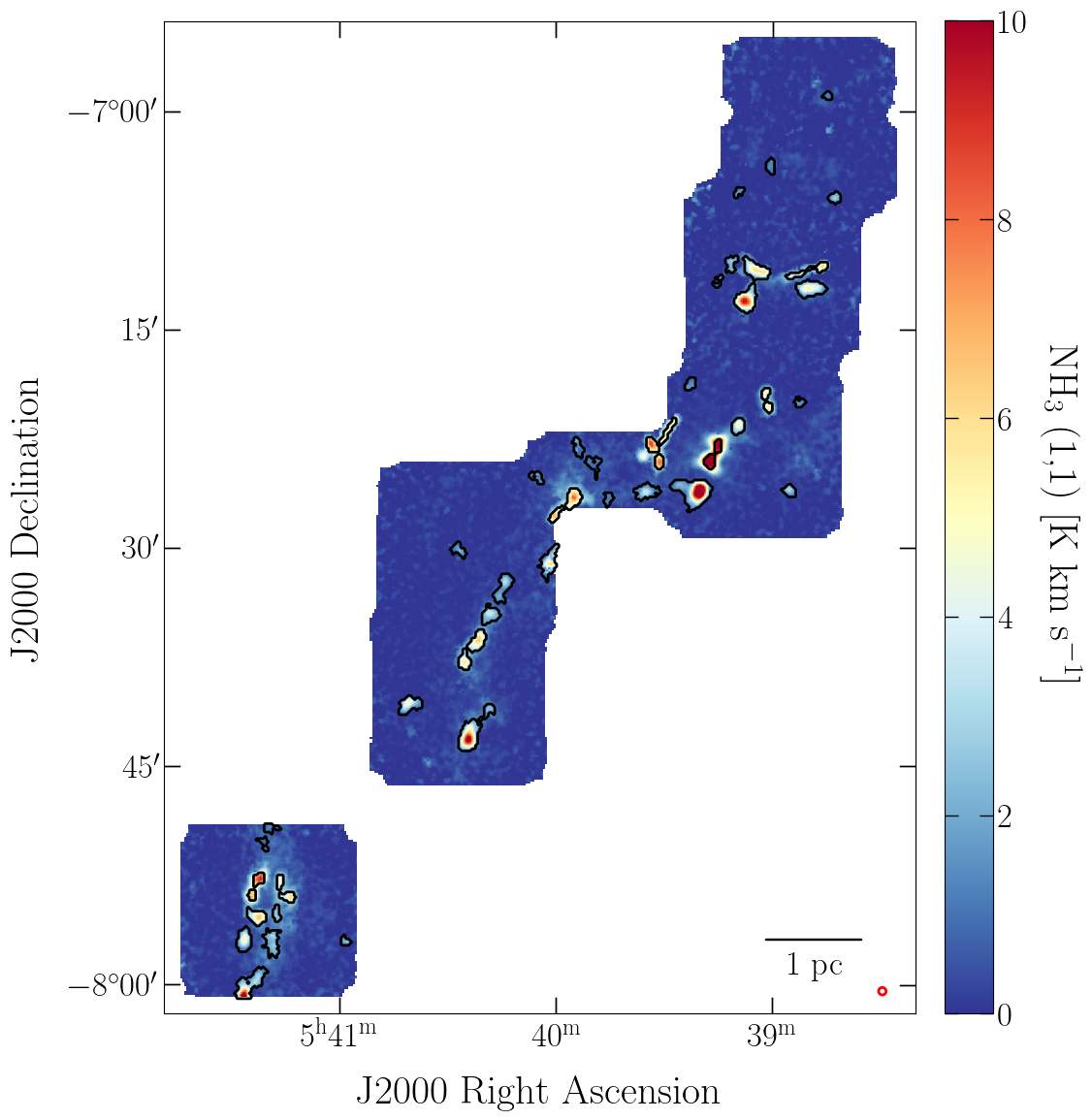}
    \includegraphics[width=0.48\textwidth]{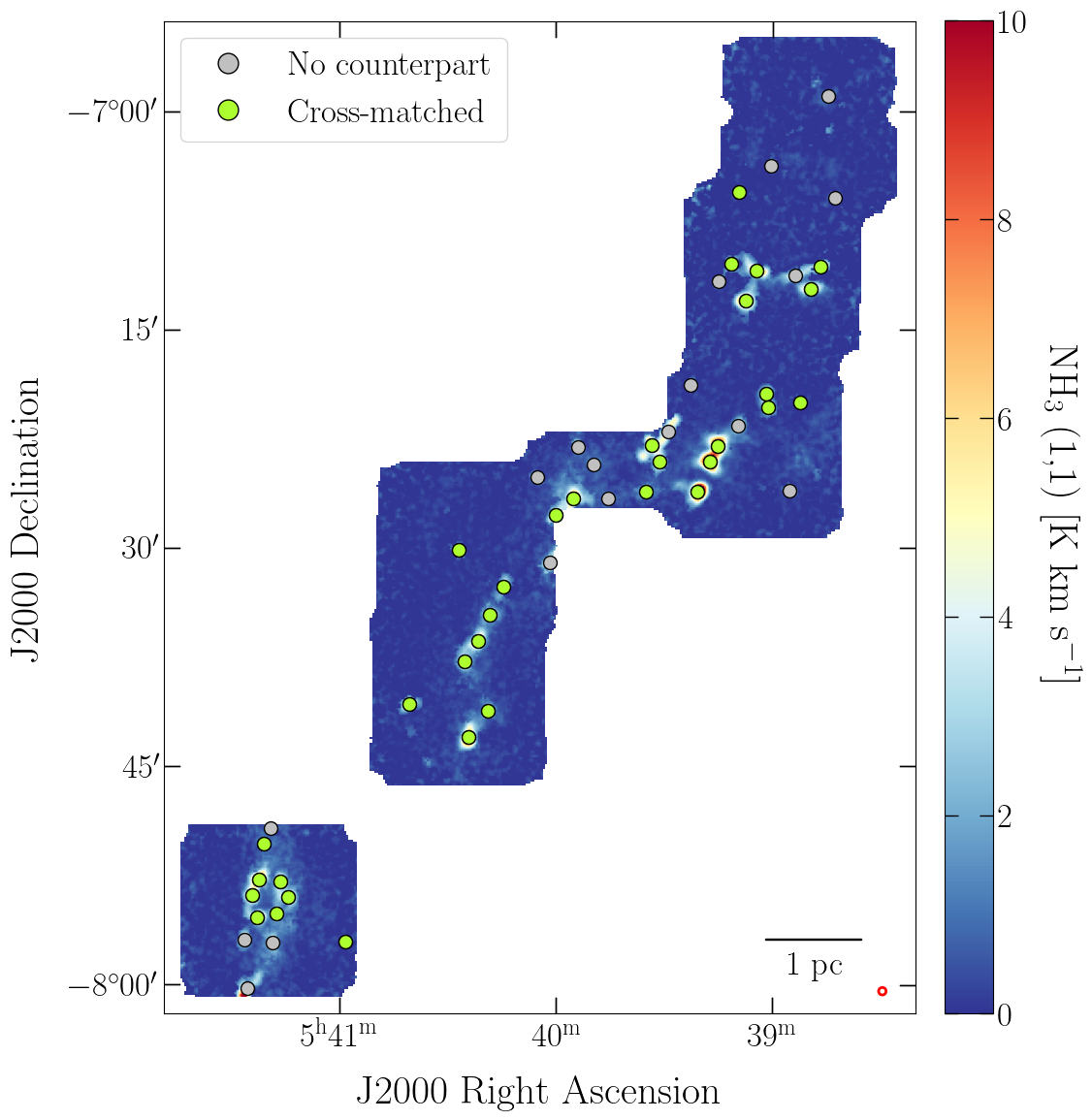}
    \caption{Analogous to Figure \ref{fig:b1_cores} but for the Orion A South region. In this region, $33$ of the $51$ cores identified in GAS were uniquely cross-matched with a continuum counterpart in the JCMT GBS data.}
    \label{fig:orionas_cores}
\end{figure*}

In Figures \ref{fig:b1e_cores}$-$\ref{fig:orionas_cores}, we present the results of the GAS NH$_3$ (1,1) integrated intensity maps for all $17$ regions listed in Table \ref{tb:cross_match}, as well as the {\tt astrodendro} core classification and continuum cross-matching results. Each figure in this section presents a pair of plots formatted in the same manner as Figure \ref{fig:b1_cores}. The left plot shows {\tt astrodendro} \textit{leaves} as black contours and the plot on the right indicates which GAS cores were successfully (as green circles) or unsuccessfully (as grey circles) cross-matched with a core in the corresponding continuum data. The 32$''$ GAS beam is presented in each figure as a red ellipse and a scale bar depicting the angular size of $1$~pc at the distance of the respective region is provided.

\end{document}